\documentclass[11pt]{article}

\newcommand{\be}{\begin{equation}}
\newcommand{\ee}{\end{equation}}
\usepackage{amsmath}
\usepackage{amssymb}
\usepackage{latexsym}
\usepackage{epsfig}

\newcommand{\half}{{\textstyle {1\over 2}}}
\newcommand{\quarter}{{\textstyle{1\over 4}}}

\newcommand{\bra}[1]{\langle #1|}
\newcommand{\ket}[1]{|#1\rangle}

  \newcommand{\DDD}{{\overleftrightarrow{D}}}
   \newcommand{\DDDB}{{\overleftrightarrow{\bar D}}}
  \newcommand{\ppp}{{\overleftrightarrow{\partial}}}

\newcommand{\RRR}{{\hbox{\rm R\kern-2.35mm R}}}
\newcommand{\p}{\partial}

\def\ZZZ{{\hbox{ Z\kern-1.6mm Z}}}

\newcommand{\nin}[1] {\underline{\phantom{h}}\hskip-6pt {#1}}

\newcommand{\sectiono}[1]{\section{#1}\setcounter{equation}{0}}

\def\oonoo#1#2#3{\vbox{\ialign{##\crcr
	\hfil\hfil\hfil{$#3{#1}$}\hfil\crcr\noalign{\kern1pt\nointerlineskip}
	$#3{#2}$\crcr}}}
\def\oon#1#2{\mathchoice{\oonoo{#1}{#2}{\displaystyle}}
	{\oonoo{#1}{#2}{\textstyle}}{\oonoo{#1}{#2}{\scriptstyle}}
	{\oonoo{#1}{#2}{\scriptscriptstyle}}}
\def\dt#1{\oon{\hbox{\bf .}}{#1}}  
\def\ddt#1{\oon{\hbox{\bf .\kern-1pt.}}#1}    
\def\slap#1#2{\setbox0=\hbox{$#1{#2}$}
	#2\kern-\wd0{\hfuzz=1pt\hbox to\wd0{\hfil$#1{/}$\hfil}}}
\def\sla#1{\mathpalette\slap{#1}}                
\def\bop#1{\setbox0=\hbox{$#1M$}\mkern1.5mu
	\lower.02\ht0\vbox{\hrule height0pt depth.06\ht0
	\hbox{\vrule width.06\ht0 height.9\ht0 \kern.9\ht0
	\vrule width.06\ht0}\hrule height.06\ht0}\mkern1.5mu}
\def\bo{{\mathpalette\bop{}}}                        
\mathcode`\*="702A                  
\def\in{\relax\ifmmode\mathchar"3232\else{\refit in\/}\fi} 
\def\f#1#2{{\textstyle{#1\over#2}}}	   
\def\half{{\textstyle{1\over{\raise.1ex\hbox{$\scriptstyle{2}$}}}}}

\def\Gamma{\mathchar"0100}
\def\Delta{\mathchar"0101}
\def\Theta{\mathchar"0102}
\def\Lambda{\mathchar"0103}
\def\Xi{\mathchar"0104}
\def\Pi{\mathchar"0105}
\def\Sigma{\mathchar"0106}
\def\Upsilon{\mathchar"0107}
\def\Phi{\mathchar"0108}
\def\Psi{\mathchar"0109}
\def\Omega{\mathchar"010A}

\catcode128=13 \def €{\"A}                 
\catcode129=13 \def {\AA}                 
\catcode130=13 \def '{\c}           	   
\catcode131=13 \def ƒ{\'E}                   
\catcode132=13 \def "{\~N}                   
\catcode133=13 \def …{\"O}                 
\catcode134=13 \def †{\"U}                  
\catcode135=13 \def ‡{\'a}                  
\catcode136=13 \def ˆ{\`a}                   
\catcode137=13 \def ‰{\^a}                 
\catcode138=13 \def Š{\"a}                 
\catcode139=13 \def ‹{\~a}                   
\catcode140=13 \def Œ{\alpha}            
\catcode141=13 \def {\chi}                
\catcode142=13 \def Ž{\'e}                   
\catcode143=13 \def {\`e}                    
\catcode144=13 \def {\^e}                  
\catcode145=13 \def '{\"e}                
\catcode146=13 \def '{\'\i}                 
\catcode147=13 \def "{\`\i}                  
\catcode148=13 \def "{\^\i}                
\catcode149=13 \def •{\"\i}                
\catcode150=13 \def –{\~n}                  
\catcode151=13 \def —{\'o}                 
\catcode152=13 \def ˜{\`o}                  
\catcode153=13 \def ™{\^o}                
\catcode154=13 \def š{\"o}                 
\catcode155=13 \def ›{\~o}                  
\catcode156=13 \def œ{\'u}                  
\catcode157=13 \def {\`u}                  
\catcode158=13 \def ž{\^u}                
\catcode159=13 \def Ÿ{\"u}                
\catcode160=13 \def  {\tau}               
\catcode162=13 \def ¢{\oplus}           
\catcode163=13 \def £{\relax\ifmmode\to\else\itemize\fi} 
\catcode164=13 \def ¤{\subset}	  
\catcode165=13 \def ¥{\infty}           
\catcode166=13 \def ¦{\mp}                
\catcode167=13 \def §{\sigma}           
\catcode168=13 \def ¨{\rho}               
\catcode169=13 \def ©{\gamma}         
\catcode170=13 \def ª{\leftrightarrow} 
\catcode171=13 \def «{\relax\ifmmode\expandafter\acute\else\expandafter\'\fi}
\catcode172=13 \def ¬{\relax\ifmmode\expandafter\ddt\else\expandafter\"\fi}
\catcode173=13 \def ­{\equiv}            
\catcode174=13 \def ®{\approx}          
\catcode175=13 \def ¯{\Omega}          
\catcode176=13 \def °{\otimes}          
\catcode177=13 \def ±{\ne}                 
\catcode178=13 \def ²{\le}                   
\catcode179=13 \def ³{\ge}                  
\catcode180=13 \def ´{\upsilon}          
\catcode181=13 \def µ{\mu}                
\catcode182=13 \def ¶{\delta}             
\catcode183=13 \def ·{\epsilon}          
\catcode184=13 \def ¸{\Pi}                  
\catcode185=13 \def ¹{\pi}                  
\catcode186=13 \def º{\beta}               
\catcode187=13 \def »{\partial}           
\catcode188=13 \def ¼{\nobreak\ }       
\catcode189=13 \def ½{\zeta}               
\catcode190=13 \def ¾{\sim}                 
\catcode191=13 \def ¿{\omega}           
\catcode192=13 \def À{\dt}                     
\catcode193=13 \def Á{\gets}                
\catcode194=13 \def Â{\lambda}           
\catcode195=13 \def Ã{\nu}                   
\catcode196=13 \def Ä{\phi}                  
\catcode197=13 \def Å{\xi}                     
\catcode198=13 \def Æ{\psi}                  
\catcode199=13 \def Ç{\int}                    
\catcode200=13 \def È{\oint}                 
\catcode201=13 \def É{\relax\ifmmode\cdot\else\vol\fi}    
\catcode202=13 \def Ê{\relax\ifmmode\,\else\thinspace\fi}
\catcode203=13 \def Ë{\`A}                      
\catcode204=13 \def Ì{\~A}                      
\catcode205=13 \def Í{\~O}                      
\catcode206=13 \def Î{\Theta}              
\catcode207=13 \def Ï{\theta}               
\catcode208=13 \def Ð{\relax\ifmmode\expandafter\bar\else\expandafter\=\fi}
\catcode209=13 \def Ñ{\overline}             
\catcode210=13 \def Ò{\langle}               
\catcode211=13 \def Ó{\relax\ifmmode\{\else\ital\fi}      
\catcode212=13 \def Ô{\rangle}               
\catcode213=13 \def Õ{\}}                        
\catcode214=13 \def Ö{\sla}                      
\catcode215=13 \def ×{\relax\ifmmode\expandafter\check\else\expandafter\v\fi}
\catcode216=13 \def Ø{\"y}                     
\catcode217=13 \def Ù{\"Y}  		    
\catcode218=13 \def Ú{\Leftarrow}       
\catcode219=13 \def Û{\Leftrightarrow}       
\catcode220=13 \def Ü{\relax\ifmmode\Rightarrow\else\sect\fi}
\catcode221=13 \def Ý{\sum}                  
\catcode222=13 \def Þ{\prod}                 
\catcode223=13 \def ß{\widehat}              
\catcode224=13 \def à{\pm}                     
\catcode225=13 \def á{\nabla}                
\catcode226=13 \def â{\quad}                 
\catcode227=13 \def ã{\in}               	
\catcode228=13 \def ä{\star}      	      
\catcode229=13 \def å{\sqrt}                   
\catcode230=13 \def æ{\^E}			
\catcode231=13 \def ç{\Upsilon}              
\catcode232=13 \def è{\"E}    	   	 
\catcode233=13 \def é{\`E}               	  
\catcode234=13 \def ê{\Sigma}                
\catcode235=13 \def ë{\Delta}                 
\catcode236=13 \def ì{\Phi}                     
\catcode237=13 \def í{\`I}        		   
\catcode238=13 \def î{\iota}        	     
\catcode239=13 \def ï{\Psi}                     
\catcode240=13 \def ð{\times}                  
\catcode241=13 \def ñ{\Lambda}             
\catcode242=13 \def ò{\cdots}                
\catcode243=13 \def ó{\^U}			
\catcode244=13 \def ô{\`U}    	              
\catcode245=13 \def õ{\bo}                       
\catcode246=13 \def ö{\relax\ifmmode\expandafter\hat\else\expandafter\^\fi}
\catcode247=13 \def÷{\relax\ifmmode\expandafter\tilde\else\expandafter\~\fi}
\catcode248=13 \def ø{\ll}                         
\catcode249=13 \def ù{\gg}                       
\catcode250=13 \def ú{\eta}                      
\catcode251=13 \def û{\kappa}                  
\catcode252=13 \def ü{\half}     		 
\catcode253=13 \def ý{\Gamma} 		
\catcode254=13 \def þ{\Xi}   			
\catcode255=13 \def ÿ{\relax\ifmmode{}^{\dagger}{}\else\dag\fi}

      \def\H{{\cal H}}   
      \def\M{{\cal M}}

\def\Ä{\varphi}  \def\¿{\varpi}	\def\Ï{\vartheta}

\def\Ç{\textstyle{Ç}}

\begin{document}

\begin{titlepage}
\rightline{July 2014} 
\rightline{\tt MIT-CTP-4566}    
\begin{center}
\vskip 2.5cm
{\Large \bf {Double Field Theory at Order $\alpha'$}}\\

 \vskip 2.0cm
{\large {Olaf Hohm and Barton Zwiebach}}
\vskip 0.5cm
{\it {${}^1$Center for Theoretical Physics}}\\
{\it {Massachusetts Institute of Technology}}\\
{\it {Cambridge, MA 02139, USA}}\\
ohohm@mit.edu, zwiebach@mit.edu

\vskip 2.5cm
{\bf Abstract}

\end{center}

\vskip 0.5cm

\noindent
\begin{narrower}

\baselineskip15pt
We investigate $\alpha'$ corrections of bosonic strings in the framework 
of double field theory.  The previously introduced
``doubled $\alpha'$-geometry"  gives $\alpha'$-deformed 
gauge transformations arising in
the Green-Schwarz anomaly cancellation mechanism but
does not apply to bosonic strings.  These require a different deformation of the
duality-covariantized Courant bracket which governs the gauge structure.
This is revealed by examining the $\alpha'$ corrections in the gauge algebra of
 closed string field theory.
We construct a  four-derivative cubic  double field theory action  invariant
under the deformed gauge transformations, 
giving a first glimpse of the gauge principle    
 underlying bosonic string $\alpha'$ corrections.  
The usual metric and $b$-field 
are related to the
duality covariant fields  
by non-covariant field redefinitions.

\end{narrower}

\end{titlepage}

\newpage

\tableofcontents
\baselineskip=16pt

\bigskip



\sectiono{Introduction}

At low energy string theory is well described by supergravity.  
Stringy corrections beyond supergravity are captured 
by higher-derivative $\alpha'$ corrections.
While Einstein's gravity and  supergravity are 
well understood 
in terms of Riemannian geometry, we have no good understanding  
of the geometry of string theory or even of \textit{classical} string theory. 
Classical string theory includes $\alpha'$ corrections. Our goal in this paper is to better understand the geometry behind these
corrections. 

Concretely, we ask whether there is a symmetry explanation for higher-derivative
$\alpha'$ corrections, i.e., a symmetry principle that
{\em requires} $\alpha'$ corrections. We know of such symmetry principles 
in some cases; for instance, in heterotic string theory Green-Schwarz 
anomaly cancellation~\cite{Green:1984sg}  requires an ${\cal O}(\alpha')$ deformation of 
the gauge transformations of the $b$-field, which in turn requires 
higher-derivative terms in the action. We have encountered this phenomenon 
as a special case of our geometrical formalism~\cite{Hohm:2014eba}.  
Building up on the work of~\cite{Hohm:2013jaa} we will give extra   
evidence that there is indeed a 
gauge principle governing the $\alpha'$ corrections of classical string theory more generally.  

Conventionally, $\alpha'$ corrections 
to the 
effective field theory of bosonic strings 
are written in terms of higher 
powers of curvature tensors,   
the three-form 
field strength $H$ of the $b$-field,  
and their covariant 
derivatives. These actions are manifestly compatible with diffeomorphism 
 invariance and the abelian $b$-field gauge invariance. 
Therefore, these corrections are not required by gauge symmetries.
In this paper we will 
invoke T-duality covariance to study $\alpha'$ corrections,  
 using closed string field theory~\cite{Zwiebach:1992ie,Kugo:1992md} and 
  double field theory (DFT)~\cite{Siegel:1993th,Hull:2009mi,Hull:2009zb,Hohm:2010jy,Hohm:2010pp}.
While T-duality results in a global continuous symmetry of the effective
theory after dimensional reduction, 
DFT features a T-duality covariance prior to any reduction.
It also features duality-covariant generalized diffeomorphisms, and the duality symmetry  
that emerges after dimensional reduction is realized as gauge symmetries \cite{Hohm:2010jy,Hohm:2012gk,Hohm:2013bwa}.  
In a duality covariant formulation,  gauge symmetries acquire $\alpha'$ corrections
and in that sense `explain'   
the origin of $\alpha'$ corrections to the effective action. 

In closed string field theory on torus backgrounds T-duality covariance is built in by  
having coordinates dual to both momentum and winding modes, thereby  
realizing the T-duality group on this doubled space. 
More precisely,  in closed string field theory we have a perturbative 
expansion in which the (fluctuating) field variables around T-dual
backgrounds are related by simple transformations    
that make  T-duality manifest~\cite{Kugo:1992md}.
String field theory enables one to read off gauge
transformations and actions, including $\alpha'$ corrections. 
String field theory was 
the starting point for the construction of DFT in \cite{Hull:2009mi}.  
While T-duality is manifest in string field theory variables, 
the gauge symmetries do not have the form expected for 
`Einstein' variables that originate from the the conventional metric
tensor.  
To ${\cal O}(\alpha')$ the field redefinitions needed to 
connect Einstein variables to T-duality covariant fields 
are \textit{not} generally covariant, 
leading to fields that transform in a non-standard way 
under gauge symmetries. 

DFT is what follows from
closed string field theory  
after restricting to the 
massless sector, performing duality-covariant field redefinitions,  
and implementing background independence.  
Moreover,  one generally
imposes a duality covariant ``strong constraint" that means that
effectively all fields depend only on half of the doubled coordinates.\footnote{While there is work on DFT without the strong constraint \cite{Hohm:2011cp,Geissbuhler:2011mx,Grana:2012rr,Geissbuhler:2013uka,Hassler:2014sba}, our understanding of
such theories is still preliminary.}   
To zeroth order in $\alpha'$, duality-covariant field  and parameter
redefinitions in closed string field theory 
(CSFT)  simplify 
the gauge transformations, which then form the algebra governed by
the C-bracket~\cite{Hull:2009zb}.  This bracket becomes the Courant bracket 
defined in \cite{Tcourant} upon reduction 
to un-doubled coordinates.

In a DFT formulation of bosonic strings
we have to describe  
the Riemann-squared term well known to appear to first order 
in $\alpha'$.
There is a duality-covariant generalized Riemann tensor, but it cannot be 
fully determined in terms of physical fields because the connection contains 
undetermined components \cite{Siegel:1993th,Hohm:2010xe,Hohm:2011si}.  
Therefore, we cannot write directly an $\alpha'$ corrected
action that preserves duality covariance.  Additionally the Riemann-squared action cannot 
be written in terms of higher derivatives
of the generalized metric~\cite{Hohm:2011si}. To cubic order, however, the 
tensor structure in Riemann-squared that causes this difficulty can 
be removed by a \textit{non-covariant} field redefinition of the metric. 
This leads to fields with non-standard gauge transformations, 
and a gauge algebra with $\alpha'$ corrections. 
This is 
in quantitative 
agreement with~\cite{Meissner:1996sa} 
that studied T-duality in reductions to 
one dimension.  In that work $\alpha'$ corrections require field redefinitions 
of ${\cal O}(\alpha')$ that are quadratic in first  
derivatives of the metric, 
and thus cannot originate from covariant field redefinitions.

Doubled $\alpha'$ geometry~\cite{Hohm:2013jaa} is also 
a formulation in which T-duality is unchanged but the gauge structure is changed.
It features a field independent deformation of the C-bracket and an action
that is exactly gauge invariant.  
This deformation, however,  does not correspond to the 
${\cal O}(\alpha')$ deformation of bosonic string theory; 
 it does not give rise to Riemann-squared terms. 
The construction of \cite{Hohm:2013jaa} was based 
on a chiral CFT introduced  in~\cite{Siegel:1993th}   
and further studied in \cite{Hohm:2013jaa}.   
This CFT has one-loop worldsheet anomalies
and captures  
some of the structure needed for heterotic string theory~\cite{Hohm:2014eba}. 
Indeed, in this theory the 
gauge transformations for the $b$-field   
make the field strength $H$ with gravitational Chern-Simons 
modification  
gauge invariant.    
Although this geometry does not describe the full $\alpha'$ corrections 
of heterotic string theory, 
it contains important ingredients.  A different approach to describe 
$\alpha'$ corrections to heterotic DFT has been discussed in~\cite{Bedoya:2014pma}, 
as we will discuss in the conclusions. 
See also \cite{Baraglia,Garcia-Fernandez,Anderson:2014xha,delaOssa:2014cia} 
for Courant algebroids in `generalized geometry' formulations of heterotic strings. 

We use closed SFT to compute the gauge algebra to first
nontrivial order in $\alpha'$. 
After simplification, the result is a 
 deformation of the C-bracket that differs from that
of doubled $\alpha'$ geometry by a sign factor linked
to the symmetry of bosonic closed strings
under orientation reversal --- a $\mathbb{Z}_2$ 
symmetry that is not part of the T-duality group.
The four-derivative  
 terms in~\cite{Hohm:2013jaa}
are in fact $\mathbb{Z}_2$ parity odd.  We call this theory DFT$^{-}$
although it has no  
 overall $\mathbb{Z}_2$ symmetry. 
We find that higher-derivative actions that respect
the $\mathbb{Z}_2$ symmetry of bosonic strings exist. 
This theory, called DFT$^{+}$, is built to cubic order.   
The correction to the C-bracket in DFT$^{+}$ features the appearance
 of background fields, as opposed to the   background independent
 deformation of DFT$^{-}$. 
In $O(D,D)$ covariant notation, with fundamental indices $M,N=1, \ldots,2D$, and gauge parameters $\xi^M$, the gauge 
algebra of  DFT$^{-}$  reads  
\be\label{Z-algebra-original-intro}
 \text{DFT$^{-}$}:\quad 
  \big[\,  \xi_{1},\,\xi_{2}\,\big]_{-}^{M}   
  \ = \  \big[\,  \xi_{1},\,\xi_{2}\,\big]_{C}^{M} \, 
 -\, \half \,    
 \,  \eta^{KL} \eta^{PQ} \, \eta^{MN}
   K_{[1\, KP}\, \p_N \hskip-1pt  K_{2]\,LQ}   \,. 
 \ee  
Here  $K_{i\,MN}=2\partial_{[M}\xi_{i\, N]} = \partial_M \xi_{i\,N} - \partial_N \xi_{i\,M}$,   $i=1,2$, antisymmetrization of indices or labels is defined by 
$A_{[1} B_{2]} \equiv \half (A_1 B_2 - A_2 B_1)$,
 and $\eta$ denotes the $O(D,D)$ invariant metric.   Moreover, 
$[ \xi_{1},\xi_{2} ]_{C}$ denotes the C-bracket
governing the gauge algebra of the two-derivative DFT:   
\be
\label{C-bracket}
\begin{split}
\big[\,  \xi_{1},\,\xi_{2}\,\big]_{C}^{M}\ = \ & \ \xi_1^K \partial_K \xi_2^M
- \xi_2^K \partial_K \xi_1^M\, - \, \half ( \xi_1^K \partial^M \xi_{2\,K}
- \xi_2^K \partial^M \xi_{1\,K} ) \\[0.8ex]
= \ & \ 2\, \xi_{[1}^K \partial_K \xi_{2]}^M
\, - \,  \eta^{KL} \eta^{MN}  \xi_{[1K} \partial_N \xi_{2]L} \,. 
\end{split}
\ee
 The second term in (\ref{Z-algebra-original-intro}) is the higher-derivative correction.
 The factor of $\alpha'$ that multiplies it is left implicit. In contrast, the gauge algebra for DFT$^{+}$ reads 
\be\label{Z-algebra-original+}
  \text{DFT$^{+}$}:\quad 
    \big[\,  \xi_{1},\,\xi_{2}\,\big]_{+}^{M} \  
     = \  \big[\,  \xi_{1},\,\xi_{2}\,\big]_{C}^{M}  \, + \,  \half \,  
 \,  \bar{\cal H} ^{KL} \eta^{PQ} \, \eta^{MN}
   K_{[1\, KP} \, \p_N \hskip-1pt  K_{2]\,LQ}   \,,  ~   
 \ee  
where $\bar{\cal H} ^{KL}$ denotes the background value 
of the generalized metric that encodes the background 
metric and b-field.   It should be emphasized that this is not the complete
algebra of DFT${}^+$ which, given the appearance of $\bar{\cal H}$, is expected to be field dependent.  At the present stage of our
perturbative calculation only the background value of the fields appear.  
 Starting from the DFT$^+$ gauge algebra  
 we are able to write $\alpha'$-deformed
 gauge transformations that realize the algebra, and we show that they are related to 
 standard tensor gauge transformations by duality-violating redefinitions of 
 precisely the expected form that absorbs the problematic structure of Riemann-squared. 

Given this discrete   
freedom in the deformation of the gauge structure of DFT, 
it is natural to ask whether one can built an `interpolating' theory 
with both $\mathbb{Z}_2$ even and $\mathbb{Z}_2$ odd contributions.
Such a theory indeed exists at this cubic level, and    
it corresponds to having both the gravitational Chern-Simons modification 
 of $H$ and a Riemann-squared term.  
The gauge algebra for the interpolating theory reads 
 \be\label{Z-algebra-gammal-intro}
 \hskip10pt   \big[\,  \xi_{1},\,\xi_{2}\,\big]_{\alpha'}^{M} \ = \  \big[\,  \xi_{1},\,\xi_{2}\,\big]_{C}^{M}  \,   +  \half \,  
 \,  \big(\gamma^+\bar{\cal H}^{KL}-\gamma^-\eta^{KL}\big)\eta^{PQ} \ 
   K_{[1\, KP}\, \p^M \hskip-1pt  K_{2]\,LQ}   \,, 
 \ee  
with 
parameters $\gamma^\pm$ that at this level are unconstrained.

To confirm the consistency of our constructions we build the
cubic action, both for DFT$^{+}$ and DFT$^{-}$, 
including all terms with four derivatives and show that
it is consistent with gauge invariance.  While the cubic DFT$^{-}$
action is simple, the DFT$^{+}$ action is quite involved, but we can 
show that it encodes Riemann-squared (or Gauss-Bonnet) 
at the cubic level.

The 
main conclusion suggested by the results 
in this paper can be summarized as follows:  
While it is always possible to write $\alpha'$ corrections in 
terms of standard `Einstein variables' $g$ and $b$, string theory 
strongly suggests 
that these are not the best variables when $\alpha'$ effects
are turned on. 
Rather, making the duality symmetries of string theory manifest requires 
field variables that  have non-covariant transformations of ${\cal O}(\alpha')$ 
under standard diffeomorphisms. This may seem a radical step since   
diffeomorphism invariance is   
the basic principle 
of Riemannian geometry,  
but   in string theory this gauge principle is replaced  
by a duality covariant one, with a gauge algebra that extends the Lie bracket 
to the duality covariant bracket (\ref{Z-algebra-gammal-intro}) with ${\cal O}(\alpha')$ contributions. 
It is to be expected that there will be a (generalized) 
geometric formulation of classical string theory that organizes the notoriously 
complicated $\alpha'$ corrections in an efficient way that is manifestly covariant 
under all symmetries. 

\medskip

This paper is organized as follows.  In section~\ref{quaactfro} we review some of
the basics of closed string field theory and then determine the gauge algebra
including terms with one and three derivatives.  This algebra is simplified
by doing duality covariant field-dependent parameter redefinitions in section~\ref{simthegaualg}.   We use the simple final form  to write field transformations
that realize this DFT$^+$ algebra.  At this stage we also note that a simple variant
gives the DFT$^-$ algebra.  In section \ref{alcoineiva} we discuss the relation between
the CSFT perturbative field variables and the `Einstein' variables.  We do this for DFT$^+$ showing that duality non-covariant field redefinitions
relate the DFT variables to Einstein variables. 
For DFT$^-$ the relation is more subtle, as reviewed here.  In section~\ref{pertheo} we develop the perturbation theory of DFT$^-$, which is a useful step to develop the same
perturbation theory for DFT${}^+$.  We discuss in detail the $\mathbb{Z}_2$ orientation reversal transformation and its action on the perturbative DFT fields.  We also explain how to relate CSFT variables to the perturbative DFT variables and   
confirm our identification of DFT$^+$ with the theory that arises from CSFT.
Finally, in section~\ref{cuacdft+-} we perform a very nontrivial check of the existence
of DFT$^+$:  we show that an invariant cubic action including four-derivative terms
exists.  The cubic terms show direct evidence of the Gauss-Bonnet terms in the
effective action.  We conclude with some additional discussion of our results
in section~\ref{concl}.

\sectiono{The gauge algebra from string field theory}\label{quaactfro}
In this section we review the facts about string field theory necessary
to extend the results of~\cite{Hull:2009mi} to include $\alpha'$ corrections
in the gauge algebra.  
We  compute the algebra of gauge transformations
directly from the string field theory, including the first nontrivial $\alpha'$
corrections.    We also review the simplification of the gauge algebra 
to zeroth order in $\alpha'$. 
This section prepares the ground for the next where we will perform 
redefinitions directly on the gauge algebra in order to obtain a simple
form of the $\alpha'$ corrections to the algebra. 

\subsection{Generalities of closed string field theory}

The string field theory action is non-polynomial and takes the form
\begin{equation}
\label{csft_action99}
(2\kappa^2) S\ = \ -\frac{4}{\alpha'}\,\Bigl( \ \f{1}{2}\langle\,  \Psi , \, 
Q\Psi\rangle + \f{1}{3!} \langle \Psi,
[\Psi\,,\Psi ] \rangle +\f{1}{4!} \langle \Psi,
[\Psi\,,\Psi \,, \Psi ] \rangle + \cdots\, ~\Bigr) \,.
\end{equation}
Here $\ket{\Psi}$ is the classical off-shell closed string field, a ghost-number two, Grassmann even state of the full matter and ghost conformal field theory that describes the 
closed string background.  The off-shell string field must satisfy
$(L_0 - \bar L_0 ) \ket{\Psi}  =  0$ and $(b_0 - \bar b_0 ) \ket{\Psi} = 0$. 
The ghost-number one operator $Q$ is the BRST operator of the 
conformal field theory and $\langle \cdot , \cdot \rangle $ denotes the (linear)
inner product:
\be
\langle  A \,, B \rangle \ \equiv \ \langle A |\, c_0^- | B \rangle \,,  \ \ \ 
c_0^\pm \equiv  \half (c_0 \pm \bar c_0)\,,
\ee
where $\bra{A}$ is the BPZ conjugate of the string field $\ket{A}$. 
The inner product vanishes unless
$\hbox{gh} (A) + \hbox{gh} (B) = 5$.
The cubic interaction
is defined in terms of a closed string bracket $[ \cdot \,, \cdot ]$ or product. This product, whose input is two string fields and its output is another string field, is graded commutative:
$[B_1, B_2 ] \ = \ (-1)^{B_1 B_2 } [B_2, B_1 ]$
where the $B_1$ and $B_2$ in the sign factor denote the Grassmanality of the 
string fields $B_1$ and $B_2$, respectively.  Moreover we have
$ \hbox{gh} ( [B_1, B_2 ] ) \ = \ \hbox{gh} (B_1) + \hbox{gh} (B_2) - 1$.
 Thus, for the Grassmann even
classical field $[\Psi, \Psi]$ does not vanish and it has ghost number three, which
is suitable for the cubic coupling of the theory not to vanish.
The quartic term in the action is defined in terms of a three-product 
$[B_1, B_2, B_3]$ that is also graded commutative.  
This product parametrizes the failure
of the bracket to be a Lie bracket, and is the next element in the $L_\infty$ structure
of the classical theory.  The dots in the action denote terms
quartic and higher order in the string field.  

The field equations ${\cal F} (\Psi) =0$ and gauge transformations
$\delta_\Lambda \Psi$ of the theory take the form
\be
\label{gauge_transformations_csft}
\begin{split}
{\cal F} (\Psi) \ \equiv  & \ \ Q \, \Psi + \f{1}{2} [\, \Psi\,, \Psi\,]  + \f{1}{3!} [\, \Psi\,,\Psi\,, \Psi\,] + \cdots\ = \ 0 \\[1.0ex]
\delta_\Lambda \Psi\  = & \ \   Q \Lambda + [ \, \Psi\,, \Lambda\, ] + \f{1}{2} [\, \Psi, \Psi, \Lambda\,]
+ \cdots \,, 
\end{split}
\ee
where $\Lambda$ is a ghost-number one string field
and  the dots denote terms with higher powers
of the string field $\Psi$.   The gauge algebra of the theory takes the form
\be
\bigl[ \delta_{\Lambda_1} \,, \delta_{\Lambda_2} \bigr] \ = \ \delta_{\Lambda_{12} (\Psi)}
+ [ \, \Lambda_1, \Lambda_2 , {\cal F} ]  + \cdots\,.  
\ee
The transformations only close on shell (using the three-product)
and the dots represent higher terms that also vanish on-shell.   The resulting gauge
parameter takes the form
\be
\label{gauge_algebra_sft}
\Lambda_{12} (\Psi) \ = \  [\,\Lambda_2\, , \Lambda_1\, ] +  [\,\Lambda_2, \Lambda_1\,, \Psi \,  ]
+ \cdots 
\ee
showing that the algebra has field-dependent structure constants.  Since the
gauge parameters are ghost number one string fields they are Grassmann odd and
thus the bracket $ [\,\Lambda_2\, , \Lambda_1\, ]$ is properly antisymmetric under the exchange of the gauge parameters. We will compute the first term on the 
above right hand side. 

The theory generically has gauge invariances of gauge invariances.  
A gauge parameter of the form $\widehat \Lambda = Q \chi$ will generate no leading order gauge
transformations in (\ref{gauge_transformations_csft}) because $Q^2=0$.  To all orders one only has
on-shell gauge invariances of gauge invariances.  Indeed, for  
\be
\label{csft_trivial_parameter}
\widehat\Lambda \ = \ Q \chi  +  [ \Psi , \chi ]  + \f{1}{2}\,   [\Psi, \Psi, \chi ] + \cdots \,,
\ee
one finds a gauge transformation that vanishes on-shell:
$\delta_{\widehat\Lambda}\,  \Psi \ = \ - [\,  {\cal F} \,, \, \chi\,  ]  -  [ {\cal F} \,, \Psi \,, \chi ] \,+\,  {\cal O}(\Psi^3)$.

 \subsection{String field and gauge parameter} 

The closed string field for the massless sector takes the form
\be
\label{the_string_field}
\begin{split}
\ket{\Psi} & =  \int dp ~ \Bigl(  - \half 
e_{ij} ( p) \,\alpha_{-1}^i \bar \alpha_{-1}^j \, c_1 \bar c_1
 + e(p)  \, c_1 c_{-1}    +  \bar e (p)  
\, \bar c_1 \bar c_{-1} 
\\[0.5ex] &\hskip60pt
+ i{\textstyle{\sqrt{\alpha'\over 2}} }\,\bigl(
\,  f_i (p) \, c_0^+ c_1 \alpha_{-1}^i 
+ \bar f_j (p) \, c_0^+ \bar c_1 \bar \alpha_{-1}^j\Bigr)\, \ket{p} \,.
\end{split}
\ee
This string field features five
component fields: $e_{ij}, e, \bar e, f_i,$ and $\bar f_i$.
The field $e_{ij}$ contains the gravity and $b$-field fluctuations
as its symmetric and antisymmetric parts, respectively.  One linear
combination of the $e$ and $\bar e$ fields (the difference) is the 
dilaton and the other linear combination (the sum) can be gauged away.
The fields $f_i$ and $\bar f_i$ are auxiliary fields and can be solved 
for algebraically.   The gauge parameter $\ket{\Lambda}$ associated to the
above string field takes the form  
\be
\label{gt-param}
\ket{\Lambda}  =  \int [dp] ~ \Bigl(  
 {i\over \sqrt{2\alpha'}} 
\lambda_i ( p) \,\alpha_{-1}^i  c_1
- {i\over \sqrt{2\alpha'}} 
\bar\lambda_i ( p) \,\bar\alpha_{-1}^i  \bar c_1
 + \mu(p)  \, c_0^+ \Bigr) \ket{p} \,.
\ee   
The string field $\Lambda$ has ghost number one and
is annihilated by $b_0^-$.  It contains two vectorial gauge parameters
$\lambda_i$ and $\bar \lambda_i$ that encode infinitesimal diffeomorphisms
and infinitesimal $b$-field gauge 
symmetries in some suitable linear
combinations.  There is also  one scalar gauge parameter $\mu$ that can be
used to gauge away the field $e+ \bar e$.
 The linearized gauge transformations are
\be
\label{collgt}
\begin{split}
\delta_\Lambda e_{ij} \ &= \  D_i
\bar\lambda_j  +\bar D_j \lambda_i \,, \\[1.0ex]
\delta_\Lambda f_i \ &= \ -\half  \, \square \,\lambda_i + D_i \mu  \,,\\[1.0ex]
\delta_\Lambda \bar f_i  \ & = \ \phantom{-}\half  \, \square \,\bar \lambda_i
+ \bar  D_i \mu\,,\\[1.0ex]
\delta_\Lambda e \ &= \  -\half D^i\lambda_i +  \mu\,, \\[1.0ex]
\delta_\Lambda \bar e \ & = \ \phantom{-}\half \bar D^i\bar \lambda_i +\, \mu\,.
\end{split}
\ee
All indices are raised and lowered with the background
metric $G^{ij}$.  The derivatives $D$ and $\bar D$ are defined as
\be
D_i \ = \ {1\over \sqrt{\alpha'}} \Big( {\partial\over \partial x^i} - E_{ik} 
{\partial\over \partial \tilde x_k} \Big) \,, \qquad 
\bar D_i \ = \ {1\over \sqrt{\alpha'}} \Big( {\partial\over \partial x^i} + E_{ki} 
{\partial\over \partial \tilde x_k} \Big) \;. 
\ee
The weak constraint means that the 
 following equality holds acting on any field or gauge parameter
\be
\square  \ \equiv \ D^2 \ = \ \bar D^2 \,.
\ee
The strong constraint is 
$D^i A D_i B \ = \ \bar D^i A \bar D_i B $ for any $A, B$. 
We can now introduce fields $d$ and $\chi$ by
\be
d= \half\, (e - \bar e)  \,, \qquad  \hbox{and}\quad 
\quad  \chi =  \half\, (e+ \bar e)\,.
\ee
The gauge transformations of $d$ and $\chi$ are
\be
\delta_\Lambda d \, = -\quarter  (D^i\lambda_i   + \bar D^i\bar \lambda_i )\,,
\qquad 
\delta_\Lambda \chi \,  = -\quarter ( D^i\lambda_i  - \bar D^i\bar \lambda_i)+ \mu\,.
\ee
It is clear that a choice of $\mu$ can be used to set $\chi=0$.  Since further
$\lambda , \bar\lambda$ gauge transformations would then reintroduce 
$\chi$,  these gauge transformations must be accompanied by compensating
$\mu$ gauge transformations with parameter  $\mu(\lambda, \bar\lambda)$
\be
\label{muchoice}
\mu (\lambda, \bar\lambda)  \ = \ \quarter (D \cdot \lambda - \bar D \cdot \bar\lambda) \;. 
\ee
Effectively, the new gauge transformations $\delta_\Lambda$  are
$\delta_\lambda + \delta_{\bar\lambda} + \delta_{\mu(\lambda, \bar\lambda)}$.  The extra term
does not affect  $d$ nor $e_{ij}$, as neither    
transforms under $\mu$ gauge transformations. 
It changes the gauge transformations of  $f$ and
$\bar f$, but this is of no concern 
as these are auxiliary fields to be eliminated.  
 We denote by $\delta_\Lambda$ the gauge transformations 
 generated by $\lambda$ and  $\bar\lambda$, and use $\delta_\lambda$
 and $\delta_{\bar \lambda}$ for the separate transformations.   We have 
\be
\label{lamtrans}
\delta_\Lambda e_{ij}  \, = ~ D_i \bar\lambda_j ~ + ~\bar D_j \lambda_i \,,\qquad
\delta_\Lambda d \,  = \, -\quarter D\cdot \lambda  -\quarter \bar D \cdot \bar \lambda \,,
\ee
The theory is invariant under the  $\mathbb{Z}_2$ symmetry
\be
\label{z2-sym}
e_{ij} ~\to~  e_{ji}\,, \quad  D_i ~\to ~ \bar D_i\,, \quad
\bar D_i  ~\to ~ D_i \,, \quad  d ~\to ~  d~ \,,  
\ee
 related
to the invariance of the closed string theory under orientation reversal.
Note that 
this relates the transformations under $\lambda$ to those under $\bar\lambda$.
Invariance under one set of gauge transformations implies invariance under the
other set. This holds both as we include field dependent terms and higher
derivatives. 

The component fields in the string field theory have simple transformations
under 
T-duality. 
Since the formulation of the theory is not background independent the theory around some background $E$ must be compared with the
theory formulated around a T-dual background $E'$.  The fluctuation fields
of the two theories, as explained in section 4.2 of~\cite{Hull:2009mi}, are related by simple matrix transformations. Schematically $e_{ij} =   M_i{}^k \bar M_j{}^l e'_{kl}$
and the dilaton $d$ is duality invariant.  Note that the first index of $e$ transforms
with the unbarred $M$ and the second with the barred $M$.  
Every expression in which indices are contracted consistently, i.e., 
unbarred with unbarred and barred with barred indices, is therefore 
T-duality covariant. 
T-duality covariant
redefinitions respect such structure in contractions of indices.

\subsection{Cubic terms and gauge transformations from CSFT}

The algebra of gauge transformations
is described by (\ref{gauge_algebra_sft}), and the field-independent
part is given by
\be
\label{gauge_algebra_sft_field-indep}
\Lambda_{12}  \ \equiv \  [\,\Lambda_2\, , \Lambda_1\, ] \,.
\ee
We use uppercase gauge parameters to  
encode all the component gauge parameters: 
\be
\Lambda_1 = (\lambda_1, \bar \lambda_1, \mu_1), \qquad
\Lambda_2 = (\lambda_2, \bar \lambda_2, \mu_2)\,, \qquad
\Lambda_{12} = (\lambda_{12}, \bar \lambda_{12}, \mu_{12}).
\ee  
The computation
of the gauge algebra is a straightforward but somewhat laborious matter in string field theory.  There are contributions with various numbers of derivatives or powers of 
$\alpha'$.  We will be interested in the terms at zero order and first order in $\alpha'$.  We will write this as
\be
\label{def_alphap}
\begin{split}
\lambda_{12}^i  \ = \ &  \lambda_{12}^{(0)i}   + \alpha'\lambda_{12}^{(1)i}  + \ldots \\[1.0ex]
\bar\lambda_{12}^i  \ = \ &  \bar\lambda_{12}^{(0)i}   + \alpha'\bar\lambda_{12}^{(1)i}  + \ldots \\[1.0ex]
\mu_{12}^i  \ = \ &  \mu_{12}^{(0)i}   + \alpha'\mu_{12}^{(1)i}  + \ldots \,,
\end{split}
\ee
where the superscripts in parenthesis denote the power of $\alpha'$. 
The result to zeroth order in $\alpha'$ is
 \be
 \label{ifnkjekkgjkdjdf98vm0}
 \begin{split}
 \lambda_{12}^{(0)i} \ =  & \ \ \  {\textstyle{1\over 2}} \, \bigl( \lambda_2\cdot  D \lambda_1^i 
 - \lambda_1\cdot D \lambda_2^i\bigr) 
\ - {\textstyle{1\over 4}}  \bigl( \lambda_2 \cdot D^i \lambda_1  -   \lambda_1 \cdot D^i \lambda_2\bigr) 
\ - {\textstyle{1\over 4}}  \bigl( \lambda_2^i \, D\cdot \lambda_1  -  \lambda_1^i \, D\cdot \lambda_2 \bigr)   \ \phantom{\Biggl(}    \\ 
& +  {\textstyle{1\over 4}}\, 
 \bigl( \bar\lambda_2 \cdot \bar D \lambda_1^i-
 \bar\lambda_1 \cdot \bar D \lambda_2^i \bigr)
   +  {\textstyle{1\over 8}}\, \bigl( \lambda_1^i  \bar D \cdot \bar\lambda_2
   - \lambda_2^i  \bar D \cdot \bar\lambda_1 \bigr)  
   - {\textstyle{1\over 4}} 
 ( \lambda_1^i \,\mu_2-  \lambda_2^i \mu_1)  \,,  \\[2.5ex]
 \bar\lambda_{12}^{(0)i} \ =  & \ \ \  {\textstyle{1\over 2}} \, 
 \bigl( \bar\lambda_2\cdot  \bar D \bar\lambda_1^i 
 - \bar\lambda_1\cdot \bar D \bar\lambda_2^i\bigr) 
\ - {\textstyle{1\over 4}}  \bigl( \bar\lambda_2 \cdot \bar D^i \bar\lambda_1  -   \bar\lambda_1 \cdot \bar D^i \bar\lambda_2\bigr) 
\ - {\textstyle{1\over 4}}  \bigl( \bar\lambda_2^i \,\bar D\cdot \bar\lambda_1  -  \bar\lambda_1^i \,\bar D\cdot \bar\lambda_2 \bigr)      \\[1.0ex]
& +  {\textstyle{1\over 4}}\, 
 \bigl( \lambda_2 \cdot  D \bar\lambda_1^i-
 \lambda_1 \cdot  D \bar\lambda_2^i \bigr)
   +  {\textstyle{1\over 8}}\, \bigl( \bar\lambda_1^i   D \cdot \lambda_2
   - \bar\lambda_2^i  D \cdot \lambda_1 \bigr)  
   + {\textstyle{1\over 4}} 
 ( \bar \lambda_1^i \,\mu_2-  \bar\lambda_2^i \mu_1) \,,  \\[2.0ex]
  \mu_{12}^{(0)} \ = \ & \  -\f{1}{8}  \,\bigl(  \lambda_1 \cdot D 
+ \bar\lambda_1 \cdot \bar D \bigr) \mu_2 
 \  - \f{1}{16}  \, \bigl(  D\cdot \lambda_1 + \bar D \cdot \bar \lambda_1 \bigr)  \, \mu_2\,. 
~~ \\[1.0ex]
\end{split} 
 \ee
A partial version of this result is given in equation (3.8) of~\cite{Hull:2009mi}. 
In that reference we only determined the contribution to $\lambda_{12}^{(0)i}$ from 
$\lambda_1$ and $\lambda_2$.  Such terms are in the first line of  $\lambda_{12}^{(0)i}$.  
There are also contributions that involve $\bar\lambda_1$ and $\bar\lambda_2$
as well as $\mu_1$ and $\mu_2$.  Such terms were not needed for~\cite{Hull:2009mi},
where the gauge algebra was recalculated after a number of parameter and field
redefinitions.   The order $\alpha'$ results will be given below in (\ref{ifnkjekkgjkdjdf98vm0xx}). 

The gauge transformations of the fields, including terms linear in fields but only two derivatives,
are somewhat complicated and were given in~\cite{Hull:2009mi}. 
Their simplification took a few steps.  One must substitute the leading values
 for the auxiliary fields $f$ and $\bar f$.  Again, one can gauge fix $e+ \bar e$ to
 zero and work with just the dilaton $d$.  
 This is followed by a redefinition of the gauge parameters: 
\be
\label{vmclgfs}
\begin{split}
\lambda'_i \ = \ & \ \lambda_i + \f{3}{4} \lambda_i d - \f{1}{4}  \bar\lambda^k e_{ik}\,,\\[1.0ex]
\bar\lambda_i' \ = \ & \ \bar\lambda_i + \f{3}{4} \bar\lambda_i d - \f{1}{4} \lambda^k e_{ki}\,,
\end{split}
\ee
and finally a duality-covariant redefinition of the fields:
\be
\label{redefine-cubic}
\begin{split}
e'_{ij}  &= ~ e_{ij} +  e_{ij} d \,, \\[0.5ex]
d'   ~ & =   ~d + \f{1}{32}\, e_{ij}e^{ij} + \f{9}{16} \, d^2\,.
\end{split}
\ee
Dropping primes, the final form of the $\alpha'$-independent
gauge transformations is
 \be
\label{vimsagrtf}
\begin{split}
\delta_{\Lambda} e_{ij}   &\ = ~~ \bar D_j\, \lambda_i  \, +  \half \bigl[\,
   (D_i \lambda^k - D^k \lambda_i)\, e_{kj}    
   + \,\lambda_k D^k e_{ij} ~\bigr] ~~~~ \\[0.8ex]
&~~+\,   D_i \,\bar\lambda_j 
+  \half \Bigl[\,
   (\bar D_j \bar\lambda^k- \bar D^k \bar\lambda_j)\, e_{ik}    
   + \,\bar\lambda_k \bar D^k e_{ij} ~\Bigr]    \,  ,
 ~     \\[1.2ex]
\delta_{\Lambda}\, d   &\ = \ 
- \quarter  D \cdot  \lambda  
+  \half (\lambda \cdot D) \,d\,  - \quarter\bar D \cdot \bar\lambda 
 + \half (\bar\lambda \cdot \bar D) \,d  \,.
 \end{split}
\ee
Trivial gauge parameters do not generate gauge transformations and 
take a simple form
\be
\lambda_i \ = \  D_i \chi  \,, ~~~  \bar\lambda_i \ = \ - \bar D_i \chi \,, 
\ee
as can be checked using the strong constraint.  
These trivial gauge parameters have no field dependence
and the resulting transformations of fields vanish without using equations
of motion.  This is simpler than what could have been expected from
 (\ref{csft_trivial_parameter}).  
We will see that such simplicity is preserved with 
$\alpha'$ corrections. The algebra of gauge transformations can be recalculated 
using (\ref{vimsagrtf}) with the conventions
\be
\bigl[ \delta_{\Lambda_1} \,, \delta_{\Lambda_2} \bigr] \ = \ \delta_{\Lambda_{c,12}}\,, 
\qquad \Lambda_{c,12} =  (\lambda_{c,12} , \bar\lambda_{c,12}) \,,\ee
with subscripts `c' for  C-bracket.  We find 
\be
\label{iwdtfttafvm99}
\begin{split}
\lambda_{c,12}^i \    =
&\, \ \ \   \half \,\bigl[ ~ (\lambda_2 \cdot D + \bar \lambda_2 \cdot \bar D) \, \lambda_1^i 
- (\lambda_1 \cdot D + \bar \lambda_1 \cdot \bar D) \lambda_2^i \  \bigr] 
\\[1.0ex] &
+\quarter \,
\bigl[ ~ \lambda_1 \cdot D^i  \lambda_2 - \lambda_2\cdot D^i  \lambda_1  \bigr] - \quarter \,
\bigl[ ~ \bar \lambda_1 \cdot D^i  \bar \lambda_2 - \bar\lambda_2\cdot D^i  \bar \lambda_1  \bigr] \,,
\\[1.3ex]
\bar \lambda_{c,12}^i \ = & \ \ 
\ \half \,\bigl[~   (\lambda_2 \cdot D + \bar \lambda_2 \cdot \bar D)\, \bar\lambda_1^i 
-\, (\lambda_1 \cdot D + \bar \lambda_1 \cdot \bar D)\, \bar \lambda_2^i  \ \bigr] 
\\[1.0ex] &
- \quarter \,\bigl[ ~ \lambda_1 \cdot \bar D^i  \lambda_2 - \lambda_2\cdot \bar D^i  \lambda_1  \bigr]
+\quarter \,\bigl[ ~ \bar\lambda_1 \cdot \bar D^i  \bar \lambda_2 - \bar \lambda_2\cdot \bar D^i  \bar \lambda_1  \bigr]~.
\end{split}
\ee
This is the gauge algebra that written in background independent language gives
the C-bracket \cite{Hull:2009zb}: $
\Lambda_{c, 12}  =   [\,\Lambda_2\, , \Lambda_1\, ]_{{}_{\rm C}}$.
This algebra is different from the zeroth-order algebra in (\ref{ifnkjekkgjkdjdf98vm0}).  It would have been convenient if  (\ref{iwdtfttafvm99}) could have been derived from 
(\ref{ifnkjekkgjkdjdf98vm0}) without recourse to the gauge transformations of fields.
We will do this in the next section as a warm-up,  before extending the analysis
to include the $\alpha'$ corrections.   As a first step, we rewrite here
$\lambda_{12}^{(0)i}$
in terms of $\lambda_{c,12}^i$.  A short calculation gives
\be
 \label{ivmvmdf98vm0}
  \begin{split}
 \lambda_{12}^{(0)i} \ =  & \ \ \lambda^i_{12\, c} +  \  {\textstyle{1\over 4}} \, \bigl( \bar\lambda_1\cdot  D^i \bar\lambda_2 \,  
 - \bar\lambda_2\cdot D^i \bar\lambda_1\bigr) 
\ - {\textstyle{1\over 4}}  \bigl( \lambda_2^i \, D\cdot \lambda_1  -  \lambda_1^i \, D\cdot \lambda_2 \bigr)   \   \\[1.0ex] 
& -  {\textstyle{1\over 4}}\, 
 \bigl( \bar\lambda_2 \cdot \bar D \lambda_1^i-
 \bar\lambda_1 \cdot \bar D \lambda_2^i \bigr)
   +  {\textstyle{1\over 8}}\, \bigl( \lambda_1^i  \bar D \cdot \bar\lambda_2
   - \lambda_2^i  \bar D \cdot \bar\lambda_1 \bigr)  
   - {\textstyle{1\over 4}} 
 ( \lambda_1^i \,\mu_2-  \lambda_2^i \mu_1)   \,.
 \end{split}
\ee

The order $\alpha'$ terms in the gauge algebra, as defined in (\ref{def_alphap})
are also calculated from the string field theory and the result is 
 \be
 \label{ifnkjekkgjkdjdf98vm0xx}
 \begin{split}
 \lambda_{12}^{(1)i} \ = 
  & \hskip10pt - {\textstyle{1\over 32}} 
\bigl( D^i_1 - D^i_2 \bigr)
\bigl(  D^j_1 + 2D^j_2\bigr)
\bigl( 2D^k_1 + D^k_2 \bigr) \lambda_{1j}  \lambda_{2k} \\[1.0ex]
& \hskip10pt -  {\textstyle{1\over 64}} 
\bigl( D^i_1 - D^i_2  \bigr)
\bigl( D^j_1 + 2 D^j_2 \bigr) \bigl( 2\bar D^k_1+ \bar D^k_2 \bigr)
(\lambda_{1\, j} \bar \lambda_{2\,k} - \lambda_{2\, j} \bar \lambda_{1\,k}) \\[1.0ex]
& \hskip10pt +  {\textstyle {1\over 32} }
\bigl( D^i_1 - D^i_2  \bigr)
\bigl( D^j_1 + 2D^j_2\bigr) (\lambda_{1\, j}\mu_2 - \lambda_{2\, j} \mu_1)  \\[1.0ex]
&\hskip10pt 
+  {\textstyle{1\over 32}} 
 \bigl( D^i_1 - D^i_2 \bigr) 
 \bigl( \bar D^j_1 +2\bar D^j_2  \bigr) 
 (\bar\lambda_{1\, j}  \,\mu_2 - \bar\lambda_{2\, j}\,  \mu_1 ) \,,
 \end{split} 
 \ee 
 \be
 \begin{split}
 \bar\lambda_{12}^{(1)i} \ =  
  & \hskip10pt - {\textstyle{1\over 32}} 
\bigl( \bar D^i_1 - \bar D^i_2 \bigr)
\bigl(  \bar D^j_1 + 2\bar D^j_2\bigr)
\bigl( 2\bar D^k_1 + \bar D^k_2 \bigr) \bar\lambda_{1j}  \bar
\lambda_{2k} \\[1.0ex]
& \hskip10pt -  {\textstyle{1\over 64}} 
\bigl( \bar D^i_1 - \bar D^i_2  \bigr)
\bigl( \bar D^j_1 + 2 \bar D^j_2 \bigr) \bigl( 2 D^k_1+  D^k_2 \bigr)
(\bar\lambda_{1\, j}  \lambda_{2\,k} - \bar\lambda_{2\, j}  \lambda_{1\,k}) \\[1.0ex]
& \hskip10pt -  {\textstyle {1\over 32} }
\bigl( \bar D^i_1 - \bar D^i_2  \bigr)
\bigl( \bar D^j_1 + 2\bar D^j_2\bigr) (\bar\lambda_{1\, j}\mu_2 - \bar\lambda_{2\, j} \mu_1)  
\\[1.0ex]
&\hskip10pt 
 - {\textstyle{1\over 32}} 
 \bigl( \bar D^i_1 - \bar D^i_2 \bigr) 
 \bigl( D^j_1 +2 D^j_2  \bigr) 
 (\lambda_{1\, j}  \,\mu_2 - \lambda_{2\, j}\,  \mu_1 ) \,,  \\[1.5ex]
  \mu_{12} \ = & \ \ \ \ \   \f{1}{32} \,  ( \,D_1^j + 2 D_2^j\,) \, (\, 2\bar D_1^k + \bar D_2^k) \, \lambda_{1,j} \, \bar\lambda_{2,k}
  \ - \  (1 \leftrightarrow 2)\,. 
 \end{split} 
 \ee
In here we have  a collection of derivatives acting on  products (or a sum of products) of gauge parameters.  The convention is that $D_1$ acts
on the first function and $D_2$ on the second function.  Thus, for example, 
$D_1^i ( f \cdot g ) =  D^i f \cdot g$,   $D^i_2 (f\cdot g) =  f \cdot D^ig$,
and $D^j_1D^i_2 (f\cdot g) =  D^jf \cdot D^ig$.  Care must be exercised not to exchange the order of  functions until all derivatives have been applied.  Note that $\bar\lambda_{12}^{i}$ is obtained from $\lambda_{12}^{i}$ by conjugating all objects
and changing the sign of any term involving $\mu_1$ or $\mu_2$.

\sectiono{Simplifying the closed string theory gauge algebra} \label{simthegaualg}
In this section we perform redefinitions of the gauge parameters 
in order to simplify the closed string field theory gauge algebra obtained
in sect.~\ref{quaactfro}. 
We begin by showing how to compute in general the change of a gauge algebra 
under a field-dependent parameter redefinition. Then we illustrate this technique by applying it  to the CSFT gauge algebra to zeroth order in $\alpha'$, to recover the 
C-bracket result~(\ref{iwdtfttafvm99}). We apply it next to the CSFT gauge algebra to first order in $\alpha'$, and 
after some steps we obtain the rather simple form  
given in (\ref{finalalpha'bracket2}). 
Based on this final form of the gauge algebra, we determine the 
associated gauge transformations to first order in $\alpha'$.  
In the last subsection, 
we point out that consistency of such a higher-derivative deformation of bracket and gauge 
transformations does not uniquely determine these transformations. Rather, there is a 
$\mathbb{Z}_2$ freedom that leaves one sign undetermined.

\subsection{General remarks on gauge parameter redefinitions}
We start with a general discussion of  (perturbative) gauge transformations and
show how field-dependent redefinitions of the gauge parameters  
can change the gauge algebra. Note that, in contrast,  
field redefinitions leave the gauge algebra unchanged,  even though
they can change the form of gauge transformations. 
We consider gauge transformations of fields, collectively denoted by $\phi$, 
with respect to a gauge parameter $\lambda$. 
They are perturbatively defined to first order in fields,
\be
\delta_\lambda \phi \ = \  f(\lambda) + g(\lambda, \phi)  + {\cal O}(\phi^2)\;, 
\ee
where $f$ is a linear function of $\lambda$, and $g$ is a linear function of 
both $\lambda$ and $\phi$.  We also write  
\be
f(\lambda) \ = \ \delta^{[0]}_\lambda \phi \,, \qquad 
g(\lambda, \phi ) \ = \ \delta^{[1]}_\lambda \phi \,, 
\ee
indicating by the superscript in brackets the power of fields. 
In general, closure of the gauge transformations requires 
\be
\label{vmka}
\Bigl[ \,\delta_{\lambda_1} \,, \delta_{\lambda_2} \Bigr] \, \phi
\ = \ {\delta_{\phantom{j}}}_{\hskip-4pt\lambda_{12} (\lambda_1, \lambda_2; \,\phi) }\, \phi  \ + \ h (\lambda_1, \lambda_2, F(\phi) ) \;, 
\ee
where $\lambda_{12} (\lambda_1, \lambda_2; \,\phi)$ are field dependent
structure constants, and $F(\phi)$ is a function of the fields such that $h=0$  on-shell. To linear order in $\phi$ we can only determine the part of
the gauge algebra (\ref{vmka}) that is independent of  $\phi$, since terms ${\cal O}(\phi)$
are affected by the $\delta^{[0]}$ variation of unknown terms ${\cal O}(\phi^2)$. 
Thus, $h$ cannot be calculated from
the field transformations to this order. Similarly, writing
\be
\lambda_{12} (\lambda_1, \lambda_2 ; \, \phi ) \ = \ \lambda_{12} (\lambda_1,
\lambda_2) + {\cal O}(\phi)\,,
\ee
we can only determine the $\phi$-independent part.  
Equation (\ref{vmka}) then reduces to   
\be
\label{vmksa}
\Bigl[ \,\delta_{\lambda_1} \,, \delta_{\lambda_2} \Bigr] \, \phi
\ = \  \ f( \lambda_{12} (\lambda_1, \lambda_2) )  \ + {\cal O}(\phi) 
\ = \ \delta^{[0]}_{\lambda_{12} (\lambda_1, \lambda_2) } \phi   \ + {\cal O}(\phi)\;.  
\ee 
On the other hand, computing the left-hand side directly from 
the transformations we find
\be
\label{vmkba}
\begin{split}
\Bigl[ \,\delta_{\lambda_1} \,, \delta_{\lambda_2} \Bigr] \, \phi
\ =  \ & \ \delta_{\lambda_1}  \bigl( f(\lambda_2 ) + g(\lambda_2, \phi) + {\cal O}(\phi^2) \bigr) - (1 \leftrightarrow 2) \\[0.5ex]
\  =  \  &  \   g(\lambda_2, f(\lambda_1) ) -  g(\lambda_1, f(\lambda_2) ) 
+ {\cal O}(\phi)\;.  \end{split}
\ee
Comparing with (\ref{vmksa})  we learn that
\be\label{GAUGEalgebra}
 f( \lambda_{12} (\lambda_1, \lambda_2) )  \ = \  g(\lambda_2, f(\lambda_1) )  -  g(\lambda_1, f(\lambda_2) ) \;. 
\ee
Now we examine how a field-dependent parameter redefinition changes the
gauge algebra.  We consider
\be
\lambda \ \rightarrow \ \lambda+\Lambda(\lambda,\phi)\,,
\ee
with $\Lambda (\lambda, \phi)$ linear in $\lambda$ and $\phi$. 
More precisely, we define new gauge transformations $\widetilde \delta_\lambda$ by
\be\label{defParared}
\widetilde \delta_\lambda \phi \ \equiv \ {\delta_{\phantom{j}}}_{\hskip-4pt\lambda + \Lambda (\lambda, \phi) } \,\phi \ = \ \delta_\lambda \phi + 
\delta_{\Lambda (\lambda, \phi)} \phi  \ = \ f(\lambda) + g(\lambda, \phi) +  f(\Lambda (\lambda, \phi)) + {\cal O}(\phi^2) \;. 
\ee
From this we next compute the new gauge algebra, which is of the form
\be
\label{vmktts}
\begin{split}
\Bigl[ \,\widetilde\delta_{\lambda_1} \,, \widetilde\delta_{\lambda_2} \Bigr] \, \phi
 \ = \ {\delta^{}}^{[0]}_{\, \widetilde \lambda_{12} (\lambda_1, \lambda_2) } \phi + {\cal O}(\phi)
 \ = \ f(\widetilde \lambda_{12} (\lambda_1, \lambda_2)) + {\cal O}(\phi) \;. 
 \end{split}
\ee
From the left-hand side we get
\be
\label{uoin}
\begin{split}
\Bigl[ \,\widetilde\delta_{\lambda_1} \,, \widetilde\delta_{\lambda_2} \Bigr] \, \phi
\ =  \ &\,   \widetilde\delta_{\lambda_1}  \Bigl(  f(\lambda_2) + g(\lambda_2, \phi) +  f(\Lambda (\lambda_2, \phi) ) + {\cal O} (\phi^2) \Bigr)  - (1 \leftrightarrow 2) 
   \\[0.5ex]
\ =  \ &\,     g(\lambda_2,  f(\lambda_1)) +  f(\Lambda (\lambda_2, f(\lambda_1) )   - (1 \leftrightarrow 2)  + {\cal O} (\phi)  
\\[0.5ex]
\ =  \ &   f(\lambda_{12} (\lambda_1, \lambda_2) ) +  f(\Lambda (\lambda_2, f(\lambda_1) )  -  f(\Lambda (\lambda_1, f(\lambda_2) ) + {\cal O} (\phi)\;,   \\
\end{split}
\ee
using (\ref{GAUGEalgebra}) in the last step. Recalling that 
$\delta^{[0]}_\lambda \phi = f(\lambda)$,  we compare with (\ref{vmktts}) and 
infer that
up to irrelevant trivial parameters 
\be\label{FinalParared}
\widetilde \lambda_{12} (\lambda_1, \lambda_2)  \ = \ 
\lambda_{12} (\lambda_1, \lambda_2)  + \,\Lambda \big(\lambda_2, \delta^{[0]}_ {\lambda_1}\phi \big)   
-  \Lambda \big(\lambda_1, \delta^{[0]}_ {\lambda_2}\phi \big) \,. 
\ee
This relation allows us to compute the modification of the gauge algebra under 
field-dependent parameter 
redefinitions generated by $\Lambda (\lambda , \phi)$ knowing only the inhomogeneous transformations $\delta^{[0]}$ of the fields.  We will apply this repeatedly below.

\subsection{Simplifying the  gauge algebra}

We first illustrate the above method by simplifying the gauge algebra following from 
CSFT to zeroth order in $\alpha'$. 
After using the gauge fixing condition (\ref{muchoice})
in (\ref{ivmvmdf98vm0}) we find that  $\lambda_{12}$ to zeroth order in $\alpha'$ can be written
as
\be
\begin{split}
\lambda_{12}^{(0)i} \ = \ & \  \lambda_{c,12}^i  + \quarter \bigl[
\  \bar\lambda_1^k\,   D^i \bar\lambda_{2,k} 
- \bar\lambda_2^k \,   D^i \bar\lambda_{1,k} \,\bigr] \,  -\quarter \bigr[ 
\ \bar\lambda_2^k  \bar D_k \lambda_1^i  - \ \bar\lambda_1^k \,\bar D_k\,  \lambda_2^i \bigr] \\[1.0ex]
& \qquad \;\, + \f{3}{4} \bigl[ - \lambda_1^i \bigl( -\quarter  ( D\cdot \lambda_2 + \bar D\cdot \bar \lambda_2) \bigl) \ 
+\  \lambda_2^i \, \bigl( - \quarter  ( D\cdot \lambda_1+ \bar D\cdot \bar \lambda_1) \bigr) \, \bigr] \;. 
\end{split} 
\ee
Next let us combine terms in the first and second line to find 
\be
\begin{split}
\lambda_{12}^{(0)i} \ = \  \  \lambda_{c,12}^i  
+ \quarter
 \bar\lambda_1^k\, (  D^i \bar\lambda_{2,k} 
+ \,\bar D_k  \lambda_2^i )  
+ \f{3}{4} \bigl[ \, - \lambda_1^i \bigl( -\quarter ( D\cdot \lambda_2 + \bar D\cdot \bar \lambda_2) \bigl)\, \bigr] 
 - \ (1\leftrightarrow 2) \;, 
\end{split} 
\ee
where  the $(1\leftrightarrow 2)$ antisymmetrization applies to all terms except
$\lambda_{c,12}^i$.  
We wrote the terms such that they take the form of the  $\delta^{[0]}$ transformations 
of $e_{ij}$ and $d$ as given  in (\ref{lamtrans}). 
Thus, using the $(1\leftrightarrow 2)$ antisymmetry, we can write 
\be
\label{sftcb}
\lambda_{12}^{(0)i}  \ =  \  \lambda_{c,12}^i  - \quarter\, \bar \lambda_2^k \,\delta_{\lambda_1}^{[0]} e^i{}_k 
+ \f{3}{4}  \,\lambda_2^i \,\delta_{\lambda_1}^{[0]} d \ - \ (1\leftrightarrow 2) \,.
\ee
Looking back at (\ref{FinalParared}) we infer that the final two terms have precisely the 
structure needed to be removable by a parameter redefinition. 
More precisely, with
\be
\Lambda^i (\lambda, \bar\lambda, e, d) \ = \  \quarter\, \bar \lambda^k \, \,e^i{}_k 
- \f{3}{4}  \,\lambda^i \,\,d \;, 
\ee
we obtain with (\ref{FinalParared}) and (\ref{sftcb}) for the redefined gauge algebra 
\be
\widetilde \lambda_{12}^i (\lambda_1, \lambda_2)  \ = \ 
\lambda_{12}^{(0)i} (\lambda_1, \lambda_2)  + \, \quarter\, \bar \lambda_2^k \, \delta^{[0]}_ {\lambda_1}\,e^i{}_k 
- \f{3}{4}  \,\lambda_2^i \, \delta^{[0]}_ {\lambda_1}\,d    -  (1\leftrightarrow 2)  \ = \ \lambda^i_{c, 12} 
 \,. 
\ee
The extra terms have cancelled and the gauge algebra reduces to the one defined by the C-bracket. 
The above parameter redefinitions are those in (\ref{vmclgfs}), 
and combined with the field redefinitions (\ref{redefine-cubic})   lead to the simplified form (\ref{vimsagrtf}) of the gauge transformations. 
Note that, more efficiently, terms in the gauge algebra of the form
$\Lambda \big(\lambda_2, \delta^{[0]}_ {\lambda_1}\phi \big)   
-  \Lambda \big(\lambda_1, \delta^{[0]}_ {\lambda_2}\phi \big)$ can be simply dropped.

\medskip

In the following we apply this strategy to the ${\cal O}(\alpha')$ corrections of the gauge algebra.  With the above simplification, the ${\cal O}((\alpha')^0)$ part of the algebra
is that of the C-bracket and the ${\cal O}(\alpha')$ terms remain unchanged 
so that, deleting tilde's, we can write
\be
\label{sftcb2}
\lambda_{12}^i  \ =  \  \lambda_{c,12}^i  \ + \ \alpha' \lambda_{12}^{(1)i} \,. 
\ee
Here $\lambda_{12}^{(1)i}$ represents the $\alpha'$ correction given
in (\ref{ifnkjekkgjkdjdf98vm0xx}) that, grouping differential operators, can be written as:
 \be
 \begin{split}
  \lambda_{12}^{(1)i} \ = \  &-\f{1}{32}(D_1^i-D_2^i)(D_1^j+2D_2^j)\Big[(2D_1^k+D_2^k)\lambda_{1j}
  \lambda_{2k}\\
  &+\f{1}{2}(2\bar{D}_{1}^k+\bar{D}_2^k)(\lambda_{1j} \bar{\lambda}_{2k}-\lambda_{2j}\bar{\lambda}_{1k}) 
  -\lambda_{1j}\mu_2+\lambda_{2j}\mu_1\Big]\\
  &  + \f{1}{32} \,(D_1^i - D_2^i \,) \, (\,\bar D_1^j + 2\bar D_2^j)\,  (\bar \lambda_{1,j}\, \mu_2 \ - \ 
\bar\lambda_{2,j}\, \mu_1)\;. 
\end{split}
\ee
Next we eliminate $\mu$ in favor of $\lambda$ using (\ref{muchoice}) 
and expand the innermost differential operator 
  \be
 \begin{split}
  \lambda_{12}^{(1)i} \ = \  &-\f{1}{32}(D_1^i-D_2^i)(D_1^j+2D_2^j)\, \Big[\, 
  2D^k\lambda_{1j}\, \lambda_{2k}+\lambda_{1j}D^k\lambda_{2k}+\bar{D}^k\lambda_{1j}\,\bar{\lambda}_{2k}
  -\bar{D}^k\lambda_{2j}\,\bar{\lambda}_{1k}\\
  &\    +\f{1}{2}\lambda_{1j}\bar{D}^k \bar{\lambda}_{2k}
  -\f{1}{2}\lambda_{2j}\bar{D}^k\bar{\lambda}_{1k}
  -\f{1}{4}\lambda_{1j}(D\cdot\lambda_2-\bar{D}\cdot \bar{\lambda}_2)
  +\f{1}{4}\lambda_{2j}(D\cdot\lambda_1-\bar{D}\cdot \bar{\lambda}_1)
  \Big]\\[0.3ex]
  &  + \f{1}{32} \,(D_1^i - D_2^i \,) \, (\,\bar D_1^j + 2\bar D_2^j)\,  \big(\f{1}{4}
  \bar \lambda_{1,j}\, (D\cdot \lambda_2-\bar{D}\cdot\bar{\lambda}_2) \ - \ 
\f{1}{4}\bar\lambda_{2,j}\, (D\cdot \lambda_1-\bar{D}\cdot\bar{\lambda}_1)\big)\;. 
\end{split}
\ee
Acting then with the second differential operator yields 
 \be\label{MassiveStep}
 \begin{split}
  \lambda_{12}^{(1)i} \ = \  &-\f{1}{32}\big(D_1^i-D_2^i\big)\Big\{ 
  2D^jD^k\lambda_{1j}\, \lambda_{2k}+D^j\lambda_{1j}\,D^k\lambda_{2k}+D^j\bar{D}^k\lambda_{1j}\,\bar{\lambda}_{2k}
  -D^j\bar{D}^k \lambda_{2j}\,\bar{\lambda}_{1k}\\
  &\qquad\qquad \qquad \quad 
  +\f{1}{2}D^j\lambda_{1j}\,\bar{D}^k\bar{\lambda}_{2k}-\f{1}{2}D^j\lambda_{2j}\,\bar{D}^k\bar{\lambda}_{1k}
  +4D^k\lambda_{1j}\,D^j\lambda_{2k}+2\lambda_{1j}D^jD^k\lambda_{2k}\\[0.4ex]
  &\qquad\qquad \qquad \quad 
  +2\bar{D}^k\lambda_{1j}\,D^j\bar{\lambda}_{2k}-2\bar{D}^k\lambda_{2j} D^j\bar{\lambda}_{1k}
  +\lambda_{1j} D^j\bar{D}^k\bar{\lambda}_{2k}-\lambda_{2j}D^j\bar{D}^k\bar{\lambda}_{1k}\\[1.0ex]
  &\qquad\qquad \qquad \quad 
  -\f{1}{4}D^j\lambda_{1j}\big(D\cdot \lambda_2-\bar{D}\cdot \bar{\lambda}_2\big)
  +\f{1}{4}D^j\lambda_{2j}\big(D\cdot \lambda_1-\bar{D}\cdot \bar{\lambda}_1\big)
  \\[0.5ex]
  &\qquad\qquad \qquad \quad 
  -\f{1}{2}\lambda_{1j}D^j\big(D\cdot \lambda_2-\bar{D}\cdot \bar{\lambda}_2\big)
  +\f{1}{2}\lambda_{2j}D^j\big(D\cdot \lambda_1-\bar{D}\cdot \bar{\lambda}_1\big)
  \\[0.5ex]
  &\qquad\qquad \qquad \quad 
  -\f{1}{4}\bar{D}^j\bar{\lambda}_{1j}\big(D\cdot \lambda_2-\bar{D}\cdot \bar{\lambda}_2\big)
  +\f{1}{4}\bar{D}^j\bar{\lambda}_{2j}\big(D\cdot \lambda_1-\bar{D}\cdot \bar{\lambda}_1\big)
  \\[0.5ex]
  &\qquad\qquad \qquad \quad 
  -\f{1}{2}\bar{\lambda}_{1j}\bar{D}^j\big(D\cdot \lambda_2-\bar{D}\cdot \bar{\lambda}_2\big)
  +\f{1}{2}\bar{\lambda}_{2j}\bar{D}^j\big(D\cdot \lambda_1-\bar{D}\cdot \bar{\lambda}_1\big)
  \Big\}\;. 
 \end{split}
 \ee  
We can now combine and simplify various terms inside the parenthesis (i.e.~before acting with the outer differential operator). We note that $(D_1-D_2)$ imposes an antisymmetry: when exchanging the first and second factor of any term we 
get a sign.  
It is then an easy calculation to show, for instance,  
that the terms quadratic in $D\cdot\lambda$ and $\bar{D}\cdot\bar{\lambda}$ 
combine into 
 \be
  \f{1}{2}\big(D\cdot \lambda_1+\bar{D}\cdot \bar{\lambda}_1\big)
  \big(D\cdot \lambda_2+\bar{D}\cdot \bar{\lambda}_2\big) \ = \ 
  -\big(\delta_{\lambda_1}^{[0]}d\big) \big(D\cdot \lambda_2+\bar{D}\cdot \bar{\lambda}_2\big)
  -\big(D\cdot \lambda_1+\bar{D}\cdot \bar{\lambda}_1\big)\big(\delta_{\lambda_2}^{[0]}d\big)\;.
 \ee
Performing similar manipulations for the remaining terms in (\ref{MassiveStep}) we find in total 
 \be\label{REwritingSTEP}
 \begin{split}
  \lambda_{12}^{(1)i}\ = \  &-\f{1}{32}\big(D_1^i-D_2^i\big)\,\Big\{
  -\big(D\cdot \lambda_1+\bar{D}\cdot \bar{\lambda}_1\big)\delta_{\lambda_2}^{[0]}d
\ + \  \f{3}{2}\lambda_1^j D_j\big(D\cdot\lambda_2+\bar{D}\cdot \bar{\lambda}_2\big)  \\[1.0ex]
  &\quad
  +\f{1}{2}\bar{\lambda}_1^j\bar{D}_{j}\big(D\cdot \lambda_2
  +\bar{D}\cdot \bar{\lambda}_2\big)
   \   +2D^k\lambda_{1j}\,D^j\lambda_{2k}
  +2 \bar{D}^k\lambda_{1j}\,D^j\bar{\lambda}_{2k}
  -(1\leftrightarrow 2)
    \Big\}\;, \\[1.0ex]
  \ = \  &-\f{1}{32}\, \big(D_1^i-D_2^i\big)\,\Big\{
  -\big(D\cdot \lambda_1+\bar{D}\cdot \bar{\lambda}_1\big)\big(\delta_{\lambda_2}^{[0]}d\big)
  -6 \lambda_1^j D_j\big(\delta_{\lambda_2}^{[0]}d \big) 
  -2\bar{\lambda}_1^j\bar{D}_{j}\big(\delta_{\lambda_2}^{[0]}d \big)
    \\[1.0ex]
  &\qquad\qquad\qquad\quad\quad  +2D^k\lambda_{1j}\,D^j\lambda_{2k}
  +2 \bar{D}^k\lambda_{1j}\,D^j\bar{\lambda}_{2k}-(1\leftrightarrow 2)
    \Big\}\;,  
 \end{split}
 \ee  
where $(1\leftrightarrow 2)$ means $(\lambda_1\leftrightarrow \lambda_2)$.
Consider  the second term on the last line. We write it in terms of $\delta^{[0]}e$ and
use the strong constraint to find  
 \be
  2\bar{D}^k\lambda_{1j}\big(\delta_{\lambda_2}^{[0]} e^{j}{}_{k}-\bar{D}_k\lambda_{2}^j\big)
  - (1\leftrightarrow 2)
  \ = \ 2\bar{D}^k\lambda_{1j}\,\delta^{[0]}_{\lambda_2} e^{j}{}_{k}
  -2D^k\lambda_{1j}\,D_{k}\lambda_{2}^j - (1\leftrightarrow 2) \;. 
 \ee 
As a result, we have
  \be\label{REwritingSTEP33}
 \begin{split}
  \lambda_{12}^{(1)i} \ = \  &-\f{1}{16}\big(D_1^i-D_2^i\big)\,\Big\{
  -\f{1}{2}\big(D\cdot \lambda_1+\bar{D}\cdot \bar{\lambda}_1\big)\big(\delta_{\lambda_2}^{[0]}d\big)
  -3 \lambda_1^j D_j\big(\delta_{\lambda_2}^{[0]}d \big) 
  -\bar{\lambda}_1^j\bar{D}_{j}\big(\delta_{\lambda_2}^{[0]}d \big)
    \\[1.0ex]
  &\qquad\qquad\qquad\quad\quad   
  + \bar{D}^k\lambda_{1j}\,\bigl(\delta_{\lambda_2}^{[0]} e^{j}{}_{k}\bigr)
  +D^k\lambda_{1j}\,D^j\lambda_{2k}
  -D^k\lambda_{1j}\,D_{k}\lambda_{2}^j-(1\leftrightarrow 2)
    \Big\}\;. 
 \end{split}
 \ee  
It is now clear that all terms that involve a $\delta^{[0]}$ have the structure
that allows them to be removed by a suitable parameter redefinition.
It thus follows that the new gauge algebra is  
 \be
  \tilde{\lambda}_{12}^{(1)i} \ = \  - \f{1}{16}\big(D_1^i-D_2^i\big)\,\big\{ ~2D^k\lambda_{1j}\,D^j\lambda_{2k}
  -2D^k\lambda_{1j}\,D_{k}\lambda_{2}^j\, \big\} \;, 
 \ee
where we noted that the antisymmetry in $\lambda_1\leftrightarrow \lambda_2$ is 
automatic under the operator $(D_1^i-D_2^i)$. 
Dropping the tilde from now on, we have simplified the gauge algebra to  
 \be
  \begin{split}
 \lambda_{12}^{(1)i} \ &=
    \  -\f{1}{16}\big(D_1^i-D_2^i\big) \big(D^k\lambda_{1j}-D_{j}\lambda_1^k\big)
   \big(D^j\lambda_{2k}-D_k\lambda_2^j\big) \\[0.5ex]
    \ &= \  \  \f{1}{16} \big(D^k\lambda_{1j}-D_{j}\lambda_1^k\big)
    \DDD^{\,i}
   \big(D^j\lambda_{2k}-D_k\lambda_2^j\big)\;, 
 \end{split}
 \ee  
with $A \DDD B\equiv AD B-DA\,B$. 
In the notation of (\ref{sftcb2}) 
we have identified the $\alpha'$-corrected gauge algebra or bracket
as 
\be\label{finalalpha'bracket}
   \lambda_{ 12}^i \ \equiv \  \lambda_{c,12}^i \ - \    
     \f{1}{16} \alpha' \big(D_j\lambda_1^k-D^k\lambda_{1j}\big)  \DDD^{\,i}
    \big(D^j\lambda_{2k}-D_{k}\lambda_2^j\big) \;. 
  \ee    
Note that for trivial parameters $\lambda^i=D^i\chi$ the full
$\alpha'$ corrected bracket vanishes. 
Moreover, this algebra is purely holomorphic. 
An exactly analogous treatment of the barred parameter
would yield 
\be\label{finalalpha'bracket-AH}
   \bar\lambda_{ 12}^i \ \equiv \  \bar\lambda_{c,12}^i \ - \ \f{1}{16}\alpha'  \big(\bar D_j\bar \lambda_1^k-\bar D^k\bar\lambda_{1j}\big)  \DDDB^{\,i}
    \big(\bar D^j\bar\lambda_{2k}-\bar D_{k}\bar\lambda_2^j\big) \;. 
  \ee    
Employing the notation   
 \be
  K_{ij} \ \equiv \ 2D_{[i}\, \lambda_{j]}\;, \qquad \bar{K}_{ij} \ \equiv \ 2\bar{D}_{[i}\, \bar{\lambda}_{j]}\;, 
 \ee 
the algebra takes the form 
\be\label{finalalpha'bracket--grouped using K}
\begin{split}
   \lambda_{ 12}^i \ \equiv \  &  \ \lambda_{c,12}^i \ - \    
     \f{1}{16} \alpha'\, \bigl(  K_1^{kl}  D^i
    K_{2kl}-  (1\leftrightarrow 2) \bigr)  \;,  \\[0.6ex]
    \bar\lambda_{ 12}^i \ \equiv \   & \  \bar\lambda_{c,12}^i \ - \ \f{1}{16}\alpha'  
    \, \bigl(\bar K_1^{kl}  \bar D^i 
    \bar K_{2kl} -  (1\leftrightarrow 2)\bigr) \;. 
    \end{split}
  \ee    

Although this holomorphic/antiholomorphic presentation of the bracket is
intriguing,
it turns out to be useful to perform one more parameter redefinition that mixes holomorphic and 
antiholomorphic parts.  In fact, the original string field theory
gauge algebra mixes holomorphic 
and antiholomorphic parameters, and the C-bracket does as well.  
Such mixing  leads to a simplified form of the gauge transformations, which we will 
discuss in the next subsection. The parameter redefinition, in the form (\ref{FinalParared}), uses
parameters $\Lambda^i$ and $\bar\Lambda^i$ given by 
 \be
  \begin{split} 
   \Lambda^i \ = \ -\f{1}{8}\alpha'\big(\bar{D}_k\bar{\lambda}_l-\bar{D}_l\bar{\lambda}_k\big)\bar{D}^k e^{il}\,, \ \ \ \ \ 
   \bar{\Lambda}^i \ = \ -\f{1}{8} \alpha' \big(D_k\lambda_l-D_l\lambda_k)D^k e^{li}\;.
  \end{split}
 \ee  
This leads to the redefined gauge algebra 
 \be
 \begin{split}
  \tilde{\lambda}_{12}^i \ &= \ \lambda^i_{12}  
  -\f{1}{8}\alpha'\big[\big(\bar{D}_k\bar{\lambda}_{2l}-\bar{D}_l\bar{\lambda}_{2k}\big)
  \bar{D}^k\big(D^i\bar{\lambda}_1^l+\bar{D}^l\lambda_1^i\big)-(1\leftrightarrow 2) \big] \\[1.5ex] 
   \ &= \ \lambda^i_{12}-\f{1}{16}\alpha'\big[\big(\bar{D}_k\bar{\lambda}_{2l}-\bar{D}_l\bar{\lambda}_{2k}\big)
  D^i\big(\bar{D}^k\bar{\lambda}_1^l-\bar{D}^l\bar{\lambda}_1^k\big)-(1\leftrightarrow 2) \big] \;, 
 \end{split}
 \ee 
where we noted that the second derivative $\bar{D}^k\bar{D}^l$ is symmetric 
in $k,l$ and so drops out of the antisymmetric contraction. 
Dropping the tilde, and combining with
(\ref{finalalpha'bracket--grouped using K}) the gauge algebra finally becomes 
\be
\label{finalalpha'bracket2}
\begin{split}
\hbox{CSFT gauge algebra:} \ \  \ \   \lambda_{12}^i \ \equiv \ & \  \lambda_{c,12}^i \ - \ \f{1}{16} \alpha'\big(\, K_1^{kl} D^i
   K_{2kl} -\bar{K}_1^{kl} D^i \bar{K}_{2kl}-(1\leftrightarrow 2)\,\big) \,, \\[1.0ex]
   \bar\lambda_{12}^i \ \equiv  \ & \  \bar\lambda_{c,12}^i \ - \ 
   \f{1}{16} \alpha'\big(\, \bar{K}_1^{kl} \bar{D}^i
   \bar{K}_{2kl} -K_1^{kl} \bar{D}^i K_{2kl}-(1\leftrightarrow 2)\,\big) \;, 
   \end{split}
  \ee    
  where we have included the corresponding antiholomorphic part. 
This is the final form of the CSFT gauge algebra that we will use next to determine the $\alpha'$-deformed 
gauge transformation.  The C${}_{\alpha'}$ bracket is read from the above
results and the definitions:
 \be
 \bigl[ \delta_{\Lambda_1} \,, \delta_{\Lambda_2} \bigr] \ = \ \delta_{\Lambda_{12}} \,, \quad \ \Lambda_{12}  \ \equiv \  [\,\Lambda_2\, , \Lambda_1\, ]_{{}_{\hbox{c}_{\alpha'}}} \quad \to \quad 
 (\lambda_{12} , \bar \lambda_{12} )  \ \equiv \  [\,(\lambda_2, \bar\lambda_2)\, , (\lambda_1, \bar\lambda_1) \, ]_{{}_{\hbox{c}_{\alpha'}}} \ \, . 
 \ee

\subsection{Gauge transformations}
We now determine the corrected gauge transformations  that close according to the 
gauge algebra (\ref{finalalpha'bracket2}). Rather than finding them 
from CSFT by performing a series of laborious field and parameter redefinitions it is 
easier to obtain them from the gauge algebra.  
To this end we consider the commutator of transformations on
the field $e_{ij}$.  If we only know the $\delta^{[0]}$ and $\delta^{[1]}$ transformations
we find 
 \be\label{recallalgebra}
 \begin{split}
  \Big[\,\delta_{\Lambda_1},\delta_{\Lambda_2}\,\Big] e_{ij} 
  \ = \ & \ \delta_{\Lambda_1}^{[0]}
  \big(\delta_{\Lambda_2}^{[1]}e_{ij}\big) -(1\leftrightarrow 2)  + {\cal O} (e) \\
 \ = \ & \  \delta_{\Lambda_{12}}^{[0]} e_{ij} + {\cal O} (e)   
  \ = \ D_i\bar{\lambda}_{12\,j}+\bar{D}_{j}\lambda_{12\,i} + {\cal O} (e)  \; , 
 \end{split} 
 \ee
 which means that 
 \be
 \label{sgbb}
 \delta_{\Lambda_1}^{[0]}\big(\delta_{\Lambda_2}^{[1]}e_{ij}\big) -(1\leftrightarrow 2) \ = \  
 D_i\bar{\lambda}_{12\,j}+\bar{D}_{j}\lambda_{12\,i}\,.
 \ee
We can look at the first order in $\alpha'$ part of this equation.
Noting that $\delta^{[0]}$ receives no $\alpha'$ correction but $\delta^{[1]}$ does, 
we will write
\be
\delta^{[1]} \ = \ \delta^{[1](0)} +  \alpha' \delta^{[1](1)}  + {\cal O} (\alpha'^2) \,, \ee
and use this to evaluate the left-hand side. 
For the right-hand side we need the parts in $\lambda_{12}$ and $\bar\lambda_{12}$ 
 (\ref{finalalpha'bracket2})  linear in $\alpha'$.  A  quick computation gives
 \be
\delta_{\Lambda_1}^{[0]}\big(\delta_{\Lambda_2}^{[1](1)}e_{ij}\big) -(1\leftrightarrow 2) 
\ = \ 
    -\f{1}{8}  
    \big(D_i\bar{K}_1^{kl}\,\bar{D}_{j}\bar{K}_{2kl}-D_iK_{1}^{kl}\,\bar{D}_{j}K_{2kl}-(1\leftrightarrow 2) \big)\;. 
 \ee   
We note that the terms of the form $KDDK$ cancelled under the $(1\leftrightarrow 2)$ antisymmetrization,  while the terms of the form $DK DK$ added up. 
We now have  to rewrite the right-hand side as
a total $\delta^{[0]}$
variation. To this end we write out one of the $K$ factors in each term, 
using the manifest antisymmetry imposed by the other factor,  and compute 
  \be
  \begin{split}
  \delta_{\Lambda_1}^{[0]}\big(\delta_{\Lambda_2}^{[1](1)}e_{ij}\big) -(1\leftrightarrow 2) 
  \ &= \ 
    -\f{1}{4}
    \big(\bar{D}^kD_i\bar{\lambda}_1^{l}\,\bar{D}_{j}\bar{K}_{2kl}-D_{i}K_{1}^{kl}
    \,D_k\bar{D}_j\lambda_{2l} -(1\leftrightarrow 2) \big) \\[1ex]
    \ &= \  -\f{1}{4} 
    \big(\delta_{\lambda_1}^{[0]}\big(\bar{D}^ke_{i}{}^{l}\big)\,\bar{D}_{j}\bar{K}_{2kl}
    +D_{i}K_{2}^{kl}
    \,\delta_{\lambda_1}^{[0]}\big(D_ke_{lj}\big) -(1\leftrightarrow 2) \big)\\[1ex] 
    \ &= \  -\f{1}{4} 
    \delta_{\lambda_1}^{[0]} 
    \big(\bar{D}^ke_{i}{}^{l} \,\bar{D}_{j}\bar{K}_{2kl}
    +D_{i}K_{2}^{kl}
    \,D_ke_{lj}  -(1\leftrightarrow 2) \big) \;, 
 \end{split}
 \ee  
where we used the $(1\leftrightarrow 2)$ antisymmetry in passing from the first
to the second line. 
Note also that while $\delta^{[0]} e_{ij}$ has two terms,  only one term 
survives due to the contraction with the antisymmetric $K$'s.  After a slight reordering of terms, we infer that closure of the gauge algebra holds  
for 
 \be\label{closinggauge}
   \delta_{\Lambda}^{[1](1)}  e_{ij} \ = \ -\f{1}{4} 
   \big[D_ke_{lj}\,D_iK^{kl}
  +\bar{D}_k e_{il}\,\bar{D}_j\bar{K}^{kl}\big]\;.
 \ee 
Writing out $K$, the result takes the form  
 \be
 \label{olafsecondversion}
  \delta_{\Lambda}^{[1](1)} e_{ij} \ = \ -\f{1}{4} 
  \big[D^ke_{lj}\,D_i\big(D_k\lambda^l-D^l\lambda_k\big)
  +\bar{D}^k e_{il}\,\bar{D}_j\big(\bar{D}_k\bar{\lambda}^l-\bar{D}^l\bar{\lambda}_k\big)\big]\;, 
 \ee 
which closes according to the $\alpha'$-deformed gauge algebra (\ref{finalalpha'bracket2})
predicted by CSFT. 
Let us finally note that for this gauge algebra, to order $\alpha'$, we have
$D\cdot\lambda_{12}+\bar{D}\cdot\bar{\lambda}_{12}=0$. 
This implies that the dilaton gauge transformations need not be deformed in order to be compatible with 
the deformed gauge algebra. Indeed, we will see below that a gauge invariant action can be constructed without 
changing the dilaton gauge transformations.

\subsection{A two-parameter freedom in the gauge algebra}
We have used CSFT to determine
a consistent deformation of the gauge algebra of the two-derivative
theory and the associated deformations of the gauge transformations.
One may have suspected that this  would be the  
unique deformation (up to parameter and field redefinitions)
of the gauge structure to first order in
$\alpha'$.  We will see, however, that there is 
more freedom, given that the gauge algebra deformation of \cite{Hohm:2013jaa} 
does not coincide with the CSFT deformation above.  
There are two possibilities with definite $\mathbb{Z}_2$ properties
under the transformation $b \to -b$  
and a continuum of possibilities with indefinite $\mathbb{Z}_2$.

The more general gauge transformation can be obtained by  
using independent coefficients 
for the two terms in (\ref{closinggauge});
the term involving  $\lambda$ and the  term
involving $\bar{\lambda}$.  Introducing parameters $\gamma^\pm$ we write
this more general transformation as  
  \be\label{closinggaugegeneral}
   \delta_{\Lambda}^{[1](1)}  e_{ij} \ = \ -\f{1}{4}
   \big[(\gamma^++\gamma^-)D_ke_{lj}\,D_iK^{kl}
  +(\gamma^+-\gamma^-)\bar{D}_k e_{il}\,\bar{D}_j\bar{K}^{kl}\big]\;. 
 \ee 
A short computation shows that these close according to the deformed gauge algebra
\be\label{finalalpha'General}
 \begin{split} 
   \lambda_{12}^i\ &\equiv \  \lambda_{c,12}^i \ - \ \f{1}{16} \alpha' \big[\, (\gamma^++\gamma^-)K_1^{kl} D^i
   K_{2kl} -(\gamma^+-\gamma^-)\bar{K}_1^{kl} D^i \bar{K}_{2kl}-(1\leftrightarrow 2)\,\big] \;, \\[1.5ex]
    \bar\lambda_{ 12}^i \ &\equiv \  \bar\lambda_{c,12}^i \ - \ 
   \f{1}{16}\alpha' \big[\, (\gamma^+-\gamma^-) \bar{K}_1^{kl} \bar{D}^i
   \bar{K}_{2kl} -(\gamma^++\gamma^-)K_1^{kl} \bar{D}^i K_{2kl}-(1\leftrightarrow 2)\,\big] \;. 
  \end{split} 
  \ee    
For $\gamma^+=1$, $\gamma^{-}=0$ this reduces to 
the CSFT transformations and gauge algebra, respectively.
The second interesting case is $\gamma^+=0$, $\gamma^{-}=1$, 
which introduces a relative sign between 
holomorphic and antiholomorphic parts and 
for which we obtain the gauge transformation 
 \be
 \label{olafthirdversion}
   {\delta}_{\Lambda}^{[1](1)-} e_{ij} \ \equiv \ -\f{1}{4}
   \big[D^ke_{lj}\,D_i\big(D_k\lambda^l-D^l\lambda_k\big)
  -\bar{D}^k e_{il}\,\bar{D}_j\big(\bar{D}_k\bar{\lambda}^l-\bar{D}^l\bar{\lambda}_k\big)\big]\;, 
 \ee 
where we indicated the new transformation by adding the superscript ${}^{-}$. 
The corresponding gauge algebra reads 
 \be\label{secondFInalalgebra}
 \begin{split}
  \lambda_{12}^{i-} \ &= \ \lambda_{12,c}^i-\f{1}{16}\alpha'\big(K_{1}^{kl}\, D^i
  K_{2kl}+
  \bar{K}_1^{kl}\,D^i
  \bar{K}_{2kl} -(1\leftrightarrow 2) \big)\;,  \\[1.0ex]
  \bar{\lambda}_{12}^{i-} \ &= \ \bar\lambda_{12,c}^i+\f{1}{16}\alpha'\big(K_{1}^{kl} \, \bar{D}^i
  K_{2kl} +
   \bar{K}_1^{kl} \,\bar{D}^i
  \bar{K}_{2kl}-(1\leftrightarrow 2)  \big)\;.  \end{split}
 \ee   
We  note that for arbitrary $\gamma^+$ and $\gamma^-$ we still have
$D\cdot\lambda_{12}+\bar{D}\cdot\bar{\lambda}_{12}=0$. Therefore, this 
deformation is also consistent with a dilaton gauge transformation that is not 
changed. 

As we will show in more detail below, the $\delta^{-}$ gauge transformation  is 
an inequivalent deformation of the two-derivative gauge structure of DFT
and is the one that arises in \cite{Hohm:2013jaa}.  In fact, while the 
$\alpha'$ deformation implied by CSFT preserves the $\mathbb{Z}_2$ symmetry
 of the two-derivative DFT, 
the deformation $\delta^{-}$  violates  $\mathbb{Z}_2$ maximally. In the following these two different theories 
are referred to as DFT$^{+}$ and DFT$^{-}$, respectively. 
 We will discuss in the next chapter their relation 
to higher-derivative deformations of Einstein gravity with 
conventional gauge transformations.

\sectiono{$\alpha'$ corrections in Einstein variables}\label{alcoineiva}
In this section we discuss the relation of the CSFT field variable $e_{ij}$, that has $\alpha'$-deformed gauge transformations, to the usual variables $h_{ij}$ in Einstein gravity, that  transform under conventional diffeomorphisms.  We first show that in order to write 
the Riemann-squared term appearing in the $\alpha'$ expansion of string theory in a 
T-duality covariant way, we have to perform a redefinition that is not diffeomorphism covariant. 
This redefinition induces an $\alpha'$ deformed gauge transformation that
in turn can be matched with that of CSFT. Finally, we discuss the $\mathbb{Z}_2$ odd gauge transformations 
of DFT$^{-}$. We find that on the $b$-field the deformed gauge transformation cannot be 
related to that of a conventional 2-form. It has an anomalous term that, however, 
is exactly as required by the familiar Green-Schwarz anomaly cancellation.

\subsection{Riemann-squared and T-duality}\label{sectionTdualRiemann}
We start with the low-energy effective action of closed bosonic string theory to 
first order in $\alpha'$ \cite{Metsaev:1987zx,Meissner:1996sa}. 
For simplicity we set for now the dilaton and the $b$-field to zero. 
The action is then given by  
 \be
  S \ = \  \int  dx\, \sqrt{g}\,  \big(R +\f{1}{4}\alpha'R_{ijkl}R^{ijkl} \big)\;, 
 \ee 
where $R_{ijkl}$ denotes the Riemann tensor.
We recall that the Riemann-squared term gives a tensor structure in $g_{ij}$ that cannot be written in a $O(D,D)$  
covariant way~\cite{Hohm:2011si}. In a perturbative expansion 
$g_{ij} = \eta_{ij} + h_{ij}$ around a constant background and  
to cubic order in fluctuations  one finds 
\be
\label{actionalpha'0}
S \ = \  \int  dx\, \sqrt{g}\,  R  
\, + \,  \f{1}{4}\alpha' \int dx  \,\, \partial^k h^{lp} ~\partial^i h_{pq} ~
\partial_i \partial_k h^q{}_l   +\cdots \;, 
\ee
where to order $\alpha'$ we indicated only the cubic structure that is problematic. 
This term can be read off from eq.~(4.41) in \cite{Hohm:2011si}, upon expanding to 
cubic order in $h$. 
The claim is that all other  cubic 
terms, indicated by dots, 
can be written in $O(D,D)$ covariant form.

Before proceeding, let us briefly explain 
why this term is problematic 
for $O(D,D)$ covariance. 
We claim that there is no $O(D,D)$ covariant term that reduces to this 
structure upon setting $\tilde{\partial}=0$ and $b=0$. Such a term would have to be written in terms of 
$e_{ij}$ and derivatives $D_i$ and $\bar{D}_i$. It is easy to convince oneself, however, 
that such a term cannot be written, for a natural candidate like 
 \be
  D^ke^{lp}\,D^ie_{pq}\,D_iD_k e_{l}{}^{q}\;, 
 \ee
violates the rules for consistent index contractions reviewed in sec.~2. Indeed,  the summation index $p$
in the first factor has to be considered barred, but in the second factor unbarred, therefore violating 
$O(D,D)$ covariance. There is no other index assignment that would be consistent. 
Thus, Riemann-squared expanded to cubic order cannot be 
written in a T-duality covariant way in terms of $e_{ij}$. 

In order to proceed we now perform a 
field redefinition that removes the problematic term. We first note that the term can be written as 
\be
 \partial^k h^{lp} ~\partial^i h_{pq} ~
\partial_i \partial_k h^q{}_l    \ = \ \f{1}{2} \partial_i \bigl( 
 \partial^k h^{lp} ~ \partial^i h_{pq} ~
\partial_k h^q{}_l \bigr)  -\f{1}{2}  \, \partial^2 h_{pq} ~
 \partial^k h^{lp} ~\partial_k h_l{}^q \;. 
\ee
Ignoring the boundary term, the action (\ref{actionalpha'0}) becomes 
\be
\label{actionalpha'}
S\big[g\big] \ = \ \int  dx\, \sqrt{g}\,  R    
-\f{1}{8}\, \alpha' \int dx  \,\, \partial^2 h_{pq} ~
 \partial^k h^{lp} ~\partial_k h_l{}^q+\cdots ~\,. 
\ee
Consider now a field redefinition of the metric fluctuation, 
\be
g'_{ij} \ = \ \eta_{ij} + h'_{ij} \ = \ \eta_{ij} + h_{ij} + \delta h_{ij} \;, 
\ee
where we view $\delta h_{ij}$ to be of first order in $\alpha'$. 
Under such a redefinition, the Einstein-Hilbert term 
is shifted by 
\be
\delta ( \sqrt{g} R) \ = \ 
\sqrt{g} \, \delta g^{ij} 
\bigl( R_{ij} - \f{1}{2} g_{ij} R \bigr) \ = \ - 
\sqrt{g} \, \delta h_{ij} 
\bigl( R^{ij} - \f{1}{2} g^{ij} R\bigr)\;.  
\ee
We thus get for the action (\ref{actionalpha'}) expressed in terms of the redefined 
fields, to first order in $\alpha'$, 
\be
\begin{split}
S \big[g\big] \ &= \ S \big[g' -\delta g\big] \ = \ S\big[g'\big] +
 \int dx \sqrt{g}~ \delta h_{ij} \bigl( R^{ij} - \f{1}{2} g^{ij} R \bigr)
 + {\cal O}(\alpha'^2) \\[1ex]
 \ &= \  \int  dx\, \sqrt{g'}\,  R(g') 
+ \int dx \sqrt{g}~ \delta h_{ij} \bigl( R^{ij} - \f{1}{2} g^{ij} R \bigr)
-\f{1}{8}\, \alpha' \int dx  \, \partial^2 h_{pq} 
 \partial^k h^{lp} \partial_k h_l{}^q + {\cal O}(\alpha'^2).
\end{split}
\ee
As this is valid up to cubic terms in $h$, we can employ the 
linearized Ricci tensor and Ricci scalar in the second term, 
\be\label{FieldREdefSTep}
\begin{split}
S\big[g\big]  \  = \  &     \int  dx\, \sqrt{g'}\,  R(g') - \int dx ~ \delta h_{ij} 
\bigl(\, \f{1}{2} \partial^2 h^{ij} -   \partial^{(i} \partial_k h^{j)k} + \f{1}{2}\partial^i \partial^j h  
 + \f{1}{2} \eta^{ij} (-\partial^2 h +  \partial^p \partial^qh_{pq})\,  \bigr)
 \\[0.5ex]
 & \qquad \qquad\qquad\quad   -\f{1}{8}\, \alpha' \int dx  \,\, \partial^2 h_{pq} ~
 \partial^k h^{lp} ~\partial_k h_l{}^q + {\cal O}(\alpha'^2)\;. 
\end{split}
\ee
We now specialize the field redefinition to be of the form 
\be
\delta h_{ij}  \ = \ -\f{1}{4} \, \alpha' \, \partial_k h_i{}^l \, \partial^k h_{jl}\;.  
\ee
This cancels precisely the undesired term in the last line of (\ref{FieldREdefSTep}). 
It is easy to see that the  
remaining terms in (\ref{FieldREdefSTep}) can be written 
in $O(D,D)$ covariant form. 

To summarize, performing the following redefinitions of the metric fluctuation
\be
\label{hprime}
h_{ij}'  \ = \ h_{ij} - \f{1}{4} \, \alpha'\,\partial_k h_i{}^p \, \partial^k h_{jp}
+ \cdots  \,, 
\ee
we removed the problematic structure in Riemann-squared, which is necessary in order 
to make T-duality manifest. 
This result is compatible 
with a similar conclusion of Meissner \cite{Meissner:1996sa}, that 
analyzed reductions  to $D=1$ of the low-energy action to 
first order in $\alpha'$ and found that field redefinitions are necessary in order to make T-duality   
manifest. 
Specifically, he found the need for a redefinition of the external 
components $g_{ij}$ of the metric by terms quadratic in the 
first derivatives  of $g_{ij}$.  This redefinition precisely reduces to (\ref{hprime})
when expanded in 
fluctuations and for zero $b$-field.  Being first order in derivatives, 
such redefinitions are not diffeomorphism covariant 
and lead to modified metric gauge 
transformations, as expected from the CSFT results.  
In the next subsection we determine the full field redefinition including 
terms involving the $b$-field.

\subsection{Relation to Einstein variables for $\mathbb{Z}_2$ even transformations}
We now aim to connect the full closed SFT field $e_{ij}$ to the (perturbative) Einstein variable $\check{e}_{ij}$
defined as the fluctuation of the field ${\cal E}_{ij}$ formed by 
adding the metric to the Kalb-Ramond field  
\be
{\cal E}_{ij}
\  = \ G_{ij}+ h_{ij}  + B_{ij} + b_{ij} \ = \ E_{ij} +\check{e}_{ij}\,.
\ee
 Here  $E_{ij}= G_{ij}+ B_{ij}$ is the sum of the background metric
 and Kalb-Ramond field  and $\check{e}_{ij} = h_{ij} + b_{ij}$ 
is the sum of their fluctuations. 
In the two-derivative DFT this field redefinition is given by \cite{Michishita:2006dr,Hull:2009zb}
 \be\label{orignaleRed}
  \check{e}_{ij} \ = \ e_{ij}+\f{1}{2}e_{i}{}^{k}e_{kj}+\cdots\;, 
 \ee
where we omitted terms of higher order in fields (that are known in closed form).  
The form of the field redefinition can be fixed from the standard gauge transformation 
of $\check{e}_{ij}$ under diffeomorphisms and $b$-field gauge transformations for $\tilde{\partial}=0$ \cite{Hull:2009zb}. 
The conventional diffeomorphism and $b$-field gauge transformations are given by 
 \be\label{conventionaldiff}
  \delta \check{e}_{ij} \ = \ \partial_i\epsilon_j+\partial_j\epsilon_i
  +\partial_i\tilde{\epsilon}_j
  -\partial_j\tilde{\epsilon}_i+\epsilon^k\partial_k e_{ij}+\partial_i\epsilon^k e_{kj}
  +\partial_j\epsilon^k e_{ik}\;, 
 \ee
where $\epsilon^i$ is the diffeomorphism parameter and $\tilde{\epsilon}_i$ the one-form 
parameter.  The relation to the DFT gauge parameter $\xi^{M}=(\tilde{\xi}_i,\xi^i)$  is given by 
\be 
 \epsilon^i \ = \ \xi^i\;, \qquad \tilde{\epsilon}_i \ = \ \tilde{\xi}_i+B_{ij}\xi^j\;. 
\ee
The parameters $\epsilon^i$ and $\tilde{\epsilon}_i$  
are related to the CSFT parameters by 
 \be\label{CSFTparam}
  \lambda_i \ = \ \epsilon_i-\tilde{\epsilon}_i \;, \qquad
  \bar{\lambda}_j \ = \ \epsilon_j+\tilde{\epsilon}_j\;.  
 \ee 
 The form of the quadratic term in the field redefinition (\ref{orignaleRed}) is such that the 
 gauge transformation of $\check{e}_{ij}$ on the left-hand side follows as required by (\ref{conventionaldiff}),  
 with the right-hand side transforming according to the CSFT gauge transformations to zeroth order in 
 $\alpha'$, as shown in detail in  \cite{Hull:2009zb}. 
 
 Let us now investigate how (\ref{orignaleRed}) generalizes when including the first 
 $\alpha'$ correction. Since in this case $\delta e_{ij}$ receives a 
 higher-derivative correction, there must be higher-derivative terms in the field redefinition 
 (\ref{orignaleRed}) so that the extra variations cancel and the Einstein variable still transforms 
 as in (\ref{conventionaldiff}). In general, the relation (\ref{CSFTparam}) between the 
 gauge parameters may also receive $\alpha'$ corrections. 
Making a general ansatz one finds that the field redefinition takes the form 
 \be\label{deformedred}
  \begin{split}
    \check{e}_{ij} \ = \ &\,e_{ij}+\f{1}{2}e_{i}{}^{k}e_{kj}+\cdots \\[0.4ex]
    &+\f{1}{4}\alpha'\big[\,\partial^k e_{i}{}^{l}\,\partial_k e_{lj}-
  \partial^le_{i}{}^{k}\,\partial_k e_{lj}\\
  &\qquad\quad  
  -\partial^ke^{l}{}_{j}\,\partial_i
  \big(e_{kl}-e_{lk}\big)+\partial^ke_{i}{}^{l}\,\partial_j\big(e_{kl}-e_{lk}\big)
  -\f{1}{2}\partial_i e^{kl}\partial_j\big(e_{kl}-e_{lk}\big)\,\big]+\cdots \;, 
  \end{split}
 \ee   
where the dots represent terms higher order in fields and higher order in $\alpha'$.  
Moreover, the relation between gauge parameters indeed gets $\alpha'$ corrected, 
 \be\label{correctparametered} 
  \begin{split}
  \lambda_i \ &= \ \epsilon_i-\tilde{\epsilon}_i -\f{1}{4}\alpha'\partial_i\big(e_{kl}-e_{lk}\big)\partial^k\epsilon^l
  +{\cal O}(\alpha'^2) \;, \\[0.5ex]
  \bar{\lambda}_j \ &= \ \epsilon_j+\tilde{\epsilon}_j+\f{1}{4}\alpha'\partial_j\big(e_{kl}-e_{lk}\big)\partial^k\epsilon^l
   +{\cal O}(\alpha'^2)\;, 
  \end{split}
 \ee
or for the inverse 
  \be\label{finalparamrel}
 \begin{split}
  \tilde{\epsilon}_i \ &= \ -\f{1}{2}\big(\lambda_i-\bar{\lambda}_i\big)
  -\f{1}{8}\alpha'\, \partial_i\big(e_{kl}-e_{lk}\big)\partial^l
  \big(\lambda^k+\bar{\lambda}^k\big) +{\cal O}(\alpha'^2)\;, \\[0.5ex]
  \epsilon_i \ &= \ \ \f{1}{2}\big(\lambda_i+\bar{\lambda}_i\big) +{\cal O}(\alpha'^2) \;. 
 \end{split}
 \ee
Note that these redefinitions are T-duality violating, as it should be.  
In order to verify the claim that the above redefinitions are the right ones 
one has to compute the gauge transformation of the right-hand side of (\ref{deformedred}) 
by means of the $\alpha'$-deformed gauge transformation (\ref{olafsecondversion})
and the inhomogeneous transformation 
$\delta^{[0]}e_{ij}$  
 in the ${\cal O}(\alpha')$ terms, setting $\tilde{\partial}=0$.
A straightforward computation yields 
 \be
  \begin{split}
  \delta^{[1](1)}  \check{e}_{ij} \ &= \ -\tfrac{1}{4}\partial_ie^{[kl]}\,\p_j\p_{k}\left(\lambda_{l}+\bar{\lambda}_{l}\right)
  +\tfrac{1}{4}\p_je^{[kl]}\,\p_i\p_{k}\left(\lambda_{l}+\bar{\lambda}_{l}\right)\\[0.5ex]
  \ &= \ \p_i\bigl(+\tfrac{1}{4}\p_je_{[kl]}\,\p^{k}\big(\lambda^{l}+\bar{\lambda}^{l}\big)\bigr)
  -\p_j\bigl(\tfrac{1}{4}\p_ie_{[kl]}\,\p^{k}\big(\lambda^{l}+\bar{\lambda}^{l}\big)\bigr)\;, 
 \end{split}
 \ee
where, as indicated by the notation on $\delta$ on the left-hand side, we included only 
the terms ${\cal O}(\alpha')$ and linear in fields.   
This is precisely of the form of the ${\cal O}(\alpha')$ terms originating in 
$\delta \check{e}_{ij}=\partial_i\tilde{\epsilon}_j-\partial_j\tilde{\epsilon}_i$ through the 
deformation of the parameter redefinition in (\ref{finalparamrel}).  
Thus, we trivialized the higher-derivative deformation.  Together with the analysis in 
\cite{Hull:2009zb} it follows that the gauge transformations reduce to  
the conventional diffeomorphism and $b$-field 
ransformations (\ref{conventionaldiff}) for $\check{e}_{ij}$. This proves 
that the field and parameter redefinitions  (\ref{deformedred}), (\ref{correctparametered}) connect 
to conventional Einstein variables and symmetries. From the leading term in the second line of 
(\ref{deformedred}) one may verify that this field redefinition indeed contains the minimal redefinition (\ref{hprime}) 
needed in order to describe Riemann-squared (note here that $e_{ij}$
has to be identified with $h'_{ij}$ and $\check{e}_{ij}$ with $h_{ij}$).

\subsection{Relation to Einstein variables for $\mathbb{Z}_2$ odd transformations}
Let us now turn to the $\mathbb{Z}_2$ violating gauge transformations of DFT$^{-}$ 
defined in (\ref{olafthirdversion}), 
 \be
 \label{olafthirdversion2}
 \delta_{\Lambda}^{[1](1)-} e_{ij} \ = \ -\f{1}{4}\,\big[D^ke_{lj}\,D_i\big(D_k\lambda^l-D^l\lambda_k\big)
  -\bar{D}^k e_{il}\,\bar{D}_j\big(\bar{D}_k\bar{\lambda}^l-\bar{D}^l\bar{\lambda}_k\big)\big]\;. 
 \ee 
We will show that in contrast to the DFT$^{+}$ transformations discussed above, these 
transformations cannot be related to those of conventional metric and $b$-field fluctuations 
upon field and parameter redefinitions. More precisely, the deformed gauge transformation 
(\ref{olafthirdversion2}) leads a gauge transformation for the antisymmetric $b$-field part of the 
fluctuation that has a non-removable higher-derivative deformation of the diffeomorphism transformation. 
 
To analyze the relation of (\ref{olafthirdversion2}) to standard gauge transformations of 
Einstein-type variables we have to set $\tilde{\partial}=0$. Useful relations between the 
different gauge parameters then follow from (\ref{CSFTparam})
 \be\label{lambdarel}
  \begin{split}
   \partial_k\lambda_l -\partial_l\lambda_k \ &= \ 2\partial_{[k}\epsilon_{l]} -2\partial_{[k}\tilde{\epsilon}_{l]} \;, \\
   \partial_k\bar{\lambda}_l -\partial_l\bar{\lambda}_k \ &= \ 2\partial_{[k}\epsilon_{l]}+2\partial_{[k}\tilde{\epsilon}_{l]} \;. 
  \end{split}
 \ee
Here $\epsilon_i$ and $\tilde{\epsilon}_i$ are the diffeomorphism and $b$-field gauge parameter, respectively. 
Thus, the linearized gauge transformations for the symmetric and antisymmetric part of 
$e_{ij}\equiv h_{ij}+b_{ij}$ read to lowest order in fields 
  \be\label{inhomogauge}
   \delta h_{ij} \ = \ \partial_i\epsilon_j+\partial_j\epsilon_i\;, \qquad
   \delta b_{ij}  \ = \  \partial_i\tilde{\epsilon}_j-\partial_j\tilde{\epsilon}_i\;.
  \ee  
Next, we evaluate the deformed gauge transformation (\ref{olafthirdversion2}) 
for $h$ and $b$ by 
using (\ref{lambdarel}) and decomposing into the symmetric and antisymmetric parts, 
 \be\label{defdelta1}
  \begin{split}
   \delta^{[1](1)-} h_{ij} \ &= \ \f{1}{2}\partial^kh^{l}{}_{j}\,\partial_i\partial_{[k}\,\tilde{\epsilon}_{l]}
   +\f{1}{2}\partial^kb_{i}{}^{l}\,\partial_j\partial_{[k}\,\epsilon_{l]}+(i\leftrightarrow j)\;, 
   \\[0.5ex]
    \delta^{[1](1)-}  b_{ij} \ &= \ -\f{1}{2}\partial^kh^{l}{}_{j}\,\partial_i\partial_{[k}\,\epsilon_{l]}
   +\f{1}{2}\partial^kb^l{}_{j}\,\partial_i\partial_{[k}\,\tilde{\epsilon}_{l]}-(i\leftrightarrow j)\;. 
  \end{split}
 \ee  
These higher-derivative deformations, which are not present for standard Einstein variables, 
were the starting point for the analysis in \cite{Hohm:2014eba}.
There we showed that these gauge transformations can be brought to the form 
of those needed for Green-Schwarz anomaly cancellation. 
Specifically, we showed that through a combined parameter and field redefinition the 
gauge transformation of $h_{ij}$ can be trivialized, so that, to this order, it reduces to 
(\ref{inhomogauge}), while the gauge transformation of $b_{ij}$ can be brought to the form
  \be\label{finaldel2}
     \delta b_{ij} \ = \ \partial_i\tilde{\epsilon}_j-\partial_j\tilde{\epsilon}_i
     +\partial^{}_{[i}\partial^{k}\epsilon^{l}\,\omega_{j]kl}^{(1)}   \;, 
  \ee   
with the linearized spin connection  $\omega_{j,kl}^{(1)} \equiv  - \partial_{[k}\, h_{l]j}$.  
To this order, this is the gauge transformation of the Green-Schwarz mechanism, 
viewed as a deformation of diffeomorphisms (as opposed to local Lorentz transformations). 
We also showed in \cite{Hohm:2014eba} that the non-linear form of these deformed 
diffeomorphisms provides an exact realization of the deformed C-bracket of DFT$^{-}$.

\sectiono{Perturbation theory of DFT$^{-}$ and DFT$^{+}$}\label{pertheo}  
In this section we compare the gauge structure discussed so far to that of the theory
developed in the context of a `doubled $\alpha'$-geometry' in \cite{Hohm:2013jaa}.
We will show that this theory corresponds, in the above terminology, to DFT$^{-}$, 
i.e., to the $\mathbb{Z}_2$ violating case. To this end we first develop the perturbation theory
for the fundamental `double metric' field ${\cal M}$ introduced in  \cite{Hohm:2013jaa}
and discuss the $\mathbb{Z}_2$ action on these fields. We finally show how to relate these
perturbative variables to those appearing in CSFT.

\subsection{Perturbative expansion of double metric in DFT$^-$}  
The theory constructed in \cite{Hohm:2013jaa} features as fundamental fields the `double metric' 
${\cal M}_{MN}$, with $O(D,D)$ indices $M,N=1,\ldots,2D$, and the dilaton density $\phi$
(which is related to the CSFT dilaton used above by $\phi=-2d$).  In contrast to the generalized 
metric formulation of double field theory in \cite{Hohm:2010pp}, the field ${\cal M}_{MN}$
is not constrained by assuming that it takes values in $O(D,D)$. Rather, it is an unconstrained field 
that does not even need to be invertible off-shell. In \cite{Hohm:2013jaa} an exactly gauge invariant 
action with up to six derivatives was constructed. Although ${\cal M}$ is unconstrained, its 
field equations read ${\cal M}_{M}{}^{K}{\cal M}_{KN}=\eta_{MN}+\cdots$, where the dots represent 
higher-derivative corrections. To lowest order this equation implies ${\cal M}\in O(D,D)$, 
from which invertibility follows, but since this equation receives higher-derivative corrections its relation to 
the usual generalized metric and thus to the conventional metric and $b$-field is subtle. 

In the following we discuss the perturbative expansion of this theory around a constant 
background $\langle {\cal M}\rangle$. 
Being constant, the higher-derivative terms in the background field equations vanish 
and so the field equations are solved for any $\langle {\cal M}\rangle\in O(D,D)$. Thus, the background 
double metric can be identified with a  background {\em generalized} metric, 
 \be\label{firstH}
  \langle {\cal M}_{MN} \rangle \ \equiv \ \bar {\cal H}_{MN} \ =    
  \  \begin{pmatrix}    G^{ij} & -G^{ik}B_{kj}\\[0.5ex]
  B_{ik}G^{kj} & G_{ij}-B_{ik}G^{kl}B_{lj}\end{pmatrix}\;,
 \ee
where $G$ and $B$ are the (constant) background metric and $B$-field. 
In the following it will be convenient to use a notation introduced in \cite{Hohm:2011si}. 
To explain this notation  
note that due to 
${\cal H}_{MN} {\cal H}^N{}_P = \eta _{MP}$ 
we may introduce the two background projectors \cite{Hohm:2010pp}
\be\label{defPRoj}
 P \ = \ \frac{1}{2}\big(\eta-\bar \H)\;, \qquad \bar{P}   
 \ = \ \frac{1}{2}\big(\eta+\bar\H\big)\;, 
\ee 
satisfying $P^2=P$, $\bar{P}^2=\bar{P}$ and $P\bar{P}=0$. Then we define projected 
$O(D,D)$ indices by 
 \be
\label{defprojindices}
\begin{split}
W_{\nin{M}} \ \equiv  \  P_M{}^N \, W_N \,,\qquad 
W_{\bar{M}} \ \equiv \  \bar P_M{}^N \, W_N\,, 
\end{split}
\ee   
and similarly for arbitrary $O(D,D)$ tensors. Note that due to the projector identity $P+\bar{P}={\bf 1}$ we
can decompose any tensor into components with projected indices, e.g., for a vector 
$W_M  =  W_{\nin{M}} +  W_{\bar M}$. 
We also use this notation for the partial derivatives, so that the strong constraint implies 
 \be
  \partial^M\partial_M \ = \ 0 \quad \Rightarrow \quad  \partial^{\,\nin{M}} \partial_{\,\nin{M}} \ = \ 
   - \partial^{\bar M} \partial_{\bar M}\;. 
 \ee 

We are now ready to set up the perturbative expansion of ${\cal M}$ around the background $\bar {\cal H}$. 
Since ${\cal M}$ is unconstrained off-shell, the expansion is simply  
 \be\label{calMexpansion}
  \M_{MN} \ = \ \bar\H_{MN}+m_{MN} \ = \ \bar\H_{MN}+m_{\bar{M}\bar{N}}+m_{\bar{M}\,\nin{N}}+m_{\,\nin{M}\bar{N}}
  +m_{\,\nin{M}\,\nin{N}}\;,
 \ee 
with unconstrained symmetric fluctuations $m_{MN}= m_{NM}$  
that we decomposed into projected indices as explained above.  
Being unconstrained, the perturbation fields $m_{MN}$ has more than 
the $D\times D$ components needed to encode the metric and $b$-field fluctuations, but we will show 
that the projections 
$m_{\bar{M}\bar{N}}$ and $m_{\,\nin{M}\,\nin{N}}$ are auxiliary fields, while the physical part is encoded in 
$m_{\,\nin{M}\bar{N}} = m_{\bar{N}\,\nin{M}}$ (symmetry properties of tensors 
imply the same properties for the projected components). 

In order to verify this claim we have to inspect the Lagrangian in a derivative expansion around 
the background. The relevant action can be straightforwardly computed from the two-derivative 
truncation, see eq.~(7.13) in  \cite{Hohm:2013jaa}, 
which reads 
 \be\label{3TermVersion}
   \begin{split}
   S 
     \ &= \  \int e^{\phi}\,\big[\,\f{1}{2} ú^{MN}( \M-\f{1}{3}\M^3)_{MN}+\f{1}{2}(\M^2-1)^{MP} \M_{P}{}^{N}
   \p_M\p_N\phi\\
   &\qquad +\f{1}{8}\M^{MN}\p_M\M^{PQ} \p_N\M_{PQ}-\f{1}{2}\M^{MN} \p_N\M^{KL}\p_L\M_{KM}
   -\M^{MN}\p_M\p_N\phi\,\big]\;. 
  \end{split}
 \ee  
Note that this action contains terms without derivatives. Inserting the expansion (\ref{calMexpansion})
and keeping all terms with no derivatives and quadratic terms with two derivatives we find the Lagrangian 
 \be
  \begin{split}
   {\cal L} \ =\   & \ \ \  \half \, m^{\nin{M}\nin{M}} \, m_{\nin{M}\nin{N}}
   \ - \half  \, m^{\nin{M}\nin{N}} \, m_{\nin{M}}{}^{\bar{P}}\, m_{\nin{N}\bar{P}}
     \ -{\textstyle{1\over 6}} \,
      m^{\nin{M}\nin{N}} \, m_{\nin{N}}{}^{\nin{P}}\, m_{\nin{N}\nin{P}}  \\[1.0ex]
     & -\half m^{\bar{M}\bar{N}}m_{\bar{M}\bar{N}}
      \ - \half  \, m^{\bar{M}\bar{N}} \, m^{\, \nin{P}}{}{}_{\bar{M}}\, m_{\nin{P}\bar{N}}
     \ -{\textstyle{1\over 6}} \,
      m^{\bar{M}\bar{N}} \, m_{\bar{N}}{}^{\bar{P}}\, m_{\bar{N}\bar{P}} \\[2ex]
   &+\half \, \partial^{\bar{M}}m^{\nin{P}\bar{Q}}\,\partial_{\bar{M}} m_{\nin{P}\bar{Q}}
  \,  +\, \half \, \partial^{\,\nin{M}}
  m^{\nin{P}\bar{Q}}\,\partial_{\,\nin{P}}m_{\, \nin{M}\bar{Q}}
 \,   -\, \half \, \partial^{\bar{M}}m^{\nin{P}\bar{Q}}\,
   \partial_{\bar{Q}}m_{\nin{P}\bar{M}}\\[1.0ex]
   &-2\, m^{\nin{M}\bar{N}}\,\partial_{\,\nin{M}}\partial_{\bar{N}}\phi-2\phi\, \partial^{\bar{M}}\partial_{\bar{M}}\phi\\[2.0ex]
   &+\quarter \, \partial^{\bar{M}}m^{\bar{P}\bar{Q}}\,\partial_{\bar{M}}m_{\bar{P}\bar{Q}}
  \,  +\, \quarter\partial^{\bar{M}}m^{\,\nin{P}\,\nin{Q}}\,\partial_{\bar{M}} m_{\,\nin{P}\,\nin{Q}}\\[1.0ex]
&    +\half\partial^{\,\nin{M}} m^{\,\nin{P}\,\nin{Q}}\,\partial_{\,\nin{Q}}m_{\,\nin{P}\,\nin{M}}
  \,  -\, \half\partial^{\bar{M}}m^{\bar{P}\bar{Q}}\,\partial_{\bar{Q}} m_{\bar{P}\bar{M}}\,.
  \\[1.0ex]
  \end{split}
 \ee  
The first two lines are the terms with no derivatives,  the next two lines contain
the physical fields, and the last two lines contain derivatives of the auxiliary fields.
Solving for the auxiliary fields to lowest order in fields and without derivatives,
the first two terms in the first and second lines give
 \be\label{lowestorder}
 \begin{split}
  m_{\,\nin{M}\,\nin{N}} \ = \ & \ \ \  \half\,m_{\nin{M}}{}^{\bar{P}} m_{\,\nin{N}\bar{P}}
  + \ldots \;, \\[1.0ex]
   m_{\bar{M}\bar{N}} \ = \ & -\half \,m^{\nin{P}}{}_{\bar{M}} m_{\,\nin{P}\bar{N}} + \ldots \;,
 \end{split}
 \ee
 where dots indicate terms with more fields or derivatives.  
Next we eliminate the auxiliary fields, which does not affect the two-derivative quadratic action 
for the physical fields. This action  is then 
 \be\label{twoderaction}
  \begin{split}
   {\cal L}^{(2)} \ =\      &\ \ \half \, \partial^{\bar{M}}m^{\nin{P}\bar{Q}}\,\partial_{\bar{M}} m_{\nin{P}\bar{Q}}
  \,  +\, \half \, \partial^{\,\nin{M}}
  m^{\nin{P}\bar{Q}}\,\partial_{\,\nin{P}}m_{\, \nin{M}\bar{Q}}
 \,   -\, \half \, \partial^{\bar{M}}m^{\nin{P}\bar{Q}}\,
   \partial_{\bar{Q}}m_{\nin{P}\bar{M}}\\[1.0ex]
   &-2\, m^{\nin{M}\bar{N}}\,\partial_{\,\nin{M}}\partial_{\bar{N}}\phi-2\phi\, \partial^{\bar{M}}\partial_{\bar{M}}\phi\;. 
      \end{split}
 \ee  
This is the quadratic approximation to the two-derivative standard
DFT action~\cite{Siegel:1993th,Hull:2009mi}. 
Beyond this approximation 
the auxiliary fields will be determined non-trivially in terms of the physical fields. 

Let us now turn to the gauge symmetries for the fluctuations $m_{MN}$. 
These can be obtained  from the gauge transformations
 in \cite{Hohm:2013jaa}, eq.~(6.39), which are\footnote{The different coefficient on the final term arises because here we use a symmetrization convention with unit weight.}
\be   
\begin{split}
¶_\xi \M^{MN} \ = & \;\;\;  
\xi^P »_P \M^{MN} + ( »^{M}\xi_{P} - »_{P}\xi^{M})\M^{PN}  
 + ( »^{N}\xi_{P} - »_{P}\xi^{N})\M^{MP}  
\\[0.8ex]
	& - ü\, \bigl[ \, »^M \hskip-2pt\M^{PQ}\, »_P (»_Q \xi^N -»^N \xi_Q )  + 2\,»_Q \M^{KM}\, »^{N} »_K \xi^Q
	+ (MªN)  \bigr]  
	\ \   \\[0.8ex]
	&  -\,  
	\tfrac{1}{2} \,  »_K »^{(M} \hskip-2pt \M^{PQ}\,  »^{N)} 
	\hskip-1pt »_P »_Q \xi^K \,.
\end{split}
\ee
Upon insertion of (\ref{calMexpansion}), and  including up to three derivatives in the transformation rules one obtains 
 \be
  \begin{split}
   \delta_{\xi}^-m_{MN}   
   \ = \ \;&\xi^{P}\partial_P m_{MN}+\big(\partial_M\xi^P-\partial^P\xi_M\big)\bar {\cal H}_{PN}  
   +\big(\partial_N\xi^P-\partial^P\xi_N\big)\bar{\cal H}_{PM}\\
   &+\big(\partial_M\xi^P-\partial^P\xi_M\big)m_{PN}+\big(\partial_N\xi^P-\partial^P\xi_N\big)m_{PM}\\[0.3ex]
   &-\f{1}{2}\partial_Mm^{PQ}\,\partial_P\big(\partial_Q\xi_N-\partial_N\xi_{Q}\big)
   -\f{1}{2}\partial_N m^{PQ}\,\partial_P
   \big(\partial_Q\xi_M-\partial_M\xi_{Q}\big)\\[0.4ex]
   &-\partial_Qm_{MK}\,\partial_N\partial^K\xi^Q-\partial_Qm_{NK}\,\partial_M\partial^K\xi^Q\;. 
  \end{split}
 \ee  
We added the minus superscript to $\delta_\xi$ to emphasize that these are the gauge
transformations for DFT$^-$.  Next we decompose the indices into their projected parts according to (\ref{defprojindices}). 
Using $\bar\H=\bar{P}-P$, which follows from (\ref{defPRoj}), we compute 
  \be\label{deltamstep}
  \begin{split}
   \delta_{\xi}^-m_{MN} \ = \ \;&\big(\partial_M\xi_{\bar{N}}-\partial_{\bar{N}}\xi_M\big)
   +\big(\partial_N\xi_{\bar{M}}-\partial_{\bar{M}}\xi_M\big)
   -\big(\partial_M\xi_{\,\nin{N}}-\partial_{\,\nin{N}}\xi_M\big)
     -\big(\partial_N\xi_{\,\nin{M}}-\partial_{\,\nin{M}}\xi_N\big) \\
   &+\xi^{P}\partial_P m_{MN}+\big(\partial_M\xi^P-\partial^P\xi_M\big)m_{PN}+\big(\partial_N\xi^P-\partial^P\xi_N\big)m_{PM}\\
   &-\f{1}{2}\partial_Mm^{PQ}\,\partial_P\big(\partial_Q\xi_N-\partial_N\xi_{Q}\big)
   -\f{1}{2}\partial_N m^{PQ}\,\partial_P\big(\partial_Q\xi_M-\partial_M\xi_{Q}\big)\\
   &-\partial_Qm_{MK}\,\partial_N\partial^K\xi^Q-\partial_Qm_{NK}\,\partial_M\partial^K\xi^Q\;. 
  \end{split}
 \ee  
We now specialize this to the external projection corresponding to the physical fluctuation $m_{\,\nin{M}\bar{N}}$
and eliminate the auxiliary fields by use of  the lowest-order 
result (\ref{lowestorder}). This yields for the gauge transformation of the physical field 
 \be\label{physicalgauge}
  \begin{split}
   \delta_{\xi}^-m_{\,\nin{M}\bar{N}} \ = \ & \ \ 
 2\left(\partial_{\,\nin{M}}\xi_{\bar{N}}-\partial_{\bar{N}}\xi_{\,\nin{M}}\right)  \\[0.5ex]
   &   +\xi^{P}\partial_P m_{\nin{M}\bar{N}}
   +\big(\partial_{\,\nin{M}}\xi^{\,\nin{P}}-\partial^{\,\nin{P}}\xi_{\,\nin{M}}\big)m_{\nin{P}\bar{N}}
   +\big(\partial_{\bar{N}}\xi^{\bar{P}}-\partial^{\bar{P}}\xi_{\bar{N}}\big)m_{\,\nin{M}\bar{P}}\\[1.5ex]
   &-\half\partial_{\,\nin{M}}m^{\nin{P}\bar{Q}}\,\partial_{\,\nin{P}}
   \left(\partial_{\bar{Q}}\xi_{\bar{N}}-\partial_{\bar{N}}\xi_{\bar{Q}}\right)
   \,  -\, \half\partial_{\bar{N}}m^{\nin{Q}\bar{P}}\,\partial_{\bar{P}}
   \left(\partial_{\,\nin{Q}}\xi_{\,\nin{M}}
   -\partial_{\,\nin{M}}\xi_{\,\nin{Q}}\right)
    \\[0.9ex]
   &-\half\partial_{\bar{N}}m^{\nin{P}\bar{Q}}\,\partial_{\,\nin{P}}
   \left(\partial_{\bar{Q}}\xi_{\,\nin{M}}-\partial_{\,\nin{M}}\xi_{\bar{Q}}\right)
   \,  -\, \half\partial_{\,\nin{M}}m^{\nin{Q}\bar{P}}\,\partial_{\bar{P}}
   \left(\partial_{\,\nin{Q}}\xi_{\bar{N}}-\partial_{\bar{N}}\xi_{\,\nin{Q}}\right)
    \\[1.5ex]
   &-\partial_{\,\nin{Q}} m_{\nin{M}\bar{K}}\,
   \partial_{\bar{N}}\partial^{\bar{K}}\xi^{\,\nin{Q}}
 \  - \  \partial_{\bar{Q}}m_{\nin{K}\bar{N}}\,\partial_{\,\nin{M}}\partial^{\,\nin{K}}\xi^{\bar{Q}}
   \\[0.9ex]
   & 
   -\partial_{\,\nin{Q}} m_{\nin{K}\bar{N}}\,
   \partial_{\,\nin{M}}\partial^{\,\nin{K}}\xi^{\,\nin{Q}}
   \  -\  \partial_{\bar{Q}}m_{\nin{M}\bar{K}}\,
   \partial_{\bar{N}}\partial^{\bar{K}}\xi^{\bar{Q}}
\;. 
  \end{split}
 \ee 
Similarly, we can compute from (\ref{deltamstep}) the gauge transformation of the auxiliary fields, using again  
the lowest-order result (\ref{lowestorder}). We find for $m_{\,\nin{M}\,\nin{N}}$
 \be\label{auxiliarygauge}
 \begin{split}
  \delta_{\xi}^-m_{\,\nin{M}\,\nin{N}} \ = \ &\;\big(\partial_{\,\nin{M}}\xi^{\bar{P}}-\partial^{\bar{P}}\xi_{\,\nin{M}}\big)m_{\,\nin{N}\bar{P}}
  +\big(\partial_{\,\nin{N}}\xi^{\bar{P}}-\partial^{\bar{P}}\xi_{\,\nin{N}}\big)m_{\,\nin{M}\bar{P}}\\
  &\hspace{-0.18cm}-\frac{1}{2}\partial_{\,\nin{M}}m^{\nin{P}\bar{Q}}\,\partial_{\,\nin{P}}\big(\partial_{\bar{Q}}\xi_{\,\nin{N}}-
  \partial_{\,\nin{N}}\xi_{\bar{Q}}\big)
  -\f{1}{2}\partial_{\,\nin{N}}m^{\nin{P}\bar{Q}}\,\partial_{\,\nin{P}}\big(\partial_{\bar{Q}}\xi_{\,\nin{M}}-
  \partial_{\,\nin{M}}\xi_{\bar{Q}}\big)\\
  &\hspace{-0.18cm}-\f{1}{2}\partial_{\,\nin{M}}m^{\bar{P}\nin{Q}}\,\partial_{\bar{P}}\big(\partial_{\,\nin{Q}}\xi_{\,\nin{N}}-
  \partial_{\,\nin{N}}\xi_{\,\nin{Q}}\big)
  -\f{1}{2}\partial_{\,\nin{N}}m^{\bar{P}\nin{Q}}\,\partial_{\bar{P}}\big(\partial_{\,\nin{Q}}\xi_{\,\nin{M}}-
  \partial_{\,\nin{M}}\xi_{\,\nin{Q}}\big)\\
  &\hspace{-0.18cm}-\partial_{\,\nin{Q}}m_{\,\nin{M}\bar{K}}\,\partial_{\,\nin{N}}\partial^{\bar{K}}\xi^{\,\nin{Q}}
  -\partial_{\bar{Q}}m_{\,\nin{M}\bar{K}}\,\partial_{\,\nin{N}}\partial^{\bar{K}}\xi^{\bar{Q}}
  -\partial_{\,\nin{Q}}m_{\,\nin{N}\bar{K}}\,\partial_{\,\nin{M}}\partial^{\bar{K}}\xi^{\,\nin{Q}}
  -\partial_{\bar{Q}}m_{\,\nin{N}\bar{K}}\,\partial_{\,\nin{M}}\partial^{\bar{K}}\xi^{\bar{Q}}\;, 
 \end{split}
 \ee 
where we made the $\nin{M}\nin{N}$ symmetrization manifest in each line.  
We observe that there is no inhomogenous term, as required for (\ref{lowestorder}) to be consistent 
with the gauge symmetries. 
The gauge transformations determine the form of the auxiliary field 
to next order, which we give here for completeness, 
 \be
 \begin{split}  
  m_{\,\nin{M}\,\nin{N}} \ = \ \, \f{1}{2}\,m_{\nin{M}}{}^{\bar{P}} m_{\,\nin{N}\bar{P}} 
 \  &-\f{1}{4}\,\partial_{(\,\nin{M}}m^{\,\nin{P}\bar{Q}}\,\partial_{\,\nin{N})} m_{\,\nin{P}\bar{Q}}
  +\partial_{(\,\nin{M}}m^{\nin{P}\bar{Q}}\,\partial_{\,\nin{P}}m_{\,\nin{N})\bar{Q}}\\[0.5ex]
  &-\f{1}{2}\partial^{\bar{P}}m_{\,\nin{M}\bar{Q}}\,\partial^{\bar{Q}}m_{\nin{N}\bar{P}}
  +\f{1}{2} \partial^{\bar{P}} m_{\nin{M}}{}^{\bar{Q}}\,\partial_{\bar{P}}m_{\nin{N}\bar{Q}}+\cdots\;. 
 \end{split}
 \ee 
It is straightforward to verify, using the gauge transformations  of the 
physical field, that this expression gives rise to the required transformations (\ref{auxiliarygauge}).  
Analogous relations  hold for 
the auxiliary field $m_{\bar{M}\bar{N}}$. 
 
\medskip

In order to relate the perturbative field variable here to that of CSFT 
we first simplify the gauge transformations (\ref{physicalgauge}) by field and parameter redefinitions.  Consider the following field redefinition
 \be\label{parameterrREdef}
  m_{\nin{M}\bar{N}}' \ = \ m_{\nin{M}\bar{N}} + \half\partial^{\bar{P}}m_{\nin{Q}\bar{N}}\,\partial_{\,\nin{M}}
  m^{\,\nin{Q}}{}_{\bar{P}}-\half\partial^{\,\nin{P}}m_{\nin{M}\bar{Q}}\,\partial_{\bar{N}}m_{\,\nin{P}}{}^{\bar{Q}}\;. 
 \ee 
With (\ref{physicalgauge}) we can compute the gauge transformation of $m'$, 
after which we drop the prime, 
 \be\label{physicalgauge--redef}
  \begin{split}
   \delta_{\xi}^-m_{\,\nin{M}\bar{N}} \ = \ & \ \ 
 2\left(\partial_{\,\nin{M}}\xi_{\bar{N}}-\partial_{\bar{N}}\xi_{\,\nin{M}}\right)  \\[0.5ex]
   &   +\xi^{P}\partial_P m_{\nin{M}\bar{N}}
   +\big(\partial_{\,\nin{M}}\xi^{\,\nin{P}}-\partial^{\,\nin{P}}\xi_{\,\nin{M}}\big)m_{\nin{P}\bar{N}}
   +\big(\partial_{\bar{N}}\xi^{\bar{P}}-\partial^{\bar{P}}\xi_{\bar{N}}\big)m_{\,\nin{M}\bar{P}}\\[1.5ex]
   &-\half\partial_{\,\nin{M}}m^{\nin{P}\bar{Q}}\,\partial_{\,\nin{P}}
   \left(\partial_{\bar{Q}}\xi_{\bar{N}}-\partial_{\bar{N}}\xi_{\bar{Q}}\right)
   \,  -\, \half\partial_{\bar{N}}m^{\nin{Q}\bar{P}}\,\partial_{\bar{P}}
   \left(\partial_{\,\nin{Q}}\xi_{\,\nin{M}}
   -\partial_{\,\nin{M}}\xi_{\,\nin{Q}}\right)
    \\[0.9ex]
  \  &+\half\partial_{\bar{N}}m^{\nin{P}\bar{Q}}\,\partial_{\,\nin{P}}
   \left(\partial_{\bar{Q}}\xi_{\,\nin{M}}-\partial_{\,\nin{M}}\xi_{\bar{Q}}\right)
   \,  +\, \half\partial_{\,\nin{M}}m^{\nin{Q}\bar{P}}\,\partial_{\bar{P}}
   \left(\partial_{\,\nin{Q}}\xi_{\bar{N}}-\partial_{\bar{N}}\xi_{\,\nin{Q}}\right) \\[0.9ex]
    &+\partial_{\bar{P}}m_{\nin{Q}\bar{N}}\,\partial_{\,\nin{M}}
   \left(\partial^{\nin{Q}}\xi^{\bar{P}}-\partial^{\bar{P}}\xi^{\nin{Q}}\right)
   \,  -\, \partial_{\,\nin{P}}m^{\nin{M}\bar{Q}}\,\partial_{\bar{N}}
   \left(\partial^{\,\nin{P}}\xi^{\bar{Q}}-\partial^{\bar{Q}}\xi^{\,\nin{P}}\right)
    \\[1.0ex]
   &-\partial_{\,\nin{Q}} m_{\nin{M}\bar{K}}\,
   \partial_{\bar{N}}\partial^{\bar{K}}\xi^{\,\nin{Q}}
 \  - \  \partial_{\bar{Q}}m_{\nin{K}\bar{N}}\,\partial_{\,\nin{M}}\partial^{\,\nin{K}}\xi^{\bar{Q}}
   \\[0.9ex]
   & 
   -\partial_{\,\nin{Q}} m_{\nin{K}\bar{N}}\,
   \partial_{\,\nin{M}}\partial^{\,\nin{K}}\xi^{\,\nin{Q}}
   \  -\  \partial_{\bar{Q}}m_{\nin{M}\bar{K}}\,
   \partial_{\bar{N}}\partial^{\bar{K}}\xi^{\bar{Q}}
\;. 
  \end{split}
 \ee  
The third and fourth lines combine and so do the fifth and sixth, giving
 \be\label{physicalgauge--redef}
  \begin{split}
   \delta_{\xi}^-m_{\,\nin{M}\bar{N}} \ = \ & \ \ 
 2\left(\partial_{\,\nin{M}}\xi_{\bar{N}}-\partial_{\bar{N}}\xi_{\,\nin{M}}\right)  \\[0.5ex]
   &   +\xi^{P}\partial_P m_{\nin{M}\bar{N}}
   +\big(\partial_{\,\nin{M}}\xi^{\,\nin{P}}-\partial^{\,\nin{P}}\xi_{\,\nin{M}}\big)m_{\nin{P}\bar{N}}
   +\big(\partial_{\bar{N}}\xi^{\bar{P}}-\partial^{\bar{P}}\xi_{\bar{N}}\big)m_{\,\nin{M}\bar{P}}\\[1.5ex]
  &+\half\partial_{\,\nin{M}}m^{\nin{P}\bar{Q}}\,\partial_{\bar{N}}
   \left(\partial_{\,\nin{P}}\xi_{\bar{Q}}-\partial_{\bar{Q}}\xi_{\,\nin{P}}\right)
   \,  +\, \half\partial_{\bar{N}}m^{\nin{P}\bar{Q}}\,\partial_{\,\nin{M}}
   \left(\partial_{\bar{Q}}\xi_{\,\nin{P}}
   -\partial_{\,\nin{P}}\xi_{\bar{Q}}\right)
     \\[1.5ex]
   &-\partial^{\bar{Q}} m_{\nin{K}\bar{N}}\,
   \partial_{\,\nin{M}}\partial_{\bar{Q}}\xi^{\,\nin{K}}
 \  - \  \partial^{\,\nin{Q}}m_{\nin{M}\bar{K}}\,\partial_{\bar{N}}\partial_{\,\nin{Q}}\xi^{\bar{K}}
   \\[0.9ex]
   & 
   -\partial_{\,\nin{Q}} m_{\nin{K}\bar{N}}\,
   \partial_{\,\nin{M}}\partial^{\,\nin{K}}\xi^{\,\nin{Q}}
    \ - \ \partial_{\bar{Q}}m_{\,\nin{M}\bar{K}}\,
   \partial_{\bar{N}}\partial^{\bar{K}}\xi^{\bar{Q}}
\;. 
  \end{split}
 \ee 
Next we use the strong constraint in the line before last and relabel both there
and in the line below to obtain
\be\label{physicalgauge--redef-new}
  ~ \begin{split}
  \delta_{\xi}^-m_{\,\nin{M}\bar{N}} \ = \ & \ \ 
 2\left(\partial_{\,\nin{M}}\xi_{\bar{N}}-\partial_{\bar{N}}\xi_{\,\nin{M}}\right)  \\[0.5ex]
   &   +\xi^{P}\partial_P m_{\nin{M}\bar{N}}
   +\big(\partial_{\,\nin{M}}\xi^{\,\nin{P}}-\partial^{\,\nin{P}}\xi_{\,\nin{M}}\big)m_{\nin{P}\bar{N}}
   +\big(\partial_{\bar{N}}\xi^{\bar{P}}-\partial^{\bar{P}}\xi_{\bar{N}}\big)m_{\,\nin{M}\bar{P}}\\[1.0ex]
  &+\half\partial_{\,\nin{M}}m^{\nin{P}\bar{Q}}\,\partial_{\bar{N}}
   \left(\partial_{\,\nin{P}}\xi_{\bar{Q}}-\partial_{\bar{Q}}\xi_{\,\nin{P}}\right)
   \,  -\, \half\partial_{\bar{N}}m^{\nin{P}\bar{Q}}\,\partial_{\,\nin{M}}
   \left(\partial_{\,\nin{P}}\xi_{\bar{Q}}  - \partial_{\bar{Q}}\xi_{\,\nin{P}}\right)
  \\[1.1ex]
   &+\partial_{\nin{P}} m_{\nin{Q}\bar{N}}\,
   \partial_{\,\nin{M}} \bigl(  \partial^{\nin{P}}\xi^{\,\nin{Q}}- \partial^{\nin{Q}}\xi^{\,\nin{P}}\bigr) 
 \  + \  \partial_{\bar{P}}m_{\nin{M}\bar{Q}}\,\partial_{\bar{N}}
 \bigl( \partial^{\bar{P}}\xi^{\bar{Q}} - \partial^{\bar{Q}}\xi^{\bar{P}}\bigr) 
 \;.  
  \end{split} 
 \ee 
It is convenient to rewrite this in terms of 
 \be
  K_{MN} \ \equiv \ \partial_M\xi_N-\partial_N\xi_M\;,
 \ee 
which yields 
\be\label{physicalgauge--redef-new}
 \begin{split}
\delta_{\xi}^-m_{\,\nin{M}\bar{N}} \ = \ & \ \ 
 2\, K_{\nin{M} \bar N}\ +\xi^{P}\partial_P m_{\nin{M}\bar{N}}
   + K_{\,\nin{M}}{}^{\nin{P}}\, m_{\nin{P}\bar{N}}
   +K_{\bar{N}}{}^{\bar{P}}\,m_{\,\nin{M}\bar{P}}\\[1ex]
  &\;\;+\half\partial_{\,\nin{M}}m^{\nin{P}\bar{Q}}\,\partial_{\bar{N}}
   K_{\,\nin{P} \bar{Q}} 
   \,  -\, \half\partial_{\bar{N}}m^{\nin{P}\bar{Q}}\,\partial_{\,\nin{M}}
   K_{\,\nin{P} \bar{Q}}      \\[1ex]
   &\;\;+\partial_{\nin{P}} m_{\nin{Q}\bar{N}}\,
   \partial_{\,\nin{M}}  K^{\,\nin{P}\,\nin{Q}}  %
 \  + \  \partial_{\bar{P}}m_{\nin{M}\bar{Q}}\,\partial_{\bar{N}}
 K^{\bar{P}\bar{Q}}   \,.
   \end{split}  
 \ee 
The final form of the gauge transformations is obtained by performing a parameter 
redefinition, which eliminates the terms in the second line.  We take
\be
 \xi_{{M}}' \ = \ \xi_{{M}}-\quarter\partial_{{M}}
    K_{\, \nin{P} \bar{Q}} \,m^{\, \nin{P}\bar{Q}}\;, 
\ee
or, more explicitly, for the different projections, 
  \be
  \label{param-redefcorr} 
   \begin{split}
     \xi_{\,\nin{M}}' \ &= \ \xi_{\,\nin{M}}-\quarter\partial_{\,\nin{M}}
    K_{\, \nin{P} \bar{Q}} \,m^{\, \nin{P}\bar{Q}} \\[1.0ex]
      \xi_{\bar{N}}' \ &= \ \xi_{\bar{N}}-\quarter\partial_{\bar{N}}
     K_{\, \nin{P} \bar{Q}} \,m^{\,\nin{P}\bar{Q}}\;. 
   \end{split}
  \ee   
Dropping primes, the final form of the gauge transformations is 
\be\label{physicalgauge--redef-new2}
\begin{split}
  \delta_{\xi}^-m_{\,\nin{M}\bar{N}} \ = \ & \ \ 
 2\, K_{\nin{M} \bar N}\ +\xi^{P}\partial_P m_{\nin{M}\bar{N}}
   + K_{\,\nin{M}}{}^{\nin{P}}\, m_{\nin{P}\bar{N}}
   +K_{\bar{N}}{}^{\bar{P}}\,m_{\,\nin{M}\bar{P}}  \\[0.5ex]
   &+\partial_{\nin{P}} m_{\nin{Q}\bar{N}}\,
   \partial_{\,\nin{M}}  K^{\,\nin{P}\,\nin{Q}}  %
 \  + \  \partial_{\bar{P}}m_{\nin{M}\bar{Q}}\,\partial_{\bar{N}}
 K^{\bar{P}\bar{Q}}    %
 \;. 
  \end{split}  
 \ee  
Summarizing, the ${\cal O} (\alpha')$ correction to the gauge transformation  
is the second line above and is linear in the fields:
 \be\label{physicalgaugenew}
    \delta_{\xi}^{[1](1)-}m_{\,\nin{M}\bar{N}} \ = \ \partial_{\,\nin{P}} m_{\,\nin{Q}\bar{N}}\,
   \partial_{\,\nin{M}}  K^{\,\nin{P}\,\nin{Q}}  %
 \  + \  \partial_{\bar{P}}m_{\,\nin{M}\bar{Q}}\,\partial_{\bar{N}}
 K^{\bar{P}\bar{Q}}\;. 
 \ee

\subsection{$\mathbb{Z}_2$ action on fields}\label{Z2section}
We will now show that the deformations of 
gauge transformations determined in the previous subsection 
are $\mathbb{Z}_2$ odd and thus belong to DFT$^{-}$. 
To this end we first have to determine 
the action of $\mathbb{Z}_2$ on the field variables $m_{\,\nin{M}\bar{N}}$,
on derivatives and on gauge parameters. 
In the generalized metric formalism, the action of 
$\mathbb{Z}_2$ has been discussed in sec.~4.1 of \cite{Hohm:2010pp}. 
This symmetry acts on 
the background fields as 
$B_{ij}\rightarrow -B_{ij}$, so 
it is easy to see that on the (background) generalized metric  (\ref{firstH}) 
it is implemented by 
the $2D \times 2D$ matrix 
\be
Z_M{}^N \  \equiv \  \begin{pmatrix}  Z^i{}_j  & Z^{ij}  \\  Z_{ij} &  Z_i{}^j \end{pmatrix}
\ = \ \begin{pmatrix} - \delta^i{}_j &  0 \\ 0 &  \delta_i{}^j  \end{pmatrix}\;, \quad 
Z^2 = {\bf 1} \,, 
\ee
satisfying $Z^2={\bf 1}$. 
More precisely, $\mathbb{Z}_2$ acts on $O(D,D)$ indices via 
\be\label{Z2Odd}
\begin{split}
\partial_M \ \to \ & \  Z_M{}^N \partial_N\;,  \\[0.4ex] 
\bar{\cal H}_{MN} \ \to \ & \  Z_M{}^P Z_N{}^Q  \bar{\cal H}_{PQ}\;,  \\[0.4ex] 
\xi^M \ \to \ & \ \xi^N\,  Z_N{}^M \;.  \\[0.2ex] 
\end{split}
\ee
On the $D$-dimensional components this indeed reduces to the expected $\mathbb{Z}_2$ action, e.g., 
 \be
    B_{ij}\ \to \  -  B_{ij}\;,  \qquad
    \tilde\partial^i \ \to \  - \tilde\partial^i \;,  \qquad
    \tilde\xi_i \ \to \  - \tilde\xi_i \;, 
 \ee
leaving all objects without tilde unchanged.     
It is important to recall that $\mathbb{Z}_2$ is not part of $O(D,D)$. Indeed, 
the $\mathbb{Z}_2$ transformation does not leave the $O(D,D)$ metric invariant, 
\be\label{Z2oddoddmetric}
\begin{split}
Z_M{}^P Z_N{}^Q  {\eta}_{PQ}\ = \ - {\eta}_{MN}   \quad  \Longleftrightarrow \quad & \  
Z_N{}^K \,\eta_{MK} \ = \ - Z_M{}^K \, \eta_{NK}  
\;, 
\end{split} 
\ee
with the analogous relation for $\eta^{MN}$ with upper indices. This has important consequences for 
the $\mathbb{Z}_2$ action on $O(D,D)$ tensors for which indices have been raised 
or lowered with $\eta$. Specifically, taking the $O(D,D)$ tensors in (\ref{Z2Odd}) as 
fundamental, the corresponding ones with raised and lowered indices transform as 
\be\label{Z2signchange}
\begin{split}
\partial^M \ \to \ & \ -  \partial^N\, Z_N{}^M\;,   \\[0.4ex] 
\bar{\cal H}_M{}^N \ \to \ & \ -  Z_M{}^P Z_Q{}^N  \bar{\cal H}_P{}^Q \;, \\[0.4ex] 
\xi_M \ \to \ & \ - Z_M{}^N\xi_N\;,      \\[0.2ex] 
\end{split}
\ee
as a direct consequence of (\ref{Z2oddoddmetric}). 

Let us now determine the $\mathbb{Z}_2$ action on the various objects of the 
perturbative formalism introduced above, starting with the background projectors 
(\ref{defPRoj}). If we view them as having index structure $P_M{}^N$ and $\bar{P}_M{}^N$
the $\mathbb{Z}_2$ action changes the sign of the $\bar{\cal H}_M{}^N$ term according to  
(\ref{Z2signchange}), thereby exchanging $P$ and $\bar{P}$. We thus find 
\be\label{PZ2TRANS}
\begin{split}
P_M{}^N \ \to  \   \ Z_M{}^P  \,\bar P_P {}^Q  \, Z_Q{}^N \;, \qquad 
\bar{P}_M{}^N \ \to \  \  Z_M{}^P  \, P_P {}^Q  \, Z_Q{}^N  \;. 
\end{split}
\ee
If we view $P$ and $\bar{P}$ as tensors with lower indices, the leading $\eta_{MN}$ 
term changes sign according to (\ref{Z2oddoddmetric}), leading to an exchange of 
$P$ and $\bar{P}$ up to a global sign, 
\be
\begin{split}
P_{MN} \ \to  \   - Z_M{}^P Z_N{}^Q  \,\bar P_{PQ}  \;,  \qquad 
\bar{P}_{MN} \ \to \   - Z_M{}^P Z_N{}^Q  \, P_{PQ}   \;. 
\end{split}
\ee
From these results we can immediately determine the transformation of the projected derivatives, 
\be
\begin{split}
\partial_{\,\nin{M}} \ &\to \ \  Z_M{}^P \partial_{\bar{P}} \;,  \qquad   \  \;\,
\partial_{\bar{M}} \ \to \  Z_M{}^P \partial_{\,\nin{P}}\;,   \\[0.5ex]
\partial^{\,\nin{M}} \ &\to \  -\partial^{\bar{P}} Z_P{}^M  \,,  \qquad   \  
\partial^{\bar{M}} \ \to \ - \partial^{\,\nin{P}} \, Z_P{}^M \;.   
\end{split}
\ee
This implies for the differential operator   
\be
\partial^{\bar M} \partial_{\bar M} \ \to \  - \partial^{\,\nin{M}} \partial_{\,\nin{M}} \ = \ 
\partial^{\bar M} \partial_{\bar M} \;, 
\ee
using the strong constraint in the last step. Thus, the operator $\partial^{\bar M} \partial_{\bar M}$, 
which reduces to the usual Laplace operator for $\tilde{\partial}=0$, is $\mathbb{Z}_2$ invariant. 
The same conclusion follows for $\partial^{\,\nin{M}} \partial_{\,\nin{M}}$. 
Note that this result 
is consistent with the fact that 
$\partial^M \partial_M=\partial^{\,\nin{M}} \partial_{\,\nin{M}}+\partial^{\bar M} \partial_{\bar M}$, containing one $\eta$, is odd under $\mathbb{Z}_2$, because by the strong constraint, which we used above, it is actually zero.

Next, we discuss the $\mathbb{Z}_2$ action on the fluctuation fields $m$. They are defined 
via  ${\cal M} = {\cal H} + m$ and so according to the rules for the $\mathbb{Z}_2$ action 
on $O(D,D)$ indices we have 
\be
m_{MN} \ \to \   Z_M{}^P  Z_N{}^Q  \, m_{PQ}  \;. 
\ee  
The analogous relations follow for any of the projections with (\ref{PZ2TRANS}),
\be
\label{cltvm}
\begin{split}
m_{\,\nin{M} \bar N} \ \to \   & \ Z_M{}^P  Z_N{}^Q  \, m_{\, \nin{Q}\bar P}\;,   \\[0.5ex]
m_{\,\nin{M} \,\nin{N}} \ \to \   & \ Z_M{}^P  Z_N{}^Q  \, m_{\, \bar{P}\bar Q}\;,   \\[0.5ex]
m_{\bar{M} \bar{N}} \ \to \   & \ Z_M{}^P  Z_N{}^Q  \, m_{\, \nin{P}\,\nin Q}\;.   
\end{split}
\ee
Similarly, the projected gauge parameters transforms as 
\be
\begin{split}
\xi^{\,\nin{M}} \ \to \ \  \xi^{\bar{P}} Z_P{}^M  \,,  \qquad   
\xi_{\,\nin{M}} \ \to  \ - Z_M{}^P \xi_{\bar{P}} \,,  
\end{split}
\ee
and completely analogously for $\xi^{\bar{M}}$. 

We are now in the position to test the $\mathbb{Z}_2$ properties of the 
gauge transformations for $m_{\,\nin{M} \bar N}$.  
On account of (\ref{cltvm}), for 
$\mathbb{Z}_2$ even transformations  
we should have
\be
\delta_\xi m_{\nin{M} \bar N} \ \to \   \ Z_M{}^P  Z_N{}^Q  \, \delta_\xi m_{\, \nin{Q}\bar P} \;.  
\ee 
In order to verify the $\mathbb{Z}_2$ parity in tensors with several (free or contracted) 
$O(D,D)$ indices it can be 
a bit laborious to insert every single $Z$ matrix, most of which drop out by $Z^2={\bf 1}$. 
Rather, one may just  apply the following simple rule which summarizes
the above results:  

{\bf Rule for $\mathbb{Z}_2$ parity:}   {\em An expression 
with free indices $\nin{M}$ and $\bar N$ is $\mathbb{Z}_2$ even/odd 
if the following action gives back the expression with the same/opposite sign.
First exchange $ \nin{M} \leftrightarrow \bar N$.    
Second, exchange bars and under-bars in 
all other indices, keeping the same letter as index label.
Third, include a minus sign factor for each index that is not in its canonical position.  For an expression without free indices, steps two and three must leave it invariant. }

The canonical positions for fluctuations, derivatives and gauge parameters are
$m_{MN}$, $\partial_M$ and $\xi^{M}$, respectively.  On the $m$ field the index
substitution is implemented as 
 $m_{\, \nin{P} \bar{Q}} \to m_{\, \nin{Q} \bar {P}}$ since, by convention, we always
 put the under barred index first.  Moreover, $\xi^P \partial_P$, for example, is
 $\mathbb{Z}_2$ even. 
 
We can verify now that  in the gauge transformation (\ref{physicalgauge--redef-new2}) 
the part with one derivative is  $\mathbb{Z}_2$ even but the higher derivative
correction is $\mathbb{Z}_2$ odd. Applying the above rule to the inhomogeneous term we find that it is left invariant
 \be
  2\big(\partial_{\,\nin{M}}\xi_{\bar{N}}-\partial_{\bar{N}}\xi_{\,\nin{M}}\big)
  \;\rightarrow\; -2\big(\partial_{\bar{N}}\xi_{\,\nin{M}}-\partial_{\,\nin{M}}\xi_{\bar{N}}\big) \ = \  2\big(\partial_{\,\nin{M}}\xi_{\bar{N}}-\partial_{\bar{N}}\xi_{\,\nin{M}}\big)\,, 
 \ee 
where the sign originated because the gauge parameters have their index in the
non-canonical position.  Similarly,  the terms homogeneous in fields
 and with one derivative are, as a whole,  $\mathbb{Z}_2$ even: 
 \be
\begin{split}
\xi^{P}\partial_P m_{\nin{M}\bar{N}} \quad \to & \ \quad \xi^{P}\partial_P m_{\nin{M}\bar{N}}\;,   \\
K_{\,\nin{M}}{}^{\nin{P}}\, m_{\nin{P}\bar{N}} \, = \, 
(\p_{\,\nin{M}}{} \xi^{\nin{P}} - \p^{\nin{P}}  \xi_{\,\nin{M}}{}  ) \, m_{\nin{P}\bar{N}} \quad \to & \ \quad (\p_{\,\bar{N}}{} \xi^{\bar{P}} - \p^{\bar{P}}  \xi_{\,\bar{N}}{}  ) \, m_{\nin{M}\bar{P}} = K_{\bar{N}}{}^{\bar{P}}\,m_{\,\nin{M}\bar{P}}\;,  \\
K_{\bar{N}}{}^{\bar{P}}\,m_{\,\nin{M}\bar{P}} \, = \, 
( \p_{\bar{N}}{} \xi^{\bar{P}} -  \p^{\bar{P}} \xi_{\bar{N}}{}) \,m_{\,\nin{M}\bar{P}}
\quad \to & \ \quad 
( \p_{\nin{M}}{} \xi^{\nin{P}} -  \p^{\nin{P}} \xi_{\nin{M}}{}) \,m_{\,\nin{P}\bar{N}}\,= \, 
K_{\,\nin{M}}{}^{\nin{P}}\, m_{\nin{P}\bar{N}}\;.  \end{split}
\ee
Note that the second and third terms were exchanged under the transformation.
Consider now the higher-derivative terms in the gauge transformation of $m_{\,\nin{M}\bar{N}}$.  For the first term
\be\begin{split}
\partial_{\,\nin{P}}\,  m_{\nin{Q}\bar{N}}\,
   \partial_{\,\nin{M}}  K^{\,\nin{P}\,\nin{Q}} \, = \, 
   \partial_{\,\nin{P}}\,  m_{\nin{Q}\bar{N}}\,
   \partial_{\,\nin{M}} ( \p^{\nin{P}}\,\xi^{\nin{Q}} - \p^{\nin{Q}} \xi^{\,\nin{P}}\, ) \ \to
   & \ \   
- \partial_{\,\bar{P}}\,  m_{\nin{M}\bar{Q}}\,
   \partial_{\,\bar{N}} ( \p^{\bar{P}}\,\xi^{\bar{Q}} - \p^{\bar{Q}} \xi^{\,\bar{P}}\, )\\
  = & \  - \ \partial_{\,\bar{P}}\,  m_{\nin{M}\bar{Q}}\,
   \partial_{\,\bar{N}} K^{\bar{P}\bar{Q}} \;, 
   \end{split}\ee
which is minus the second term.  Similarly, the transformation of the second term
is minus the first.  Thus the $\alpha'$ terms in (\ref{physicalgauge--redef-new2}) are $\mathbb{Z}_2$ odd.

Let us finally point out that also the field redefinition (\ref{parameterrREdef}) was
$\mathbb{Z}_2$ violating. This is as it should because it eliminates 
$\mathbb{Z}_2$ odd terms through variations of  inhomogeneous 
$\mathbb{Z}_2$ even terms in $\delta_{\xi}m$.  Similarly, the parameter redefinitions 
(\ref{param-redefcorr}) are $\mathbb{Z}_2$ odd. 
Summarizing, the gauge transformations of order $\alpha'$, determined for the 
theory constructed in \cite{Hohm:2013jaa}, are  $\mathbb{Z}_2$ odd, and so this 
theory actually corresponds to DFT$^{-}$. In the next subsection we relate 
the field variables here to those in the CSFT language, confirming 
explicitly this conclusion.

\subsection{Relating CSFT and DFT frameworks} \label{DFTSFTRel}
We now relate in detail the gravitational field variable $e_{ij}$ of CSFT  
to the double metric fluctuation $m_{\,\nin{M}\bar{N}}$. On the face of it they appear to be 
rather different: the former carries $D$-dimensional indices as in standard gravity, 
and the latter
carries doubled $O(D,D)$ indices. Since the $O(D,D)$ indices 
are projected,  
however, they are effectively $D$-dimensional.  
The two formalisms are essentially equivalent, as we will 
show in the following. 

The most efficient way to establish this relation 
is in terms of a frame or vielbein formalism \cite{Siegel:1993th,Hohm:2010xe}, 
see \cite{Hohm:2011dz}.  
More specifically, here we employ a frame formalism for the constant background fields. 
The `tangent space' symmetry in this case reduces to a global $GL(D)\times GL(D)$ symmetry, 
indicated by flat frame indices $A=(a,\bar{a})$, so that a tangent space tensor is decomposed 
as $U_A = (U_a, U_{\bar a})$. Next we define the background vielbein for a 
particular `gauge choice',  
 \be\label{fixedEbein}
  {\cal E}_{A}{}^{M} \ = \ \begin{pmatrix}    {\cal E}_{ai} & {\cal E}_a{}^i \\[0.5ex]
 {\cal  E}_{\bar{a}i} & {\cal E}_{\bar{a}}{}^{i} \end{pmatrix}
  \ = \  \begin{pmatrix}    -E_{ai} & \delta_a{}^i \\[0.5ex]
  E_{i\bar{a}} & \delta_{\bar{a}}{}^{i} \end{pmatrix}\;. 
 \ee
Some components have been fixed to be Kronecker deltas, which in turn  
allows us to identify $i,j$ indices with $a,b$ indices. 
The matrix $E$ describes, as in sec.~2, the sum of background metric and $B$-field. 
For completeness we also give 
the inverse frames ${\cal E}_M {}^A$, satisfying ${\cal E}_M {}^A {\cal E}_{A}{}^{N} = \delta_M{}^N$ as well as   
$ {\cal E}_{A}{}^{M}{\cal E}_M {}^B = \delta_A{}^B$:
 \be
  {\cal E}_{M}{}^{A} \ = \ \begin{pmatrix}    
  {\cal E}^{ia} & {\cal E}^{i\bar a} \\[0.5ex]
 {\cal  E}_i{}^a  & {\cal E}_i{}^{\bar a}  \end{pmatrix}
  \ = \  \begin{pmatrix}    -{1\over 2} G^{ia}  & {1\over 2} G^{i\bar a}  \\[0.9ex]
 {1\over 2} E_{bi} G^{ab}  & {1\over 2} E_{i\bar b} G^{\bar a \bar b}  \end{pmatrix}\;. 
\ee
Next, we inspect the tangent space metric, defined from the metric $\eta_{MN}$ by   
 \be
  {\cal G}_{AB} \ \equiv \   
  \begin{pmatrix}    {\cal G}_{ab} & {\cal G}_{a\bar b} \\[0.5ex]
  {\cal G}_{\bar a b}  & {\cal G}_{\bar{a}\bar{b}} \end{pmatrix} \ \equiv \ {\cal E}_{A}{}^{M} {\cal E}_{B}{}^{N}\eta_{MN} \ = \ 
{\cal E}_{A}{}^{i} \,{\cal E}_{B\, i} + {\cal E}_{A\, i}\, {\cal E}_{B}{}^i \ = \  \begin{pmatrix}    -2G_{ab} & 0 \\[0.5ex]
  0 & 2G_{\bar{a}\bar{b}} \end{pmatrix}\;, 
 \ee 
where the last equality follows by a direct calculation from (\ref{fixedEbein}).
Consequently, the inverse metric ${\cal G}^{AB}$ is given by
 \be\label{inverseG}
  {\cal G}^{AB} \ \equiv \ 
  \begin{pmatrix}    {\cal G}^{ab} & {\cal G}^{a\bar b} \\[0.5ex]
  {\cal G}^{\bar a b}  & {\cal G}^{\bar{a}\bar{b}} \end{pmatrix}  \  \equiv \ {\cal E}_{M}{}^{A} {\cal E}_{N}{}^{B}\eta^{MN}  
  \ = \   \begin{pmatrix}    -{1\over 2} G^{ab} & 0 \\[0.5ex]
  0 & {1\over 2} G^{\bar{a}\bar{b}} \end{pmatrix}\;.
 \ee 
 These tangent space metrics are used to raise and lower frame indices $A, B$. 
 Due to the factors of $\pm 2$ and $\pm \tfrac{1}{2}$ appearing in the metric ${\cal G}$
 and its inverse, respectively, there is an ambiguity regarding which metric is used when 
 $D$-dimensional indices are contracted. 
 Here we follow the conventions in which
 
 \begin{itemize}
  
 \item the tangent space metric ${\cal G}$ (and its inverse) is used to 
 contract indices whenever they are written with latin letters from the beginning of the 
 alphabet, i.e., $a$, $b$ or $\bar{a}$, $\bar{b}$,  but
 
 \item the metric $G$ (and its inverse) is 
 used to contract indices whenever they are written with latin letters from the middle of the 
 alphabet, i.e., $i$, $j$, etc. 
 
 \end{itemize}

\noindent
The background projectors (\ref{defPRoj}) are defined in terms of 
the frame fields as 
 \be\label{projDEF}
  P_M{}^N \ = \ {\cal E}_{M}{}^a \,{\cal E}_{a}{}^{N}\;, \qquad 
  \bar{P}_{M}{}^N \ = \  {\cal E}_M{}^{\bar{a}}\, {\cal E}_{\bar{a}}{}^{N}\;. 
 \ee
Alternatively, we have
\be\label{projDEF-var}
  P^{MN} \ = \ {\cal G}^{ab}\,  {\cal E}_{a}{}^M \,{\cal E}_{b}{}^{N}\;, \qquad 
  \bar{P}_{M}{}^N \ = \ {\cal G}^{\bar a\bar b}\, {\cal E}_{\bar a}{}^M \,{\cal E}_{\bar b}{}^{N}\;. 
 \ee

Using the frame field and its inverse we now can introduce various 
`flattened' objects. The partial derivatives in flat indices, 
 \be\label{flatD}
  D_{A} \ \equiv \  \ {\cal E}_{A}{}^{M}\partial_M \ = \ (D_a, D_{\bar{a}})\;, 
 \ee
take the following explicit form for the choice (\ref{fixedEbein}), 
 \be
 \begin{split}
  D_a \ = \  \  \partial_a - E_{ai} \tilde \partial^i \,, \qquad 
  \ D_{\bar a} \ = \  \ 
 \partial_{\bar a} + E_{ia} \tilde\partial^i  \;. 
 \end{split}
 \ee
Looking back at sec.~2, we infer that these operators coincide with the differential operators 
introduced there under the same name (recalling that for (\ref{fixedEbein})
we can identify flat and curved indices). 
 Similarly, for the flattened gauge parameters we identify 
 \be\label{flatparaM}
  \Lambda_A \ \equiv \  \  {\cal E}_{A}{}^{M}\xi_M \ = \ 
  (\Lambda_a, \Lambda_{\bar a} ) \ = \ (-\lambda_a, \bar{\lambda}_{\bar{a}} )\;, 
 \ee
so that we find with  (\ref{fixedEbein}) 
 \be
 \lambda_a \ = \  \  -\tilde\xi_a + E_{ai} \,  \xi^i  \;,   \qquad 
 \bar\lambda_{\bar a}  \ = \ 
 \tilde \xi_{\bar a}\, + \, E_{ia} \, \xi^i  \;. 
 \ee
This coincides with the gauge parameters $\lambda_i$, $\bar{\lambda}_i$ of CSFT
as discussed in sec.~3 of \cite{Hull:2009zb}.  
Note that 
 we introduced a relative sign in (\ref{flatparaM}) in order to comply with the 
conventions of CSFT. Note that in CSFT we view the parameters $\lambda_i$, $\bar{\lambda}_i$ 
with lower indices as fundamental, while in the $O(D,D)$ covariant language $\xi^{M}$ 
with upper indices is fundamental. 
 This requires some care when translating expressions from a frame-like basis to the 
 CSFT basis. For instance, the contraction of two $O(D,D)$ vectors $U$ and $V$, 
 whose fundamental indices are lower, barred indices, reads 
  \be
  U^{\bar{P}} V_{\bar{P}} \ = \ \bar{P}^{PQ} U_{P} V_{Q} \ = \ {\cal G}^{\bar a \bar b}
  {\cal E}_{\bar a}{}^{P}{\cal E}_{\bar b}{}^Q \, U_P V_Q \ =  \  {\cal G}^{\bar{a}\bar{b}}  U_{\bar a} V_{\bar b} 
\ = \ \f{1}{2} G^{ij} \bar U_i \bar V_j    \ = \ \f{1}{2}\bar U_{{i}} \bar V^{{i}}\;, 
 \ee  
using (\ref{projDEF-var}) and ${\cal G}^{\bar{a}\bar{b}}=\tfrac{1}{2}G^{\bar{a}\bar{b}}$.
Here we indicated the barred nature of the indices on $U$ and $V$ by barring the objects, as it is customary in DFT.  We also have,
in completely analogous fashion 
\be
  U^{\nin{P}} V_{\nin{P}} \ = \ {P}^{PQ} U_{P} V_{Q} \ = \ {\cal G}^{ab}\, {\cal E}_{ a}{}^{P}
  {\cal E}_b{}^Q\, U_P V_Q \ = \ 
     {\cal G}^{{a}{b}}  U_{ a} V_{ b} 
     \ = \ - \f{1}{2} G^{ij} U_i V_j 
   \ = \ -\f{1}{2}U_{{i}} V^{{i}}\;. 
 \ee
 As a general translation tool, these are most useful in the form   
 \be
 \label{most-useful}
  U^{\bar{P}} V_{\bar{P}} \ = \ U^{\bar a} V_{\bar a}\,, \qquad  U^{\nin{P}} V_{\nin{P}} \ = \ 
  U^{a} V_{ a}\,, 
 \ee  
where the flat indices are raised with the appropriate ${\cal G}$.  It is a simple matter to verify that flattening of projected indices works according to the association $\nin{M} \leftrightarrow a $ of under-barred indices with normal latin indices and $\bar M \leftrightarrow \bar a$ of barred indices :
 \be
 {\cal E}_a{}^M B_{\nin{M}} \ = \  B_a\,, \quad  {\cal E}_a{}^M B_{\bar{M}} \ = \  0\,, \qquad
  {\cal E}_{\bar a}{}^M B_{\nin{M}} \ = \  0\,, \quad  {\cal E}_{\bar a}{}^M B_{\bar{M}} \ = \  B_{\bar a} \,.  
 \ee
The transport operator $\xi^{M}\partial_{M}$
has a simple translation into frame objects:
 \be
  \xi^{M}\partial_{M} \ = \ \Lambda^A D_A \ = \ {\cal G}^{ab}\Lambda_a D_b
  \, +\, {\cal G}^{\bar a\bar b}\Lambda_{\bar{a}}D_{\bar{b}}\ = \ 
  - {\cal G}^{ab}\lambda_a D_b
  \, +\, {\cal G}^{\bar a\bar b}\bar\lambda_{a}\bar D_{{b}}
 \ = \ 
  \f{1}{2}\big(\lambda^iD_i+\bar{\lambda}^i\bar{D}_i\big)\;. 
 \ee 
In the above we converted curved into flat indices 
and decomposed into $\lambda$ and $\bar{\lambda}$ components
according to (\ref{flatparaM}). 
In the last step we used the metric components $G$ according to (\ref{inverseG}). This introduced a factor of $\tfrac{1}{2}$ and cancelled the minus sign 
from the frame definition of $\lambda_i$. 

\medskip
Our main goal in this formalism is to  translate the gauge variation of the double
metric fluctuation $m_{\,\nin{M}\bar{N}}$ to that in terms of 
the CSFT fluctuation $e_{ij}$ in order to compare results. We 
claim that these fluctuations are related by 
 \be\label{emRelation} 
  e_{a\bar{b}} \ = \ \tfrac{1}{2}\,{\cal E}_{a}{}^{M}{\cal E}_{\bar{b}}{}^{N} m_{\,\nin{M}\bar{N}}
  \, ,  \qquad \hbox{or} \qquad   
m_{\,\nin{M}\bar{N}} \ = \ 2  \, {\cal E}_M{}^a {\cal E}_N{}^{\bar b}  \, e_{a\bar b} \,,   
 \ee
as we will 
show  that it relates the 
gauge transformations to leading order in derivatives
\be
   \delta^-_{\Lambda}e_{a\bar{b}} \ = \  \tfrac{1}{2}\,{\cal E}_{a}{}^{M}{\cal E}_{\bar{b}}{}^{N} 
  \delta_\xi^- m_{\,\nin{M}\bar{N}} \,.
  \ee
We evaluate the right-hand side using (\ref{physicalgauge--redef-new2}): 
 \begin{equation*}
 \begin{split}  
  \delta^-_{\Lambda}e_{a\bar{b}}   \ = \ & \tfrac{1}{2}\,{\cal E}_{a}{}^{M}{\cal E}_{\bar{b}}{}^{N} 
  \Bigl( 2 (\p_{\nin{M}} \xi_{\bar N} - \p_{\bar N} \xi_{\nin{M}}) \, +\xi^{P}\partial_P m_{\nin{M}\bar{N}}
   + (\p_{\nin{M}} \xi^{\nin{P}} \,  - \, \p^{\nin{P}} \xi_{\,\nin{M}}) 
   \, m_{\nin{P}\bar{N}}
   +(\p_{\bar{N}} \xi^{\bar{P}} \, - \, \p^{\bar{P}}  \xi_{\bar{N}}) 
   \,m_{\,\nin{M}\bar{P}}  \\[0.1ex]
   &\hskip50pt +\,
   \partial_{\,\nin{M}} 
   ( \p_{\,\nin{P}} \xi^{\nin{Q}} \, - \, \p^{\nin{Q}} \xi_{\,\nin{P}}\, )
   \partial^{\nin{P}} m_{\nin{Q}\bar{N}} %
 \  + \  \,\partial_{\bar{N}}
 (\p_{\bar{P}} \xi^{\bar{Q}}  \, - \, \p^{\bar{Q}} \xi_{\bar{P}} ) 
 \partial^{\bar{P}}m_{\nin{M}\bar{Q}}\Bigr)  \\[1.0ex]
  \ = \ & \ \  D_a\Lambda_{\bar{b}}-D_{\bar{b}}\Lambda_{a}\, 
  +\, \xi^{P}\partial_{P}e_{a\bar{b}}
  \, +\, \big(D_{a}\Lambda^c-D^c\Lambda_a\big)e_{c\bar{b}}
  \, +\, \big(D_{\bar{b}}\Lambda^{\bar{c}}
  -D^{\bar{c}}\Lambda_{\bar{b}}\big)e_{a\bar{c}} \\[1.0ex]
  & +  D_a(D_c\Lambda^d-D^d\Lambda_c)D^c e_{d\bar{b}}+D_{\bar{b}}(D_{\bar{c}}\Lambda^{\bar{d}}-D^{\bar{d}}\Lambda_{\bar{c}})
  D^{\bar{c}}e_{a\bar{d}} \,,
 \end{split}\end{equation*}  
where we used repeatedly (\ref{most-useful}). 
We pass to $D$-dimensional curved indices letting $a \to i $  and $\bar b \to j$.
Note that then $\Lambda_a \to - \lambda_i$ and $\Lambda_{\bar b} \to \bar\lambda_j$.
A short calculation then gives 
 \be
 \begin{split}
  \delta_{\Lambda}^-e_{ij} \ = \ & \  \ \ D_i\bar{\lambda}_j+\bar{D}_{j}\lambda_i\\
 &  +\f{1}{2}\big(\lambda^iD_i+\bar{\lambda}^i\bar{D}_i\big)e_{ij}+\f{1}{2}\big(D_i\lambda^k-D^k\lambda_i\big)
  e_{kj}+\f{1}{2}\big(\bar{D}_j\bar{\lambda}^k
  -\bar{D}^{k}\bar{\lambda}_j\big)e_{ik}\\[0.5ex]
  & -\f{1}{4}D_i(D_k\lambda^l-D^l\lambda_k)D^k e_{lj}
  +\f{1}{4}\bar{D}_{j}(\bar{D}_{k}\bar{\lambda}^{l}-\bar{D}^{l}\bar{\lambda}_{k})
  \bar{D}^{k}e_{il}\;, 
\end{split}
 \ee 
The factors of $\f{1}{4}$ on the last line
originate from the two inverse metrics ${\cal G}^{-1}$ required
by the two index contractions. Since the same type metric is used
twice on each term the sign difference between ${\cal G}^{ab}$ 
and ${\cal G}^{\bar{a}\bar{b}}$ is immaterial.  The first two lines
on the above equation are the familiar CSFT gauge transformations of
$e_{ij}$~\cite{Hull:2009mi}.  This confirms the correctness of the identification
(\ref{emRelation}) of $e$ with $m$. 

The $\alpha'$ correction of the gauge transformation is on the last line.
It differs from the CSFT result (\ref{olafsecondversion}) 
in the sign of the second term, but agrees precisely with 
 the DFT$^{-}$ transformation (\ref{olafthirdversion}). 
Thus, in agreement with the previous section, the theory studied
so far in this section is DFT$^{-}$.

\medskip

We close this subsection by verifying the above conclusion at the level of the gauge algebra.  
We first recall the gauge algebra for the background-independent DFT constructed in \cite{Hohm:2013jaa}
 \be\label{BIDFTminus}  
 - \xi_{12}^M \ = \  \big[\xi_1, \xi_2 \big]^M \ = \ \xi_1^N \partial_N \xi_2^M   -  \xi_2^N \partial_N \xi_1^M  
  - \, {\textstyle{1\over 2}}   \xi_1^K  \ppp^M \xi_{2K} 
   +  {\textstyle{1\over 2}}   \,\p_K \xi_1^L  \ppp^M \p_L \xi_2^K\;, 
 \ee 
where the last term encodes the ${\cal O}(\alpha')$ correction.  
We relate this algebra to the CSFT one by converting to flat indices. 
One finds for the flattened parameter (\ref{flatparaM}) 
 \be
 \label{from-DFT}  
 \begin{split}
 \Lambda_{12\,A} \ = \ & \ \  {\textstyle{1\over 2}}  (\bar\lambda_1 \cdot D + \bar\lambda_1 \cdot \bar D )  \Lambda_{2A} \, - \, (1 \leftrightarrow 2) \\[1.3ex]
 &  + {\textstyle{1\over 4}} \bigl(  \lambda_1 \cdot \DDD_A  \lambda_2
 - \bar \lambda_1 \cdot \DDD_A  \bar\lambda_2 \bigr) \\[1.3ex]
 &  - {\textstyle{1\over 16}} \Bigl(  {K}_{1\,kl}  \, \DDD_A  \,
 { K}_2^{\ kl}   +   \bar{ K}_{1\,kl}  \, \DDD_A  \,
 \bar{ K}_2^{kl} 
 - 2  { L}_{1\, kl} \,  \DDD_A  \,
 { L}_2^{\  kl}  \Bigr) \;, 
\end{split}
\ee
where as before: ${ K}_{kl} =  D_k \lambda_l - D_l \lambda_k$, 
$\bar{ K}_{kl} = \bar D_{k} \bar\lambda_{ l} - \bar D_{l} \bar\lambda_{k}$ and we 
defined  ${ L}_{kl}  \equiv    D_{k} \bar\lambda_{l} + \bar D_{l} \lambda_{ k}   =  \delta_\lambda e_{kl}$.
As in various previous examples, the last term in (\ref{from-DFT}) can thus be removed by a parameter redefinition. 
Doing this and converting the external flat index the gauge algebra reads  
 \be\label{secondFInalalgebra99}
 \begin{split}
  \lambda_{12}^i \ &= \ \lambda_{12,c}^i-\f{1}{16}\, \alpha'\,\big({ K}_1^{kl}\, \DDD^i
  { K}_{2\,kl}\ + \
  {\bar  K}_1^{ k l} \ \DDD^i
  { \bar K}_{2\, k  l} \big)\;,  \\[0.7ex]
  \bar{\lambda}_{12}^i \ &= \ \bar\lambda_{12,c}^i
  +\f{1}{16}\, \alpha'\, \big({ K}_1^{kl}\ \DDDB^i 
  { K}_{2\,kl}\  +\
   {\bar K}_1^{ k  l}\,\DDDB^i
   {\bar K}_{2\,  k  l}\big)\;.  \end{split}
 \ee   
The sign difference between the ${\cal O}(\alpha')$ contributions is due to the relative sign 
in the frame definition of $\lambda$ and $\bar{\lambda}$ in (\ref{flatparaM}). 
This agrees with the DFT$^-$ gauge algebra anticipated in (\ref{secondFInalalgebra}).

\subsection{Direct comparison of gauge algebras}
We have seen  that the background-independent gauge algebra (\ref{BIDFTminus}) 
introduced in \cite{Hohm:2013jaa} corresponds to DFT$^{-}$. Since this is the unique 
field-independent deformation of the C-bracket there is no analogous background-independent form for the DFT$^{+}$ 
algebra. 
It is illuminating, however, to give a form in which every tensor is written with un-projected $O(D,D)$ indices.  

To this end it is convenient to rewrite the DFT$^{+}$ gauge transformations and algebra in terms of 
objects with (doubled) $O(D,D)$ indices, using $m_{\,\nin{M}\bar{N}}$ as field variable, 
which is straightforward using the map (\ref{emRelation}) between the 
two formalisms.  We should start from the form of the DFT$^{-}$ gauge transformations that gave 
the background independent gauge algebra (\ref{BIDFTminus}) directly, without 
further parameter redefinitions, which is given in (\ref{physicalgauge--redef-new}). 
Changing the relative signs in the last two lines
of (\ref{physicalgauge--redef-new}) in order to make it  
$\mathbb{Z}_2$ invariant, one finds  
\be\label{physicalgauge-ef-new33}
 \begin{split}
 \delta_{\xi}^{+}m_{\,\nin{M}\bar{N}} \ = \ & \ \ 
 2\, K_{\nin{M} \bar N}\ +\xi^{P}\partial_P m_{\nin{M}\bar{N}}
   + K_{\,\nin{M}}{}^{\nin{P}}\, m_{\nin{P}\bar{N}}
   +K_{\bar{N}}{}^{\bar{P}}\,m_{\,\nin{M}\bar{P}}\\[0.5ex]
   &\;\;+\half\partial_{\,\nin{M}}m^{\,\nin{P}\bar{Q}}\,\partial_{\bar{N}}
   K_{\,\nin{P} \bar{Q}} 
   \,  + \, \half\partial_{\bar{N}}m^{\,\nin{P}\bar{Q}}\,\partial_{\,\nin{M}}
   K_{\,\nin{P} \bar{Q}} 
     \\[0.9ex]
   &\;\;+\partial_{\,\nin{P}} m_{\,\nin{Q}\bar{N}}\,
   \partial_{\,\nin{M}}  K^{\,\nin{P}\,\nin{Q}}  %
 \  - \  \partial_{\bar{P}}m_{\nin{M}\bar{Q}}\,\partial_{\bar{N}}
 K^{\bar{P}\bar{Q}}    %
 \;, 
  \end{split}  
 \ee
where we indicated by a super-script ${}^{+}$ that this describes the DFT$^{+}$ transformations. 
In the DFT$^{-}$ case  
the corresponding terms in the second line could be removed by the parameter redefinition (\ref{param-redefcorr}),
but this introduces a background dependence in the gauge algebra; in contrast, the terms here  
are removable by a field redefinition, which does not change the algebra and so does not affect the 
background dependence. Specifically, the terms in the second line of (\ref{physicalgauge-ef-new33}) 
equal a total variation, 
 \be
  \quarter\partial_{\,\nin{M}}m^{\,\nin{P}\bar{Q}}\,\partial_{\bar{N}}
   \big(\delta_{\xi}m_{\,\nin{P}\bar{Q}}\big)
   \,  + \, \quarter\partial_{\bar{N}}m^{\,\nin{P}\bar{Q}}\,\partial_{\,\nin{M}}
   \big(\delta_{\xi}m_{\,\nin{P}\bar{Q}}\big) \ = \ 
  \quarter \delta_{\xi}^{[0]}\big(\partial_{\,\nin{M}}m^{\,\nin{P}\bar{Q}}\,\partial_{\bar{N}}m_{\,\nin{P}\bar{Q}}\big)\;, 
  \ee 
and are thus removable by a field redefinition.  
Computing the gauge algebra directly from  (\ref{physicalgauge-ef-new33}) one finds 
 \be\label{EVENFinalgebra}
  \xi_{12}^M \ = \ -\f{1}{4} K_2^{\,\nin{P}\,\nin{Q}}\partial^MK_{1\,\nin{P}\,\nin{Q}}
  + \f{1}{4} K_2^{\bar{P} \bar{Q}}\partial^MK_{1\bar{P}\bar{Q}}-(1\leftrightarrow 2)\;.  
 \ee  
Next, we eliminate the background projectors by (\ref{defPRoj}) in order to find the 
$O(D,D)$ covariant form without projected indices. A straightforward computation yields 
\be\label{Z-algebra-original+}
  \text{DFT$^{+}$}:\qquad 
   \xi_{12}^{M} \ = \  \xi_{12\,C}^M \, -\half
 \,  \bar {\cal H} ^{KL} \eta^{PQ} \,   
   K_{[1\, KP} \p^M \hskip-1pt  K_{2]\,LQ}   \,. ~   
 \ee  
This is to be contrasted with the DFT$^{-}$ algebra, which in the same notation reads  
\be\label{Z-algebra-original}
 \text{DFT$^{-}$}:\qquad 
   \xi_{12}^{M} \ = \  \xi_{12\,C}^M \, +\half
 \,  \eta^{KL} \eta^{PQ} \ 
   K_{[1\, KP} \p^M \hskip-1pt  K_{2]\,LQ}   \,. 
 \ee  
Note the background field dependence $\bar {\cal H}$ in the  DFT$^{+}$ algebra.
This strongly suggests that in a 
manifestly background independent formulation of  DFT$^{+}$
 the gauge algebra will be field dependent. 

We close this section by discussing the general gauge algebra for arbitrary $\gamma^+$, $\gamma^{-}$. 
To this end it is convenient to start from the gauge transformations 
that follow
 from (\ref{physicalgauge--redef-new}) and (\ref{physicalgauge-ef-new33})  
\be\label{physicalgauge-ef-newGamma}
 \begin{split}
 \delta_{\xi}^{\gamma}m_{\,\nin{M}\bar{N}} \ = \ & \ \ 
 2\, K_{\nin{M} \bar N}\ +\xi^{P}\partial_P m_{\nin{M}\bar{N}}
   + K_{\,\nin{M}}{}^{\nin{P}}\, m_{\nin{P}\bar{N}}
   +K_{\bar{N}}{}^{\bar{P}}\,m_{\,\nin{M}\bar{P}}\\[0.5ex]
   &\;\;+\half(\gamma^++\gamma^-)\partial_{\,\nin{M}}m^{\,\nin{P}\bar{Q}}\,\partial_{\bar{N}}
   K_{\,\nin{P} \bar{Q}} 
   \,  + \, \half(\gamma^+-\gamma^-)\partial_{\bar{N}}m^{\,\nin{P}\bar{Q}}\,\partial_{\,\nin{M}}
   K_{\,\nin{P} \bar{Q}} 
     \\[0.9ex]
   &\;\;+(\gamma^++\gamma^-)\partial_{\,\nin{P}} m_{\,\nin{Q}\bar{N}}\,
   \partial_{\,\nin{M}}  K^{\,\nin{P}\,\nin{Q}}  %
 \  - \  (\gamma^+-\gamma^-) \partial_{\bar{P}}m_{\nin{M}\bar{Q}}\,\partial_{\bar{N}}
 K^{\bar{P}\bar{Q}}    %
 \;.
  \end{split}  
 \ee
The terms in the second line proportional to $\gamma^+$ are removable by a field redefinition 
(and can thus be ignored for the sake of computing the gauge algebra); the terms in the second line proportional 
to $\gamma^-$ are removable by a parameter redefinition (and thus should be kept as in (\ref{physicalgauge--redef-new})). 
A direct computation then shows closure with the effective parameter
 \be
 \begin{split}
   \xi_{12}^{M} \ = \  \xi_{12\,C}^M & -\quarter (\gamma^++\gamma^-)K_2^{\,\nin{P}\,\nin{Q}}\partial^MK_{1\,\nin{P}\,\nin{Q}}
  + \quarter(\gamma^+-\gamma^-) K_2^{\bar{P} \bar{Q}}\partial^MK_{1\bar{P}\bar{Q}}\\[0.5ex]
  &-\half\, \gamma^- K_2^{\,\nin{P}\bar{Q}}\partial^M K_{1\,\nin{P}\bar{Q}} \; -\,(1\leftrightarrow 2)\;. 
 \end{split}
 \ee 
Eliminating now the projectors by (\ref{defPRoj}) we find the gauge algebra 
 \be\label{Z-algebra-gammal}
   \xi_{12}^{M} \ = \  \xi_{12\,C}^M \, -\half
 \,  \big(\gamma^+{\cal H}^{KL}-\gamma^-\eta^{KL}\big) \ 
   K_{[1\, K}{}^{P} \p^M \hskip-1pt  K_{2]\,LP}   \,, 
 \ee  
which interpolates between the background-dependent (\ref{Z-algebra-original+}) and 
the background-independent (\ref{Z-algebra-original}).

\sectiono{Cubic actions for DFT$^-$ and DFT$^+$}\label{cuacdft+-}
We now explicitly construct cubic actions of order $\alpha'$, i.e., with three fields and four derivatives, for both 
DFT$^{-}$ and DFT$^{+}$ and thereby also for the interpolating theories. It is convenient to cast the cubic 
action into a semi-geometric form, partially written in terms of linearized connections and 
curvatures that have simple transformation rules under the lowest-order gauge symmetries. 
In the first subsection we introduce these objects and use them to define the $O(D,D)$ covariant form of the Gauss-Bonnet term (to quadratic order in fields). 
Then we define the cubic DFT$^{-}$ and DFT$^{+}$ actions and discuss their respective differences
as well as the interpolating case related to the heterotic string.

\subsection{Linearized connections, curvatures and Gauss-Bonnet}\label{ConnCurvGB} 
The two-derivative DFT can be cast into a geometric form, with generalized connections and 
curvatures, but an important difference to standard geometry is that not all connection 
components can be determined in terms of the 
physical fields \cite{Siegel:1993th,Hohm:2010xe,Hohm:2011si,Hohm:2012mf,Jeon:2011cn}. (This is the 
very reason that $\alpha'$ corrections are non-trivial and require an extension of the 
framework, c.f.~the discussion in \cite{Hohm:2011si,Hohm:2012mf}.) 
In the following, however,  
it is sufficient  to work with the linearized version of the determined connections, 
which are given by 
\be\label{lincoNN}
  \begin{split}
   \Gamma_{\bar{M}\,\nin{N}\,\nin{K}} \ &\equiv \ \partial_{\,\nin{N}}m_{\,\nin{K}\bar{M}} -\partial_{\,\nin{K}}m_{\,\nin{N}\bar{M}}\;,\\
    \Gamma_{\,\nin{M}\bar{N}\bar{K}} \ &\equiv \ \partial_{\bar{N}}m_{\,\nin{M}\bar{K}} -\partial_{\bar{K}}m_{\,\nin{M}\bar{N}}\;, \\
    \Gamma_{\,\nin{M}} \ &\equiv \ \partial^{\bar{N}} m_{\,\nin{M}\bar{N}}-2\partial_{\,\nin{M}}\phi\;, \\
    \Gamma_{ \bar{M}} \ &\equiv \ \partial^{\,\nin{N}} m_{\,\nin{N}\bar{M}}+2\partial_{\bar{M}}\phi\;. 
  \end{split}
 \ee 
It is convenient to record the $\mathbb{Z}_2$ properties 
of the connections. These are easily found applying the rules spelled out in sec.~\ref{Z2section}
(recalling that  the dilaton is $\mathbb{Z}_2$ invariant):
 \be\label{Z2ACTION}
 \begin{split}
 \Gamma_{\bar{M}\,\nin{N}\,\nin{K}}  \quad \xrightarrow{\mathbb{Z}_2}  & \ \quad
 \Gamma_{\,\nin{M}\bar{N}\bar{K}}\;,  \\
 \Gamma_{\bar{M}}  \quad \xrightarrow{\mathbb{Z}_2}  & \ \quad
 -\, \Gamma_{\,\nin{M}}\;,  \\
 K_{\bar M \bar N}   \quad \xrightarrow{\mathbb{Z}_2}  & \ \quad
 -\, K_{\nin{M} \,\nin{N}}\;. 
 \end{split}
 \ee 
The gauge variations of these connections under the lowest-order gauge transformation, 
c.f.~(\ref{physicalgauge--redef-new}), 
\be\label{deltaZero}
\delta_\xi^{[0]}  
  m_{\,\nin{M} \bar N} \ = \ 
2 (\p_{\,\nin{M}} \xi_{\bar N}  - \p_{\bar N} \xi_{\,\nin{M} } )  \,, \qquad  \delta_\xi^{[0]} \phi 
\ = \ \partial_{\,\nin{M}}\xi^{\,\nin{M}}+\partial_{\bar{M}}\xi^{\bar{M}}\;, 
\ee
can be conveniently written in terms of the gauge parameters:
\be
\begin{split}
K_{\bar M \bar N}  \ \equiv \   \p_{\bar M} \xi_{\bar N} - \p_{\bar N} \xi_{\bar M} \;, \qquad 
K_{\nin{M} \,\nin{N}}  \ \equiv \   \p_{\nin{M}} \xi_{\nin{N}} - \p_{\nin{N}} 
\xi_{\, \nin{M}} \;,  
\end{split}
\ee
and read 
 \be\label{linnGaugeVar}
  \begin{split}  
   \delta_{\xi}^{[0]} \Gamma_{\bar{M}\,\nin{N}\,\nin{K}} \ &= \ -2\partial_{\bar{M}}K_{\,\nin{N}\,\nin{K}}\;, \qquad 
  \delta_{\xi}^{[0]} \Gamma_{\,\nin{M}\bar{N}\bar{K}}  \ = \ 2\partial_{\,\nin{M}}K_{\bar{N}\bar{K}}\;,  \\[1.0ex]
   \delta_{\xi}^{[0]}  \Gamma_{\,\nin{M}} \ &= \ 2\partial^{\,\nin{N}} K_{\,\nin{N}\,\nin{M}}\;, \qquad\qquad 
   \delta_{\xi}^{[0]}\Gamma_{\bar{M}} \ = \ -2\partial^{\bar{N}}K_{\bar{N}\bar{M}}\;. 
  \end{split}
 \ee  
Note, in particular, that  $\Gamma_{\bar{M}\,\nin{N}\,\nin{K}}$ and $\Gamma_{\,\nin{M}}$ are 
gauge invariant under $\xi^{\bar{M}}$ transformations. Similarly, $\Gamma_{\,\nin{M}\bar{N}\bar{K}}$
and  $\Gamma_{\bar{M}}$ are gauge invariant under $\xi^{\,\nin{M}}$ transformations. 
This fact simplifies the construction of gauge invariant actions below.  

Next, we can define the linearized Ricci tensor and scalar curvature: 
 \be\label{linRicci}
 \begin{split}
  {\cal R}_{\,\nin{M}\bar{N}} \ &\equiv \  
  \partial^{\,\nin{K}}\Gamma_{\bar{N}\,\nin{K}\,\nin{M}}+\partial_{\bar{N}} \Gamma_{\,\nin{M}}
  \ = \  -\partial^{\bar{K}}\Gamma_{\,\nin{M}\bar{K}\bar{N}}-\partial_{\,\nin{M}}\Gamma_{\bar{N}}\;, \\[0.5ex]
    {\cal R} \ &\equiv \   \partial^{\,\nin{M}}\Gamma_{\,\nin{M}} \ = \ \partial^{\bar{M}}\Gamma_{\bar{M}}\;.  
\end{split}
 \ee 
The equivalence of the two definitions in each case can be verified with the 
explicit form of the connections (\ref{lincoNN}). These tensors are gauge invariant as can be easily verified 
with (\ref{linnGaugeVar}). Inserting (\ref{lincoNN}) the explicit form of the linearized curvatures is given by 
 \be\label{explicitcurv}
  \begin{split}
   {\cal R}_{\,\nin{M}\bar{N}} \ &= \  \square\, m_{\,\nin{M}\bar{N}}-\partial_{\,\nin{M}}\partial^{\,\nin{K}}m_{\,\nin{K}\bar{N}}
   +\partial_{\bar{N}}\partial^{\bar{K}}m_{\,\nin{M}\bar{K}}-2\partial_{\,\nin{M}}\partial_{\bar{N}}\phi \;, \\[1.0ex]
     {\cal R} \ \ &= \  \partial^{\,\nin{M}}\partial^{\bar{K}}m_{\,\nin{M}\bar{K}}-2\,\square\phi \;, 
  \end{split}
 \ee  
where $\square=\partial^{\,\nin{M}}\partial_{\,\nin{M}}$. 
These curvatures appear in the general variation of the quadratic two-derivative action (\ref{twoderaction}),   
  \be\label{generaldelta2}
  \delta  S^{(2)} \ = \  \int  \delta  m^{\,\nin{M}\bar{N}}\
   {\cal R}_{\,\nin{M}\bar{N}}-2\,\delta\phi\,{\cal R}\;. 
 \ee
It is also interesting to note that, up to boundary terms,  
the two-derivative action can be written in terms of connections, 
  \be
  \begin{split}
   {\cal L}^{(2)} \ = \   \f{1}{4}  \,\Gamma^{\,\nin{M} \bar P \bar Q } \Gamma_{\,\nin{M} \bar P \bar Q }  \, + \, \f{1}{2}  \,\Gamma^{\bar M} \Gamma_{\bar M} \;. 
      \end{split}
 \ee  
 
\medskip
There is no $O(D,D)$ covariant Riemann tensor that 
is fully determined in terms of the physical fields  \cite{Siegel:1993th,Hohm:2011si} 
or that even encodes the physical Riemann tensor among undetermined components  \cite{Hohm:2012mf}. 
However, there is a linearized gauge invariant Riemann tensor (that encodes
the linearized physical Riemann tensor for vanishing $b$-field), as noted in 
\cite{Siegel:1993th}.\footnote{This tensor does not have a non-linear completion:  there is no tensor that is covariant under the non-linear (un-deformed) gauge transformations 
of DFT and reduces to it  
upon expansion around a background.}  
It is defined by 
 \be\label{ODDRiemann}
  {\cal R}_{\,\nin{M}\,\nin{N}\bar{K}\bar{L}} \ = \ 2\partial_{[\,\nin{M}}\Gamma_{\,\nin{N}]\bar{K}\bar{L}}
  \ \equiv \ 2\partial_{[\bar{K}}\Gamma_{\bar{L}]\,\nin{M}\,\nin{N}}\ \equiv \ {\cal R}_{\bar{K}\bar{L}\,\nin{M}\,\nin{N}}\;. 
 \ee 
Its explicit form is given by 
 \be
  {\cal R}_{\,\nin{M}\,\nin{N}\bar{K}\bar{L}} \ = \ \partial_{\,\nin{M}}\partial_{\bar{K}}\, m_{\,\nin{N}\bar{L}}
  -\partial_{\,\nin{N}}\partial_{\bar{K}}\, m_{\,\nin{M}\bar{L}}
  -\partial_{\,\nin{M}}\partial_{\bar{L}}\, m_{\,\nin{N}\bar{K}}
  +\partial_{\,\nin{N}}\partial_{\bar{L}}\, m_{\,\nin{M}\bar{K}}\;. 
 \ee 
It is easily seen with (\ref{linnGaugeVar}) or (\ref{deltaZero}) 
that this tensor is indeed gauge invariant.  
The linearized Riemann and Ricci tensor and the curvature scalar satisfy differential Bianchi 
identities,   
 \be\label{linBianchi}
 \begin{split}
  \partial_{\,\nin{M}}{\cal R}^{\,\nin{M}\,\nin{N}\bar{K}\bar{L} } \ &= \ 2\partial^{[\bar{K}}{\cal R}^{|\,\nin{N}|\bar{L}]}\;, 
  \qquad
  \partial_{\bar{M}}{\cal R}^{\bar{M}\bar{N}\,\nin{K}\,\nin{L}} \ = \ -2\partial^{[\,\nin{K}}{\cal R}^{\,\nin{L}\,]\bar{N}}\;, \\
  \partial_{\,\nin{M}}{\cal R}^{\,\nin{M}\bar{N}} \ &= \ \partial^{\bar{N}}{\cal R}\;, \qquad \qquad \qquad 
  \partial_{\bar{M}}{\cal R}^{\,\nin{N}\bar{M}} \ = \ -\partial^{\,\nin{N}}{\cal R}\;. 
 \end{split}
 \ee 
These are easily verified using the definition of these curvatures in terms of connections.

\medskip
We close this subsection by giving an $O(D,D)$ covariant form of the Gauss-Bonnet combination 
(to quadratic order in fields), because this will be important below when relating to the usual 
${\cal O}(\alpha')$ actions of string theory that are conveniently written in terms of Gauss-Bonnet~\cite{Zwiebach:1985uq}. 
Using the linearized $O(D,D)$ covariant curvatures above, 
the Gauss-Bonnet combination is defined by 
  \be\label{ODDcovGB}
   {\rm GB} \ \equiv \   {\cal R}_{\,\nin{M}\,\nin{N}\bar{K}\bar{L}}\,   {\cal R}^{\,\nin{M}\,\nin{N}\bar{K}\bar{L}} 
   +4\, {\cal R}_{\,\nin{M}\bar{N}}\,  {\cal R}^{\,\nin{M}\bar{N}} +4\, {\cal R}^2\;. 
  \ee 
This combination is a total derivative (as is the conventional Gauss-Bonnet combination at the 
quadratic level). Indeed, we can write   
   \be\label{GBTerm}
   {\rm GB}
   \ = \ \partial_{\,\nin{M}}B^{\,\nin{M}}+\partial_{\bar{M}}B^{\bar{M}}\;, 
  \ee  
where 
 \be
  \begin{split}
   B^{\,\nin{M}} \ &= \ \Gamma_{\,\nin{N}\bar{K}\bar{L}}\,{\cal R}^{\,\nin{M}\,\nin{N}\bar{K}\bar{L}} 
   +2\Gamma^{\bar{K}\,\nin{M}\,\nin{N}}\,{\cal R}_{\,\nin{N}\bar{K}}
   -2\Gamma_{\bar{N}} \,{\cal R}^{\,\nin{M}\bar{N}}+2\Gamma^{\,\nin{M}}\,{\cal R}\;, \\
   B^{\bar{M}} \ &= \  \Gamma_{\bar{N}\,\nin{K}\,\nin{L}}\,{\cal R}^{\bar{M}\bar{N}\,\nin{K}\,\nin{L}} 
   -2\Gamma^{\,\nin{K}\bar{M}\bar{N}}\,{\cal R}_{\,\nin{K}\bar{N}}
   +2\Gamma_{\,\nin{N}}\, {\cal R}^{\,\nin{N}\bar{M}}+2\Gamma^{\bar{M}}\,{\cal R}\;. 
  \end{split}
 \ee 
In order to check that the divergence of these 
vectors leads to the Gauss-Bonnet combination (thereby proving that the latter is a total derivative)
one has to use repeatedly the Bianchi identities (\ref{linBianchi}). 
  
It is instructive to investigate the gauge transformations of $ B^{\,\nin{M}} $ and $B^{\bar{M}}$, because 
they play a role analogous to the Chern-Simons three-forms whose exterior 
derivatives define the conventional Gauss-Bonnet term ${\rm tr}(R\wedge R)$. 
These Chern-Simons forms are not gauge invariant 
but transform into exterior derivatives, and it is interesting to find the DFT analogue of this fact.   
Using again the Bianchi identities, one finds that the gauge variations of the $B^M$ 
can be written as 
 \be
  \begin{split}
   \delta_{\xi} B^{\,\nin{M}} \ = \ \partial_{\bar{N}}C^{\,\nin{M}\bar{N}}+\partial_{\,\nin{N}}C^{\,\nin{M}\,\nin{N}}\;, \qquad 
   \delta_{\xi}B^{\bar{M}} \ = \   -\partial_{\,\nin{N}}C^{\,\nin{N}\bar{M}}-\partial_{\bar{N}}C^{\bar{M}\bar{N}}\;, 
 \end{split}
 \ee
 where 
 \be
  \begin{split}
   C^{\,\nin{M}\bar{N}} \ &= \ -4K^{\,\nin{M}\,\nin{K}}\,{\cal R}_{\,\nin{K}}{}^{\bar{N}}+4K^{\bar{N}\bar{K}}\,
   {\cal R}^{\,\nin{M}}{}_{\bar{K}}\;, \\
   C^{\,\nin{M}\,\nin{N}} \ &= \ 2K_{\bar{K}\bar{L}}\,{\cal R}^{\,\nin{M}\,\nin{N}\bar{K}\bar{L}}-4K^{\,\nin{M}\,\nin{N}}\,{\cal R}\;, \\
   C^{\bar{M}\bar{N}} \ &= \ 2K_{\,\nin{K}\,\nin{L}}\,{\cal R}^{\bar{M}\bar{N}\,\nin{K}\,\nin{L}}-4K^{\bar{M}\bar{N}}\,{\cal R}\;. 
  \end{split}
 \ee 
As $C^{\,\nin{M}\,\nin{N}}$  and $C^{\bar{M}\bar{N}}$ are by definition antisymmetric, this makes it manifest 
that the total divergence of the $B^M$ is gauge invariant.

\subsection{Cubic action for DFT at order $\alpha'$}
We now turn to the construction of the cubic action to first order in $\alpha'$, i.e., with 
four derivatives.  We will denote this action by $S^{(3,4)}$ where the first superscript denotes the number of fields and the second the number of derivatives. 
For DFT${}^+$ we will call this action $S^{(3,4)}_+$ and 
for DFT${}^-$ we will call it  $S^{(3,4)}_-$.  
The quadratic action, which is known, is written as $S^{(2,2)}$; it has two fields and two derivatives. 
The cubic action $S^{(3,4)}$ is determined by gauge invariance, which to this order in fields requires  
 \be\label{S3Inv}
  \delta^{[1](1)}S^{(2,2)}+\delta^{[0]}S^{(3,4)} \ = \ 0\;,   
 \ee
where we recall that the superscripts on $\delta$ indicate the number of fields
in brackets and the power of $\alpha'$ in parenthesis. 
Here we assumed 
that the action does not contain  
terms quadratic in fields with four derivatives. This assumption is justified, 
because one can always choose a field basis in which the curvature-squared invariants enter 
in the Gauss-Bonnet combination, which reduces to a total derivative at the quadratic level. 
In fact, in CSFT there are no such terms. 
Sometimes it may be more convenient to work with another field basis, and we will return to this 
case below. 
 
Let us now discuss the invariance condition in a little more detail. 
It turns out to be convenient to write 
the variation of order $\alpha'$ in terms of  linearized connections. In fact, 
(\ref{physicalgaugenew}) can be written as 
 \be\label{delta1Gamma}
  \delta_{\xi}^{[1](1) \sigma} 
  m_{\,\nin{M}\bar{N}} \ = \ \f{1}{2}\big(\partial_{\,\nin{M}}K^{\,\nin{K}\,\nin{L}}\,
  \Gamma_{\bar{N}\,\nin{K}\,\nin{L}}\, -\,\sigma\, \partial_{\bar{N}}K^{\bar{K}\bar{L}}\,\Gamma_{\,\nin{M}\bar{K}\bar{L}}\big)\;, 
 \ee 
with $\sigma=+1$ for DFT$^{+}$ and $\sigma=-1$ for DFT$^{-}$. 
In constructing the cubic action it is sufficient to focus on one projection of the gauge parameter, 
provided the action has a definite $\mathbb{Z}_2$ parity. Indeed, for DFT${}^+$  
gauge invariance under 
$\underline{\xi}$ implies gauge invariance under $\bar{\xi}$:
 \be\label{plusinvariance}
  \delta_{\underline{\xi}}^{[1](1) +}\, S^{(2,2)} + \delta_{\underline{\xi}}^{[0]}\, S_{+}^{(3,4)} \ = \ 0\quad \xrightarrow{\mathbb{Z}_2} \quad 
  \delta_{\bar{\xi}}^{[1](1) +}\,S^{(2,2)}+\delta_{\bar{\xi}}^{[0]}\,S_{+}^{(3,4)} \ = \ 0\;.
 \ee 
Similarly, for DFT$^{-}$ we have 
 \be\label{minusinvariance}
  \delta_{\underline{\xi}}^{[1](1)-}\, S^{(2,2)} + \delta_{\underline{\xi}}^{[0]}\, S_{-}^{(3,4)}
   \ = \ 0\quad \xrightarrow{\mathbb{Z}_2} \quad 
  -\delta_{\bar{\xi}}^{[1](1)-}\,S^{(2,2)} -\delta_{\bar{\xi}}^{[0]}\, 
 S_{-}^{(3,4)}  \ = \ 0\;. 
 \ee 
As before,  $\bar{\xi}$ invariance follows from $\underline{\xi}$ invariance. 
More generally, given cubic, four-derivative actions $S_{-}^{(3,4)}$ and $S_{+}^{(3,4)}$ 
we can construct an invariant action for linear combinations. In fact, the gauge transformations with parameters
$\gamma^+$ and $\gamma^-$ in (\ref{physicalgauge-ef-newGamma}) are equivalent to 
 \be
  \delta_{\xi}^{[1](1)\gamma} \ = \ \gamma^+\,\delta_{\xi}^{[1](1) +} + \gamma^-\,\delta_{\xi}^{[1](1) -}\;.
 \ee
Then the cubic action 
 \be\label{generalgammaaction}
  S_{\gamma}^{(3,4)}   
   \ \equiv \ \gamma^+\,S_{+}^{(3,4)} + \gamma^-\, S_{-}^{(3,4)} \;, 
 \ee
leads to a gauge invariant action as a direct consequence of (\ref{plusinvariance}) and (\ref{minusinvariance}). 

We now discuss the specific construction for DFT$^{-}$. As explained
 above, it is sufficient to focus on, say,  
the $\underline{\xi}$ variation, which is given by 
\be\label{delta1Gamma-under}
 \delta_{\,\nin{\xi}}^{[1](1)} m_{\,\nin{M}\bar{N}} \ = \ \half\,\partial_{\,\nin{M}}K^{\,\nin{K}\,\nin{L}}\,
  \Gamma_{\bar{N}\,\nin{K}\,\nin{L}} \, . 
 \ee 
(Note that, as long as we ignore $\bar{\xi}$, $\delta^{+}$ and $\delta^{-}$ coincide.)  
Inserting this variation into the general form (\ref{generaldelta2}) we compute 
 \be
 \label{delta1s2}
 \  \delta_{\nin{\xi}}^{[1](1)} S^2 \ = \ -\half \,\partial^{\,\nin{M}}K^{\,\nin{P}\,\nin{Q}}\Gamma^{\bar{N}}{}_{\,\nin{P}\,\nin{Q}}
  \big(\partial^{\bar{K}}\Gamma_{\,\nin{M}\bar{K}\bar{N}}+\partial_{\,\nin{M}}\Gamma_{\bar{N}}\big)\;, 
 \ee 
using the (second) definition of the linearized Ricci tensor in (\ref{linRicci}). 
In order to determine the cubic action we have to find cubic coupling whose $\delta^{[0]}$ variations cancel 
these terms.  It turns out that these terms can be naturally written in terms of the connections (\ref{lincoNN}). 
After some 
manipulations, discarding total derivatives and using the strong constraint, 
one can show that the cubic couplings are  
  \be\label{oddcubics}
   \begin{split}
  S^{(3,4)}_- \ = \ -\,  \f{1}{8}  \big(&\,\Gamma^{\,\nin{P}\bar{M}\bar{N}}\,
    \Gamma_{\bar{M}}{}^{\,\nin{K}\,\nin{L}}\ \partial_{\,\nin{P}}\Gamma_{\bar{N}\,\nin{K}\,\nin{L}}
  \  +\ \Gamma^{\bar{P}\,\nin{M}\,\nin{N}}\,
    \Gamma_{\,\nin{M}}{}^{\bar{K}\bar{L}}\ \partial_{\bar{P}}
    \Gamma_{\,\nin{N}\bar{K}\bar{L}}  \ \ \\[1ex]
    &-\Gamma^{\bar{M}}{}_{\,\nin{K}\,\nin{L}}\,\Gamma^{\bar{N}\,\nin{K}\,\nin{L}}\,\partial_{\bar{M}}\Gamma_{\bar{N}}
  \   - \ \Gamma^{\,\nin{M}}{}_{\bar{K}\bar{L}}\,\Gamma^{\,\nin{N}\bar{K}\bar{L}}\,\partial_{\,\nin{M}}\Gamma_{\,\nin{N}}\ \big)\;. 
   \end{split}
  \ee  
With the $\mathbb{Z}_2$ action (\ref{Z2ACTION}) on the connections and the rules explained in sec.~\ref{Z2section} 
it follows that this action is $\mathbb{Z}_2$ odd. Indeed, in the first line there are five $\eta$ implicit, leading 
to a sign change under $\mathbb{Z}_2$; in the second line there are four $\eta$ implicit, but 
$\mathbb{Z}_2$ acts as $\Gamma_{\bar{M}} \rightarrow - \Gamma_{\,\nin{M}}$, which also leads  
to a sign change.

\subsection{Cubic action for DFT$^{+}$}

We now turn to the cubic action for DFT$^{+}$.  
It can be written in various different forms, all related by total derivatives or 
covariant field redefinitions. Here we give two forms, one for a field-basis with Riemann-squared,
one for the Gauss-Bonnet combination. 

The Riemann-squared case turns out to be a little simpler, so we start with this one. 
We now  have to include an $S^{(2,4)}$ term quadratic in fields and with
four-derivatives, 
namely the square of the Riemann tensor (\ref{ODDRiemann}). The full gauge invariance 
requires 
\be\label{InvarianceVar}
  \delta_{\xi}^{[1](1)+}\,S^{(2,2)}+\delta_{\xi}^{[1](0)}\,S^{(2,4)}+\delta_{\xi}^{[0]}S^{(3,4)} \ = \ 0\;. 
 \ee
A gauge invariant action to this order is then given by 
 \be\label{RiemannSquaredCubic} 
  \begin{split}
   S \ = \ &\, S^{(2,2)}+S^{(3,2)} \\[0.5ex]
   &+ \f{1}{4}\,{\cal R}^{\,\nin{M}\,\nin{N}\bar{K}\bar{L}}\, {\cal R}_{\,\nin{M}\,\nin{N}\bar{K}\bar{L}} 
   + \f{1}{4}\,\phi\, {\cal R}^{\,\nin{M}\,\nin{N}\bar{K}\bar{L}}\, {\cal R}_{\,\nin{M}\,\nin{N}\bar{K}\bar{L}}   \\[0.8ex]
   &-\f{1}{8}\big(\,\Gamma^{\,\nin{P}\bar{M}\bar{N}}\,
    \Gamma_{\bar{M}}{}^{\,\nin{K}\,\nin{L}}\ \partial_{\,\nin{P}}\Gamma_{\bar{N}\,\nin{K}\,\nin{L}}
  \  -\ \Gamma^{\bar{P}\,\nin{M}\,\nin{N}}\,
    \Gamma_{\,\nin{M}}{}^{\bar{K}\bar{L}}\ \partial_{\bar{P}}
    \Gamma_{\,\nin{N}\bar{K}\bar{L}}  \ \  \\[0.5ex]
    &\qquad \; \;-\Gamma^{\bar{M}}{}_{\,\nin{K}\,\nin{L}}\,\Gamma^{\bar{N}\,\nin{K}\,\nin{L}}\,\partial_{\bar{M}}\Gamma_{\bar{N}}
  \   + \ \Gamma^{\,\nin{M}}{}_{\bar{K}\bar{L}}\,\Gamma^{\,\nin{N}\bar{K}\bar{L}}\,\partial_{\,\nin{M}}\Gamma_{\,\nin{N}}\ \big) \\[0.8ex]
  &-\f{1}{2}\,{\cal R}_{\,\nin{M}\,\nin{N}\bar{K}\bar{L}}\,\Gamma^{\bar{K}\,\nin{M}\,\nin{P}}\,\Gamma^{\bar{L}\,\nin{N}}{}_{\,\nin{P}}
  +\f{1}{2}\,{\cal R}_{\,\nin{K}\,\nin{L}\bar{M}\bar{N}}\,\Gamma^{\,\nin{K}\bar{M}\bar{P}}\,\Gamma^{\,\nin{L}\bar{N}}{}_{\bar{P}}
  \\[0.5ex]
  &-\f{1}{2}m_{\,\nin{M}\bar{N}}\,{\cal R}^{\,\nin{M}\,\nin{K}\bar{P}\bar{Q}}\,\partial^{\bar{N}}\Gamma_{\,\nin{K}\bar{P}\bar{Q}}
  +\f{1}{2}m_{\,\nin{M}\bar{N}}\,{\cal R}^{\,\nin{P}\,\nin{Q}\bar{N}\bar{K}}\,\partial^{\,\nin{M}}\Gamma_{\bar{K}\,\nin{P}\,\nin{Q}}\\[0.5ex]
  &+\f{1}{2}\,{\cal R}_{\,\nin{M}\,\nin{N}\bar{K}\bar{L}}\,\partial^{\,\nin{P}}m^{\,\nin{M}\bar{K}}\,\partial_{\,\nin{P}}m^{\,\nin{N}\bar{L}}\;. 
  \end{split}  
 \ee
Here $S^{(2,2)}$ and $S^{(3,2)}$ are the quadratic and cubic couplings of the two-derivative theory,   $S^{(2,4)}$ is the term quadratic in the Riemann tensor, and all remaining terms belong to $S^{(3,4)}$, 
the cubic couplings with four derivatives. 
Note that the explicit form of $S^{(3,2)}$ is not needed for the ${\cal O}(\alpha')$ proof of gauge invariance. 
The gauge invariance can be verified systematically by computing the variation in (\ref{InvarianceVar}) 
and integrating by parts so that all terms appear with an undifferentiated 
gauge parameter $\xi$. These terms have to cancel, 
without any total derivative ambiguities. We have verified this (and in fact constructed (\ref{RiemannSquaredCubic})) 
with the help of a {\tt Mathematica} code.

Next we turn to the field basis with Gauss-Bonnet combination (\ref{ODDcovGB}). 
In this case the quadratic terms with four derivatives  contribute only a boundary term 
and can thus be ignored.   A gauge invariant action to this order is then given by 
   \be\label{gaussbonnetCubic} 
  \begin{split}
   S \ = \ &\, S^{(2,2)}+S^{(3,2)} \\[0.5ex]
   &+ \f{1}{4}\,\phi\, \Big({\cal R}^{\,\nin{M}\,\nin{N}\bar{K}\bar{L}}\, {\cal R}_{\,\nin{M}\,\nin{N}\bar{K}\bar{L}} 
   +4\, {\cal R}_{\,\nin{M}\bar{N}}\,  {\cal R}^{\,\nin{M}\bar{N}} +4\, {\cal R}^2  \Big) \\[0.5ex]
   &-\f{1}{8}\big(\,\Gamma^{\,\nin{P}\bar{M}\bar{N}}\,
    \Gamma_{\bar{M}}{}^{\,\nin{K}\,\nin{L}}\ \partial_{\,\nin{P}}\Gamma_{\bar{N}\,\nin{K}\,\nin{L}}
  \  -\ \Gamma^{\bar{P}\,\nin{M}\,\nin{N}}\,
    \Gamma_{\,\nin{M}}{}^{\bar{K}\bar{L}}\ \partial_{\bar{P}}
    \Gamma_{\,\nin{N}\bar{K}\bar{L}}  \ \  \\[0.5ex]
    &\qquad \; \;-\Gamma^{\bar{M}}{}_{\,\nin{K}\,\nin{L}}\,\Gamma^{\bar{N}\,\nin{K}\,\nin{L}}\,\partial_{\bar{M}}\Gamma_{\bar{N}}
  \   + \ \Gamma^{\,\nin{M}}{}_{\bar{K}\bar{L}}\,\Gamma^{\,\nin{N}\bar{K}\bar{L}}\,\partial_{\,\nin{M}}\Gamma_{\,\nin{N}}\ \big) \\[0.5ex]
  &+4\, m_{\,\nin{M}\bar{N}}\,\partial^{\,\nin{M}}\partial^{\bar{N}}\phi\, \square \phi 
  - 4\, m_{\,\nin{M}\bar{N}}\,\partial^{\,\nin{M}}\partial_{\,\nin{K}}\phi\, \partial^{\bar{N}}\partial^{\,\nin{K}}\phi
  +4\,\square\phi\,\partial^{\,\nin{K}}\phi\,\partial_{\,\nin{K}}\phi\\[1ex]
  &+\partial_{\,\nin{P}}\partial_{\,\nin{Q}}m_{\,\nin{M}\bar{N}}\,\partial^{\,\nin{P}}m^{\,\nin{M}\bar{K}}\,\partial_{\bar{K}}
  m^{\,\nin{Q}\bar{N}}+\partial_{\,\nin{P}}\partial_{\bar{Q}}m_{\,\nin{M}\bar{N}}\,\partial^{\,\nin{P}}m^{\,\nin{K}\bar{N}}\,\partial_{\,\nin{K}}
  m^{\,\nin{M}\bar{Q}}\\[1ex]
  &+\partial_{\,\nin{M}}\partial^{\,\nin{L}}m_{\,\nin{L}\bar{N}}\,\partial^{\bar{N}}m^{\,\nin{K}\bar{P}}\,\partial_{\,\nin{K}}m^{\,\nin{M}}{}_{\bar{P}}-\partial_{\bar{M}}\partial^{\bar{L}}m_{\,\nin{N}\bar{L}}\,\partial^{\,\nin{N}}m^{\,\nin{P}\bar{K}}\,\partial_{\bar{K}}m_{\,\nin{P}}{}^{\bar{M}}\\[0.5ex]
  &-\f{1}{2}\, \partial^{\,\nin{M}}\partial^{\bar{N}}m_{\,\nin{M}\bar{N}}\,
  \partial^{\,\nin{K}}m_{\,\nin{K}\bar{P}}\,\partial_{\,\nin{L}}m^{\,\nin{L}\bar{P}}
  +\frac{1}{2}\, \partial^{\,\nin{M}}\partial^{\bar{N}}m_{\,\nin{M}\bar{N}}\,
  \partial^{\bar{K}}m_{\,\nin{P}\bar{K}}\,\partial_{\bar{L}}m^{\,\nin{P}\bar{L}}\\[0.5ex]
  &+\f{1}{2}\,\partial^{\,\nin{M}}\partial^{\bar{N}}m_{\,\nin{M}\bar{N}}\,\partial_{\,\nin{K}}m_{\,\nin{L}\bar{P}}\,
  \partial^{\,\nin{K}}m^{\,\nin{L}\bar{P}} \\[0.5ex]
  &+\f{1}{2}\,{\cal R}_{\,\nin{M}\,\nin{N}\bar{K}\bar{L}}\,\partial^{\,\nin{P}}m^{\,\nin{M}\bar{K}}\,\partial_{\,\nin{P}}m^{\,\nin{N}\bar{L}}\;.  
  \end{split}  
 \ee
Note that we obtained cubic couplings of the form dilaton times Gauss-Bonnet. This is consistent with
the conventional spacetime action of ${\cal O}(\alpha')$  in string frame where such terms arise.  
Again, we proved the gauge invariance condition (\ref{S3Inv}) by computing the variation
and integrating by parts to show that all terms cancel. 

Let us stress that there is a large field-redefinition ambiguity and total derivative
ambiguitiy,  
 so the forms given in (\ref{RiemannSquaredCubic})
and (\ref{gaussbonnetCubic}) are not unique. The Riemann-squared completion in 
(\ref{RiemannSquaredCubic}) takes a `semi-geometric' form, written in terms of (linearized) 
connections and curvatures. We did not manage to find a similarly geometric form of (\ref{gaussbonnetCubic}). 
It would be interesting, however, 
to further elucidate the geometrical content of this action, thus arriving 
at a DFT-extended form of the Gauss-Bonnet action discussed at the linearized level in sec.~\ref{ConnCurvGB}. 

\medskip

We close this section by briefly mentioning the `interpolating' heterotic case (for vanishing gauge vectors). 
The corresponding action is given by (\ref{generalgammaaction}) with both $\gamma^+$
and $\gamma^-$ switched on, thus containing  a linear combination of the cubic action (\ref{oddcubics})
and, depending on the field basis, (\ref{RiemannSquaredCubic}) or (\ref{gaussbonnetCubic}). 
Dropping $\tilde{\partial}$ derivatives and writing the action in terms of conventional 
perturbative variables, it encodes both a Riemann-squared term and the gravitational Chern-Simons modification 
of the $b$-field field strength.

\sectiono{Conclusions and Outlook} \label{concl}
In this paper we have developed DFT$^+$, the double field theory for bosonic string theory to 
first order in $\alpha'$ and compared it to the `doubled $\alpha'$ geometry'  in  \cite{Hohm:2013jaa}. 
As reviewed here and discussed in more detail in \cite{Hohm:2014eba}, 
the latter theory, DFT$^-$, has elements of heterotic string theory. Indeed, the gauge algebra for 
DFT$^+$ 
differs from that for DFT$^{-}$.  
We computed the gauge algebra for the cubic DFT$^{+}$ from closed string field theory to first order 
in $\alpha'$. Then we computed  
the gauge transformations that close according to this gauge algebra and determined the cubic action. 
While the cubic action for DFT$^{-}$ describes (part of)  
the Chern-Simons modifications of the three-form curvature needed for Green-Schwarz 
anomaly cancellation, the cubic action for  DFT$^{+}$ describes the T-duality invariant 
extension of the Riemann-squared term that is known to appear in bosonic string theory. 
The claim that DFT$^{+}$ encodes Riemann-squared requires a justification.\footnote{The fact that the cubic action (\ref{RiemannSquaredCubic}) includes the square of the 
linearized Riemann tensor does not suffice: up to field redefinitions this term may be 
replaced by the Gauss-Bonnet combination (\ref{GBTerm}), which is a total derivative~\cite{Zwiebach:1985uq}.}  
Therefore we summarize in the following three independent arguments that imply this result: 
\begin{itemize}
 \item[(1)] As explained in sec.~\ref{sectionTdualRiemann}, 
 writing the cubic terms of Riemann-squared in a T-duality invariant 
 way requires a non-covariant field redefinition  \cite{Hohm:2011si}. This leads to modified 
 diffeomorphism transformations that agree with the 
 gauge transformations of DFT$^{+}$. 
 \item[(2)] The results 
 in \cite{Meissner:1996sa} imply that, upon reduction to one dimension, writing 
 the ${\cal O}(\alpha')$ terms in bosonic string theory
 in an $O(d,d)$ covariant way  requires field redefinitions 
 that are in quantitative agreement with those discussed under (1). 
 \item[(3)] The gauge algebra (and therefore, indirectly, the gauge transformations) of DFT$^{+}$
 have been determined from bosonic closed string field theory and thus must lead to a theory 
 that encodes the known Riemann-squared correction. In fact, taken together with the 
 results under (1) our analysis determined the coefficient of the Riemann-squared term 
 as predicted from string field theory and agrees perfectly with the literature.  

\end{itemize}
A final observation supporting the conclusion that DFT$^{+}$ describes Riemann-squared 
is that the cubic actions (\ref{RiemannSquaredCubic}) or (\ref{gaussbonnetCubic}) 
contain cubic couplings involving the dilaton times Riemann-squared or Gauss-Bonnet, 
exactly as expected for the cubic couplings of the ${\cal O}(\alpha')$ terms in the string frame.

So far we constructed only the cubic action for DFT$^{+}$ or, more generally, 
for the interpolating theory relevant for heterotic string theory, describing both gravitational 
Chern-Simons modifications and Riemann-squared. 
It is clearly desirable to construct the background-independent theory, i.e., to 
all orders in fluctuations. As the DFT$^{+}$ gauge algebra (\ref{Z-algebra-original+}) is 
background-dependent this requires a further extension to a \textit{field-dependent} gauge algebra. Most 
likely, this extension goes beyond replacing the background generalized metric  
by the full generalized metric. 

Very recently an interesting   
proposal appeared \cite{Bedoya:2014pma}  
that aims to describe the complete ${\cal O}(\alpha')$ corrections of heterotic string theory 
in DFT.  
It starts from the heterotic DFT \cite{Siegel:1993th,Siegel:1993bj,Hohm:2011ex,Hohm:2011ex2} that incorporates 
$n$ gauge vectors in an enlarged generalized metric taking values in 
$O(D,D+n)$. The theory is  
defined on a further extended space with $n$ new coordinates and subject to additional constraints. 
By declaring part of the connections to be (torsionful) Lorentz connections 
one obtains the desired ${\cal O}(\alpha')$ corrections of heterotic string theory as in  \cite{Bergshoeff:1989de}.
It is asserted  in \cite{Bedoya:2014pma} that this procedure leads to an $O(D,D)$ covariant result.  
It would be interesting to investigate if there is a relation to the recent constructions in \cite{Polacek:2013nla} 
that also extend further the coordinates  
to encode  Lorentz algebra directions.

\section*{Acknowledgments}
This is work is supported by the 
U.S. Department of Energy (DoE) under the cooperative 
research agreement DE-FG02-05ER41360. 
The work of O.H. is supported by a DFG Heisenberg fellowship. 
We thank Marc Henneaux, Chris Hull, Dieter L\"ust and Warren Siegel 
for discussions. We are particularly indebted to Ethan Dyer for 
assistance with programming the Mathematica code needed 
for the construction of the cubic actions. 
 Finally, 
we happily acknowledge
several useful comments and suggestions by Ashoke Sen.


\end{document}